\documentclass[twocolumn, emulateapj]{aastex61}
\usepackage[utf8]{inputenc}
\usepackage{gensymb}

\usepackage{natbib}
\usepackage{soul}
\usepackage{float}
\pdfoutput=1

\newcommand{\HaLbol}{$\left[L_{\mathrm{H_{\alpha} }}/L_{\mathrm{bol}}\right]$}

\begin{document}

\title{The Strongest Magnetic Fields on the Coolest Brown Dwarfs}

\author{Melodie M. Kao}
\affil{California Institute of Technology, Department of Astronomy, 1200 E California Blvd, MC 249-17, Pasadena, CA 91125, USA}
\affil{Arizona State University, School of Earth and Space Exploration, 550 E Tyler Mall, PSF 686, Tempe, AZ 85287, USA}
\email{mkao@asu.edu}
\author{Gregg Hallinan}
\affil{California Institute of Technology, Department of Astronomy, 1200 E California Blvd, MC 249-17, Pasadena, CA 91125, USA}
\author{J. Sebastian Pineda}
\affil{University of Colorado Boulder, Laboratory for Atmospheric and Space Physics, 3665 Discovery Drive, Boulder CO 80303, USA}
\author{David Stevenson}
\affil{California Institute of Technology, Division of Geological \& Planetary Sciences,1200 E California Blvd, MC 150-21, Pasadena, CA 91125, USA}
\author{Adam Burgasser}
\affil{University of California San Diego, Center for Astrophysics and Space Sciences, 9500 Gilman Drive, MC 0424, La Jolla, CA 92093, USA}

\begin{abstract}
We have used NSF's Karl G. Jansky Very Large Array (VLA) to observe a sample of five known radio-emitting late L and T dwarfs ranging in age from $\sim$0.2--3.4~Gyr. We observed each target for seven hours, extending to higher frequencies than previously attempted and establishing proportionally higher limits on maximum surface magnetic field strengths.  Detections of circularly polarized pulses at 8--12~GHz  yield measurements of 3.2--4.1~kG localized magnetic fields on four of our targets, including the archetypal cloud variable and likely planetary-mass object T2.5 dwarf SIMP~J01365663+0933473. We additionally detect a pulse at 15--16.5 GHz for the T6.5 dwarf 2MASS~10475385+2124234, corresponding to a localized 5.6~kG field strength.  For the same object, we tentatively detect a 16.5--18~GHz pulse, corresponding to a localized 6.2~kG field strength. We measure rotation periods between $\sim$1.47--2.28~hr for 2MASS~J10430758+2225236, 2MASS~J12373919+6526148, and SDSS~J04234858-0414035, supporting (i) an emerging consensus that rapid rotation may be important for producing strong dipole fields in convective dynamos and/or (ii) rapid rotation is a key ingredient for driving the current systems powering auroral radio emission.  We observe evidence of variable structure in the frequency-dependent time series of our targets on timescales shorter than a rotation period, suggesting a higher degree of variability in the current systems near the surfaces of brown dwarfs. Finally, we find that age, mass, and temperature together cannot account for the strong magnetic fields produced by our targets.
 
\end{abstract}

\section{Introduction}\label{sec.Intro_15a374}

Characterizing magnetic fields in the coolest dwarfs and eventually exoplanets can provide valuable insight into the formation, emission, and evolution of planets through stars. For instance, they are key players in disk accretion onto pre-main-sequence T~Tauri stars \citep{hartmann2016}, affecting planet formation mechanisms.  Plasma flow across magnetic field lines drive large-scale currents in brown dwarf and planetary systems, producing auroral emission that likely contributes to the optical and infrared variability traditionally attributed to atmospheric clouds \citep[e.g.][]{artigau2009, radigan2014a, hallinan2015, badman2015, kao2016}. Magnetic fields have been invoked to explain fundamental properties such as inflated radii in planets and stars \citep{batygin2010, kervella2016}.  Finally, they can mitigate the erosion of planetary atmospheres from strong stellar winds and coronal mass ejections, a special concern for planets in the habitable zones of M dwarfs and young stars \citep{vidotto2013, brain2015, leblanc2015}.

To characterize such magnetic fields, it is important to understand the physical principles driving field generation in fully convective objects, which remains an open question in dynamo theory.  Applications of convective dynamos span a wide breadth of cases, including rocky planet inner cores, gas giant planets, brown dwarfs, and low-mass stars. Fully convective objects cannot rely on strong differential rotation occurring between radiative and convective zones to help drive their dynamos.  However they still exhibit magnetic activity like H$\alpha$, X-ray, and radio emission \citep[e.g.][]{berger2001, burgasser2003_redOpticalData, berger2005, mclean2012, schmidt2015, pineda2016}, and kilogauss fields have been confirmed for M, L, and T dwarfs \citep[e.g.][]{reinersBasri2007, reinersBasri2009, morin2010, hallinan2006, hallinan2007, hallinan2008, route2012, routeWolszczan2016, kao2016, shulyak2017}. Turbulence dissipates fossil fields within $\sim$10--100~years \citep{chabrierKuker2006}, implying that a dynamo must continuously regenerate these strong fields.

Efforts to elucidate magnetic behaviors of fully convective objects have included many fruitful investigations into the role of rotation.  For instance, H$\alpha$ and X-ray emission are both tracers of hot chromospheres and coronae in F through mid-M stars heated in part by magnetic processes \citep{vernazza1981, schmittRosso1988, ulmschneider2003}.  Rotation appears to affect such magnetic processes, as H$\alpha$ and X-ray emission scale with increasing surface rotation or decreasing Rossby\footnote{Quantified as $\mathrm{Ro} \sim P/\tau_c$, where $P$ is the stellar rotation period and $\tau_c$ is the convective turnover time.} number Ro, which measures the effect of the Coriolis force in the inertial part of the fluid flow (the convective time derivative of velocity). At $\mathrm{Ro}\sim 0.1$, the activity-rotation scaling appears to saturate at a constant $\log L_{\mathrm{X, H\alpha}} / L_{\mathrm{bol}}$ \citep{mclean2012}, indicating a possible saturation of the influence of rotation on dynamo activity in mid-M and earlier type dwarfs.  However, the neutral atmospheres of dwarfs $\gtrsim$M7 may preclude magnetic heating processes of similar nature from occurring in the coolest brown dwarfs \citep{mohanty2002}, underscoring the need for an alternative way to evaluate magnetism on the coolest brown dwarfs.

Indeed, $\gtrsim$M7 dwarfs exhibit systematically weaker H$\alpha$ emission while $L_{\mathrm{X}} / L_{\mathrm{bol}}$ decreases with increasing $v\sin i$ or decreasing Ro \citep{mohantyBasri2003, reinersBasri2008, reinersBasri2010, berger2010, mclean2012}, and the G{\"u}del-Benz relation appears to break down for objects later than M7 due to a suppression of X-ray luminosities, even when taking activity-rotation saturation into account \citep{berger2010, williams2014}.  The precipitous drop-off of X-ray emission from M7 and later objects indicate that such objects lack hot coronae.  Consequently, previously established relationships between magnetic flux and tracers of coronal and chromospheric magnetic activity may not apply.  This calls for comparisons of direct magnetic field measurements rather than observational proxies to rotation rates.  Pulsing radio brown dwarfs in particular provide a rich probe of rotationally dependent magnetism, since their radio emission frequencies map to field strengths, while rotational modulation of the emission can provide rotation period measurements.   

Models explore how different parameters quantifying competing forces such as Lorentz, buoyancy, and Coriolis affect energy exchange mechanisms at play in the magnetohydrodynamics occurring in fully convective dynamo regions.  These models observe various dependencies between global magnetic field behaviors such as field topologies, magnetic energy, and time variation to observable object parameters such as luminosity, rotation, and age \citep[e.g.,][]{browning2008, christensen2009, gastine2013, yadav2016}.  Testing them requires a means to probe magnetism in the coolest objects: planets and brown dwarfs.  

The unexpected detection of quiescent and flaring radio emission from the M9 brown dwarf  LP~944-20 at 4.9 and 8.5~GHz with NSF's Karl G. Jansky Very Large Array (VLA) at the beginning of this millennium heralded an unexpected new window into brown dwarf magnetism \citep{berger2001}.  This discovery paved the way to the subsequent detection of rotationally modulated and highly circularly polarized radio pulses attributed to the electron cyclotron maser (ECM) instabilty \citep{hallinan2006, hallinan2007}, which is the same process driving auroral radio emissions in the magnetized Solar System planets \citep{zarka1998}.

The identification of auroral ECM emission from brown dwarfs was a crucial step to probing magnetic field strengths on the coolest brown dwarfs.  For cool brown dwarfs with largely neutral atmospheres where collisions are negligible (the ratio of the plasma frequency to the electron cyclotron frequency is very small), emission occurs very near the electron cyclotron fundamental frequency $\nu_{\mathrm{MHz}} \sim 2.8 \times B_{\mathrm{Gauss}}$ \citep[and references therein]{treumann2006}.  While auroral ECM emission cannot provide detailed insight into global magnetic field properties and its absence does not necessarily imply the absence of strong magnetic fields, detections provide powerfully direct measurements of field strengths at emitting regions within the magnetosphere. 

In contrast, magnetic field measurements from the Zeeman broadening of magnetically sensitive spectral lines can return filling factor and surface-averaged field strengths with $\sim$15\%--30\% uncertainties \citep{valenti1995,johnskrullValenti1996, johnskrull2000ASPC, reinersBasri2007, reiners2012, shulyak2010}.  Zeeman Doppler imaging adds the ability to spatially distinguish different regions of different field strengths and reconstruct surface field topologies by fitting spectropolarimetric observations to those synthetically generated from test magnetic maps.   Structure of opposite polarity on scales smaller than a spatial resolution element can cancel out, so ZDI is preferentially sensitive to the largest scales \citep{reinersBasri2009, yadav2015}, with significant confusion between the dipole and quadrupole components, and $\sim$10--30\% uncertainties in dipole energies \citep{morin2010}.  Observations only probing some and not all of the Stokes parameters are further constrained in their abilities to fully capture complex field topologies \citep{rosen2015}.  Finally, known Land\'e factors remain limited and prevent Zeeman broadening and ZDI techniques from accessing L and later dwarfs \citep{berdyugina2002, shulyak2010}.

While auroral ECM emission is likely only sensitive to large-scale fields, a careful interpretation of the measurements allows for comparison to Zeeman broadening measurements and paves the way to extending observational tests of fully convective dynamos to the coolest brown dwarfs \citep{kao2016}.

However, efficient detection of brown dwarf auroral radio emission eluded astronomers for over a decade, with an overall detection rate of just $\sim$10\% in previous volume-limited surveys \citep{antonova2013, route2016}.  Moreover, only one detection out of $\sim$60 L6 or later targets had been achieved before 2016 \citep{route2012}, seriously hindering the application of ECM emission to testing dynamos mechanisms in the mass and temperature gap between planets and stars.  Yet, the unprecedented discovery of a T6.5 dwarf emitting at $\sim$4~GHz demonstrated that such emission could indeed extend to objects probing the substellar-planetary boundary \citep{route2012}.

We previously developed and tested a selection strategy for identifying likely ECM-emitting brown dwarf candidates by making use of an emerging connection between ECM emission and possible tracers of aurora \citep{kao2016}.  We selected targets with known H$\alpha$ emission and/or optical/infrared variability, leading to the detection of ECM emission in four out of five new L7--T6.5 brown dwarf pilot targets at 4--8~GHz, confirming $>$2.5~kG magnetic fields.  A subsequent study confirmed detectable levels of H$\alpha$ emission for all but one of these targets \citep[e.g.][]{burgasser2003_redOpticalData,pineda2016}.  

The addition of this collection of radio brown dwarf magnetic field measurements to the single previous measurement from the T6.5 dwarf 2MASS~10475385+2124234 \citep{route2012, williams2015} provided strong observational evidence that very cold brown dwarfs can generate kilogauss fields, as well as a means for initial tests of dynamo theory at $\sim$1000 K temperatures.  Comparisons of ECM-derived magnetic field measurements to Zeeman-based measurements tentatively suggested that dynamos operating in the coolest brown dwarfs may in fact produce fields that differ from values predicted by the luminosity-driven \citet{christensen2009} model. 

Higher frequency measurements of these objects can provide yet tighter constraints, motivating this work.  Observations of ECM auroral emissions in the solar system planets demonstrate that the emission drops off sharply at a cutoff frequency corresponding to the strength of the field near the surface of the object.  The persistence of highly circularly polarized and pulsing emission in our targets throughout the previously observed 4--8 GHz bandwidth suggested that the emitting electrons were still traversing the magnetospheres of our targets toward increasing magnetic flux.  A detection of a cutoff in the ECM emission would provide the tightest radio-derived constraints on brown dwarf magnetic fields, and in fact none has yet been detected in any brown dwarfs to date.  

Finally, the rotational modulation of auroral ECM emission provides a means of measuring rotational periods and eventually testing dynamo models that examine the role of rotation by observing our known auroral radio emitters for longer time blocks to achieve full rotational phase coverage. Previous studies verified that pulse periods are consistent with rotational broadening from spectral lines \citep{berger2005, hallinan2006, hallinan2008, berger2009}.

In this work, we present new 8--12~GHz and 12--18~GHz observations of targets detected in our previous 4--8~GHz pilot survey (\S\ref{sec.Observations_15a374}, \S\ref{sec.Imaging_15a374}).  We carefully trace the evolution of auroral ECM pulses through 1 or 1.5~GHz sub-bands (\S\ref{sec.Timeseries_15a374}, \S\ref{sec.Cutoff_15a374}) and measure rotation periods (\S\ref{sec.Rotation_15a374}).  Finally, we comment on implications for dynamo theory (\S\ref{sec.Discussion_15a374}).

\section{Targets}\label{sec.Targets_15a374}

\setlength{\tabcolsep}{0.05in}
\begin{deluxetable*}{lllllllll}[htp]
\tabletypesize{\footnotesize}
\tablecaption{Survey Targets \label{table:properties_15a374}}
\tablehead{
	\colhead{Object Name}               &
	\colhead{Abbrev.}                   & 
	\colhead{SpT}                       &
	\colhead{Parallax}                  &
	\colhead{Distance}                  &
	\colhead{$\mu_{\alpha}\cos\delta$}  &
	\colhead{$\mu_{\delta}$}            &
	\colhead{Notes}                     &
	\colhead{Ref. \tablenotemark{a}}                  
	    \\
	\colhead{}                          & 	
	\colhead{Name}                      & 
	\colhead{}                          &
	\colhead{(mas)}                     &	
	\colhead{(pc)}                      &
	\colhead{(mas yr$^{-1}$)}           &  
	\colhead{(mas yr$^{-1}$)}           &
	\colhead{}                          &
	\colhead{}                     
}
\startdata	
2MASS~10475385+2124234 & 2M1047   & T6.5 				& \phn94.73$\pm$3.81& 10.56\phn$\pm$0.52\phn		& \phn\phn-1714\phd$\pm$7       & -489\phd$\pm$4			& H$\alpha$, detected prior					& \textbf{1} \textit{2} \uline{3} 4--8	\\ 
SIMP~J01365662+0933473 & SIMP0136 & T2.5 				& 162.32$\pm$0.89	& \phn6.139$\pm$0.037			& \phm{-}1222.70$\pm$0.78		& \phn0.5\phd$\pm$1.2		& IR var, no H$\alpha$\tablenotemark{b}		& \textbf{10} \textit{10} \uline{9} 8 11 12 \\ 
2MASS~J10430758+2225236& 2M1043   & L8   				& \nodata   	 		& \phn\phn16.4$\pm$3.2\phn\phn	& \phn\phn-134.7$\pm$11.6       & \phn-5.7$\pm$17.0  		& H$\alpha$ emission 						& \textbf{13} \textit{13} \uline{14} 8 15 		    \\ 
2MASS~J12373919+6526148& 2M1237   & T6.5 				& \phn96.07$\pm$4.78& 10.42\phn$\pm$0.52\phn 		& \phn\phn-1002\phd$\pm$8      	& -525\phd$\pm$6 			& H$\alpha$, IR var?\tablenotemark{c}		& \textbf{1} \textit{16}  \uline{3} 4 16-18 		\\ 
SDSS~J04234858-0414035 & SDSS0423 & L7\tablenotemark{d}	& \phn65.93$\pm$1.7	& 15.17\phn$\pm$0.39\phn		& \phn\phn\phn-331\phd$\pm$49  	& \phm{-}\phn76\phd$\pm$11	& H$\alpha$, IR var, binary\tablenotemark{c}& \textbf{19} \textit{3 20} \uline{8} 21-28 		\\ 
\enddata
\tablenotetext{a}{Citation legend: \textbf{Discovery}; \textit{SpT}; \uline{Parallax, Distance, Proper Motion}; Notes }
\tablenotetext{b}{(8) reported upper limits \HaLbol~$<-6.6$.}
\tablenotetext{c}{(16) and (18) report conflicting evidence of $J$-band variability. }
\tablenotetext{d}{Secondary is spectral type T2.5 at orbital separation 0$\farcs$16 (26, 27, 28). }
\tablerefs{ 
(1) \cite{burgasser1999};
(2) \cite{burgasser2006a};
(3) \cite{vrba2004};
(4) \cite{burgasser2003_redOpticalData}; 
(5) \cite{route2012};
(6) \cite{williams2013};
(7) \cite{williams2015};  
(8) \cite{pineda2016};
(9) \cite{weinberger2016};
(10) \cite{artigau2006}; 
(11) \cite{artigau2009};
(12) \cite{apai2013};
(13) \cite{cruz2007};
(14) \cite{schmidt2010};
(15) \cite{miles2017};
(16) \cite{burgasser2002b};
(17) \cite{burgasser2000b};
(18) \cite{artigau2003}; 
(19) \cite{geballe2002};
(20) \cite{cruz2003};
(21) \cite{kirkpatrick2008};  
(22) \cite{enoch2003};        
(23) \cite{clarke2008};		  
(24) \cite{radigan2014a};     
(25) \cite{burgasser2007};    
(26) \cite{carson2011};		  
(27) \cite{burgasser2005b};   
(28) \cite{burgasser2006b}}   
\end{deluxetable*}

Our sample of targets is discussed in \citet{kao2016} but is again summarized here with updated literature for completeness.  All targets are known to emit ECM emission at 4--8~GHz \citep{kao2016}.

\textbf{2MASS~10475385+2124234}.
2M1047 is a T6.5 dwarf with known weak \HaLbol $\sim-5.5$ \citep{burgasser2003_redOpticalData} and was the first T-dwarf detected at radio frequencies \citep{route2012}.  The detected emission was highly circularly polarized ($\gtrsim$72\%) at 4.75~GHz.  Follow-up observations detected detected both quiescent and ECM emission up to 10~GHz \citep{williams2013, williams2015}, the latter of which was used to measure a $\sim$1.77~hr rotation period up through 10~GHz.  We included 2M1047 in our pilot survey to examine long-term variability and detected both pulsed and quiescent emission through 8~GHz.  Using H$_2$O and $K/H$ indices, \cite{kao2016} derived T$_{\mathrm{eff}} = 869^{+35}_{-29}$~K, $>$0.026~M$_{\odot}$ estimated mass, and $>$2.5~Gyr age.

\textbf{SIMP~J01365662+0933473}.
SIMP0136 is a T2.5 dwarf well known for periodic ($P=2.3895\pm0.0005$~hr) and high-amplitude ($>$5\%) J- and $K_s$-band photometric variability \citep{artigau2009, croll2016}.  High-amplitude infrared variability appears to occur at a higher rate in L/T transition dwarfs \citep{radigan2014a, radigan2014b} and has been attributed to the onset of patchy clouds \citep{ackermanMarley2001, burgasser2002clouds, marley2010, apai2013, radigan2014a} to explain wavelength-dependent variability.  No H$\alpha$ emission has been detected down to \HaLbol~$<-6.6$ but it  has anomalously strong Li~I at EW = $6.6\pm1.0$ and $7.8\pm1.0$~\AA \,for two different nights and is the latest-type object with a clear lithium detection, indicative of a young age \citep{pineda2016}.  \citet{kao2016} derived T$_{\mathrm{eff}} = 1089^{+62}_{-54}$, 0.022$^{+0.015}_{-0.012}$~M$_{\odot}$ estimated mass, and 0.6$^{+1.1}_{-0.3}$~Gyr age.  Recently, \citet{gagne2017} reported that SIMP0136 may be a member of the $\sim$200 Myr-old Carina-Near moving group.  Using an empirical measurement of its bolometric luminosity and the the \cite{saumonMarley2008} models, they inferred $R = 1.22\pm0.01$~R$_{\mathrm{J}}$, which together predicted T$_{\mathrm{eff}} = 1098\pm6$K and $M=12.7\pm1.0$~M$_{\mathrm{J}}$.  New $v \sin i$ measurements and its photometric periodicity further constrained $R>1.01\pm0.02$~R$_{\mathrm{J}}$ and  $M<42.6^{+2.5}_{-2.4}$~M$_{\mathrm{J}}$.

\textbf{2MASS~J10430758+2225236}.
2M1043 is an unusually red L8 dwarf with previously reported tentative H$\alpha$ emission \citep{cruz2007}.  \citet{pineda2016} confirmed \HaLbol~$ = -5.8\pm0.2$ as well as a tentative Li~I absorption line with EW = $10\pm3$~\AA.   \citet{kao2016} derived T$_{\mathrm{eff}} = 1390 \pm 180$~K, 0.011$^{+0.011}_{-0.005}$~M$_{\odot}$ estimated mass, and 0.6$^{+4.6}_{-0.3}$~Gyr age.

\textbf{2MASS~J12373919+6526148}.
2M1237 is a T6.5 dwarf with anomalously hyperactive H$\alpha$ emission at \HaLbol~$\sim -4.2$ \citep{burgasser2000b, burgasser2003_redOpticalData} with conflicting evidence of $J$-band variability \citep{burgasser2002b, artigau2003}.  \citet{kao2016} derived T$_{\mathrm{eff}} = 831^{+31}_{-27}$~K, $>$0.028~M$_{\odot}$ estimated mass, and $>$3.4~Gyr age.

\textbf{SDSS~J04234858-0414035}.
SDSS0423 is an L6/T2 binary with 0$\farcs$16 separation \citep{burgasser2005b, carson2011} and strong H$\alpha$ emission ($\mathrm{EW} = 3$~\AA) and Li~I absorption ($\mathrm{EW}=11$~\AA) \citep{kirkpatrick2008}. \citet{pineda2016} confirmed H$\alpha$ $\mathrm{EW}=2.95\pm 0.3$~\AA\, and Li~I $\mathrm{EW}=11.1\pm 0.4$~\AA.  It additionally exhibits $J$- and $K$-band but no $I_c$ photometric variability \citep{enoch2003,clarke2008, wilson2014}. \citet{kao2016} derived T$_{\mathrm{eff}} = 1678^{+174}_{-137}$~K,  0.015$^{+0.021}_{-0.006}$~M$_{\odot}$ estimated mass, and 0.49$^{+0.62}_{-0.17}$~Gyr age, although these values are uncertain given that they are based on blended light spectra.

\section{Observations}\label{sec.Observations_15a374}
\setlength{\tabcolsep}{0.05in}
\begin{deluxetable*}{lccclccclcc}[htp]
\tablecaption{Summary of observations\label{table:obs_15a374}}
\tablehead{
	\colhead{}                  &
	\colhead{}                  &
	\colhead{Obs.}              &
	\colhead{Obs.}              &
	\colhead{Time on}           &
	\colhead{VLA}               &
	\colhead{Synthesized Beam}  &
	\colhead{}                  &
	\colhead{Phase}             &
	\colhead{Flux}              &
	\colhead{Ref. Set}
	    \\
	\colhead{Object}            &
	\colhead{Band}              &
	\colhead{Date}              &
	\colhead{Block}             &
	\colhead{Source}            & 
	\colhead{Configuration}     &
	\colhead{Dimensions}        &
	\colhead{RMS}               &
	\colhead{Calibrator}        &
	\colhead{Calibrator}        &
	\colhead{Frequency}
	    \\
	\colhead{}                  &
	\colhead{(GHz)}             &
	\colhead{(2015)}            &
	\colhead{(h)}               & 
	\colhead{(s)}               & 
	\colhead{}                  &
	\colhead{(arcsec $\times$ arcsec)}            &
	\colhead{($\mathrm{\mu}$Jy)}& 
	\colhead{}                  &
	\colhead{}                  &
	\colhead{(GHz)}
}
\startdata
2M1047      & 12.0--18.0	& 05/18 & 7.0 & 20870   & BnA & \phn 0\farcs62 $\times$ 0\farcs50     & 1.7 , 1.8    	& J1051+2119 & 3C295 & 14.064\\
SIMP0136    & 8.0--12.0 	& 05/17 & 7.0 & 20870   & BnA & \phn 0\farcs66 $\times$ 0\farcs37     & 1.3 , 1.1		& J0149+0555 & 3C48  & \nodata   \\
2M1043      & 8.0--12.0 	& 05/20 & 7.0 & 20612   & BnA & \phn 0\farcs60 $\times$ 0\farcs33     & 1.0 , 1.0    	& J1051+2119 & 3C295 & 11.064\\
2M1237      & 8.0--12.0 	& 05/18 & 7.0 & 21484   & BnA & \phn 0\farcs69 $\times$ 0\farcs43     & 1.0 , 1.1 		& J1339+6328 & 3C295 & 8.464\\
SDSS0423    & 8.0--12.0 	& 05/30 & 7.0 & 17234   & BnA & \phn 0\farcs68 $\times$ 0\farcs37  	& 1.2 , 1.4		& J0423-0120 & 3C147 & \nodata\\
\enddata
\end{deluxetable*}

We observed four of our sources with previous C band (4--8 GHz) detections at X band (8--12 GHz) and one source (2M1047) which had a previous X band detection at Ku band (12--18 GHz) with the full VLA.  We used the WIDAR correlator in 3-bit observing mode for 4~GHz or 6~GHz bandwidth observations with 2s integrations in 7-hour time blocks for 35 total program hours.  Observations took place during May 2015 in BnA configuration. Table~\ref{table:properties_15a374} and Table~\ref{table:obs_15a374} summarize target properties and observations, respectively.

\subsection{Calibrations}\label{sec.Calibrations_15a374}
\setlength{\tabcolsep}{0.05in}
\begin{deluxetable}{lcccc}[htp]
\tablecaption{Comparison of phase calibrator flux densities\label{table:3C295cal}}
\tablehead{
	\colhead{}            &
	\colhead{Ref. Freq}   &    
	\colhead{Ref. Freq}   &
	\colhead{Ref. Freq}   &
	\colhead{Ref. Freq}   
	    \\
	\colhead{Object}            &
	\colhead{8.464~GHz}         &
	\colhead{11.064~GHz}        &
	\colhead{14.064~GHz}        &
	\colhead{16.564~GHz}        
	    \\
	\colhead{}                  &
	\colhead{(mJy)}             &
	\colhead{(mJy)}             & 
	\colhead{(mJy)}             &
	\colhead{(mJy)} 	           
}
\startdata
2M1047   & \nodata          & \nodata          &   $603.7 \pm 0.4$    &   $561.1 \pm 0.2$  \\ 
2M1043   & $466.4 \pm 1.2$  & $469.0 \pm 1.3$  &   \nodata            &  \nodata            \\ 
2M1237   & $173.3 \pm 1.0$  & $185.0 \pm 1.0$  & \nodata              &  \nodata            \\
\enddata
\end{deluxetable}

For SIMP0136 and SDSS0423, we calibrated our measurement sets using standard VLA flux calibrators 3C48 and 3C147, respectively, and nearby phase calibrators.  Flux calibrators were observed at the beginning and end of each observing block and interpolated.  After initially processing raw measurement sets with the VLA Calibration Pipeline, we manually flagged remaining radio frequency interference (RFI).  Strong time-dependent RFI resulted in $\sim$71 minutes of data loss near the end of the observing block for SDSS0423.  Typical full-bandwidth sensitivity at BnA configuration for 7-hour observing blocks ($\sim$5.5 hours and $\sim$4 hours on source) is 1.2\,$\mu$Jy and 2.1\,$\mu$Jy for X and Ku bands, respectively. Typical 3-bit observations reach an absolute flux calibration accuracy of $\sim$5\% by bootstrapping flux densities with standard VLA flux calibrators.  To correct for flux errors resulting from gain phase variation over our observing window, we alternated between target and phase calibrator integrations, with 15- and 6-minute cycle times for X and Ku bands, respectively.  Our gain solutions varied slowly and smoothly over time and without any ambiguous phase wraps, suggesting that this source of error is negligible.  

For 2M1047, 2M1043, and 2M1237, we observed the flux calibrator 3C295, which is typically recommended only for low-frequency observations in compact configurations.  This calibrator was fully resolved at both X and Ku bands for our observations.  For targets observed at X bands (2M1043 and 2M1237), we modified the VLA scripted pipeline to use A configuration 8.464~GHz and 11.064~GHz model images observed on 02/16/2016 by VLA staff to set flux levels and determine bandpass solutions.  The emission from 3C295 is stable within 1$\%$ over 24--28 years for X and Ku bands \citep{perleyButler2013}.  Because the lobed structure of 3C295 is resolved at our observing frequencies and the VLA sky sensitivity fringes are wavelength-dependent, we expect there to be a discrepancy in flux densities bootstrapped using these different images of 3C295.  To estimate the additional uncertainty in flux densities introduced by calibrating with 3C295, we compared the flux densities of each target's phase calibrator as bootstrapped by the different model images of 3C295.  We list these flux densities in Table~\ref{table:3C295cal}.  These comparisons suggest that the flux densities of 2M1043 and 2M1237 have an additional $\sim$1--7\% uncertainty.  We repeated the same process for our Ku band target (2M1047) but instead used model images of 3C295 at 14.064~GHz and 16.564~GHz, which we expect to introduce an additional $\sim$8\% uncertainty. 

We flagged all data from 12--12.8~GHz during the first $\sim$34 minutes of our target observing scans for 2M1047 due to strong RFI.  After manually flagging remaining RFI, we average all of the measurements sets down in time from 2s integrations to 10s for faster processing.

\subsection{Source Motion\label{sec.Motion_15a374}}
We corrected the 2MASS coordinates \citep{2mass2006} of our targets using the proper motion measurements listed in Table~\ref{table:properties_15a374} to obtain expected source positions.  For the known binary SDSS0423, we did not correct for orbital motion because its 0$\farcs$16 orbital separation is well within the synthesized beam resolution.

\section{Methods}\label{sec.Methods_15a374}
In this section, we describe our general approach to analyzing the data.  In \S \ref{sec.Results_15a374}, we detail specific challenges encountered in the analysis of data for each target.

\setlength{\tabcolsep}{0.03in}
\begin{deluxetable}{lccccc}[t]
\tablecaption{Summary of initial imaging detections\label{table:imageDetections}}
\tablehead{
	\colhead{}                  &
	\colhead{}                  &
	\colhead{}                  & 
	\colhead{}                  &
	\colhead{}          			&
	\colhead{}               
	    \\
	\colhead{Object}            &
	\colhead{RA}   &
	\colhead{Dec}       &
	\colhead{Stokes I}          &
	\colhead{Stokes V}          &
	\colhead{S/N}               
	    \\
	\colhead{}                  &
	\colhead{(hh mm ss.ss)}     &
	\colhead{(dd mm ss.ss)}     &
	\colhead{($\mu$Jy)}         &
	\colhead{($\mu$Jy)}         &
	\colhead{($I$, $V$)}       
}
\startdata
2M1047      & 10 47 51.78 		& +21 24 14.90  & 21.9$\pm$1.3  & 3.9$\pm$1.5   & 16.8, 2.6    \\
SIMP0136    & 01 36 57.86		& +09 33 47.00  & 85.7$\pm$1.3  & -23.8$\pm$ 1.1& 65.9, 21.6    \\
2M1043      & 10 43 07.44		& +22 25 23.31  & 9.5$\pm$1.0   & -4.7$\pm$1.0  & 9.5, 4.7    \\
2M1237      & 12 37 36.58		& +65 26 05.70  & 35.0$\pm$1.0  & 16.9$\pm$1.2  & 35.0, 14.1   \\
SDSS0423    & 04 23 48.23 		& -04 14 02.15  & 15.4$\pm$1.2  & -0.5$\pm$1.4  & 12.8, 0.4  \\
\enddata
\end{deluxetable}

\subsection{Imaging}\label{sec.Imaging_15a374}
We produced Stokes I and Stokes V images of each object (total and circularly polarized intensities, respectively) with the Common Astronomy Software Applications (CASA) \texttt{clean} routine, modeling the sky emission frequency dependence with one term and using natural weighting. Pixel sizes were 0\farcs04$\times$0\farcs04.  We searched for a point source at the proper motion-corrected coordinates of each target.  For our targets calibrated with 3C295, we selected a single calibrated measurement set as a reference set, noted in Table~\ref{table:imageDetections}.  We performed all subsequent reduction and analysis on this reference set. 

Flux densities and source positions were determined by fitting an elliptical Gaussian point source to the cleaned image of each object at its predicted coordinates using the CASA task \texttt{imfit}.

\subsection{Time series: Detecting ECM Pulses}\label{sec.Timeseries_15a374}
We used the \texttt{clean} routine to model all sources within a primary beam of our targets and subtract these sources from the UV visibility data using the CASA \texttt{uvsub} routine to prevent sidelobe contamination in our targets' time series.  We then added phase delays to our visibility data using the CASA \texttt{fixvis} routine to place our targets at the phase center. 

We checked all targets for highly circularly polarized flux density pulses to confirm the presence of ECM emission. Rather than searching for pulsed emission in Stokes I and V, we elected to search for pulses in the rr and ll correlations (right- and left-circularly polarized, respectively), where signal to noise is a factor of $\sqrt{2}$ higher in cases where the pulsed emission is 100\% circularly polarized, as is expected in an ideal case of ECM emission. 

Using the CASA plotting routine \texttt{plotms} to export the real UV visibilities averaged across all baselines, channels, and spectral windows of the rr and ll correlations at 10s, 60s, and 120s time resolutions, we created rr and ll time series for all X-band targets at 8--9~GHz, 9--10~GHz, 10--11~GHz, 11-12~GHz, 8--10~GHz, 10--12~GHz, and 8--12~GHz bandwidths to check for frequency-dependent ECM emission cutoff.   We repeat the same procedure for 2M1047 but divide the total bandwidth into 12--13.5~GHz, 13.5--15~GHz, 15--16.5~GHz, 16.5--18~GHz, 12--15~GHz, 15--18~GHz, and 12--18~GHz.  Figures~\ref{fig:2m1047timeseries}, \ref{fig:simp0136timeseries}, and \ref{fig:2m1237_sdss0423timeseries} show the time series for each object. 

We identify pulses using the following method: we smooth each time series with a locally weighted first degree polynomial regression (LOESS) and a smoothing window of 2.5\% of the on-target time to prevent anomalous noise spikes, typically very narrow with $\sim$single time resolution element widths, from erroneously being identified as a pulse while also preventing the smearing out of slightly wider legitimate pulses.  We then identify 2$\sigma$ outlier peaks in the smoothed time series  and measure the full width half maximum (FWHM) of the smoothed pulse, where we use the rms of the time series as a proxy for any quiescent emission.  In reality, these peaks lie above twice the quiescent emission, since the rms includes the peaks. Approximating each pulse as Gaussian, we define the full width of each pulse as three times the FWHM and remove each pulse from the raw time series.  These initial steps remove the strongest pulses present in the time series that may cause weaker pulses from being automatically identified.  Finally, we repeat the process once more to identify any other pulse candidates.  Because sensitivity can be a concern at narrow time resolutions and bandwidths in the time series, we elected to conservatively set the detection threshold for this second iteration at 2$\sigma$ and separately verify the pulses by imaging each candidate pulse in Stokes I and V and comparing flux densities with that of the non-pulsed (quiescent) emission. 

We confirm pulses with Stokes I and V imaging over the 60s FWHM of each candidate pulse and measuring integrated Stokes I and Stokes V flux densities using the CASA routine \texttt{imfit}.  In an initial set of fits, we allow the peak location to float and fix the semi-major and semi-minor axes to the dimensions of a synthesized beam, and our fitting region is a 100$\times$100 pixel region centered at the target location measured in \S\ref{sec.Imaging_15a374}. We select the highest signal-to-noise pulse as a benchmark and perform a second iteration of fits while also holding the benchmark peak location constant. We list measurements for pulses with unambiguous imaging and rms noise limits for frequency sub-bands with no detection.  Imaging for some sub-bands show evidence for a possible point source at the expected target location that is not clearly distinguishable by eye from the noise in the image.  We classify flux density measurements for these sub-bands as tentative detections and bootstrap the significance of the possible point source by randomly drawing 10,000 pointings in a 4096$\times$4096 pixel (2.7\arcmin$\times$2.7\arcmin) image and measuring the flux densities for a point source centered on these pointings.  

We calculate the highest likelihood percent circular polarization, where negative and positive percentages correspond to left and right circular polarizations, respectively.  We report uncertainties that correspond to the upper and lower limits of the 68.27\% confidence interval and record the evolution of pulse flux densities across sub-bands in Table~\ref{table:2m1047} (2M1047), Table~\ref{table:simp0136} (SIMP0136 \& 2M1043), Table~\ref{table:2m1237} (2M1237), and Table~\ref{table:sdss0423} (SDSS0423).  Some pulses appear to have Stokes V fluxes that are higher than the Stokes I fluxes, which is not physically possible.  However, these anomalous excess flux densities are within the rms noise.  For objects with 100\% circular polarization, we give the lower-bounds of the 68.27\%  and 99.73\% confidence intervals on the circular polarization. 

\begin{figure*}[h]
\epsscale{1.1}
\plotone{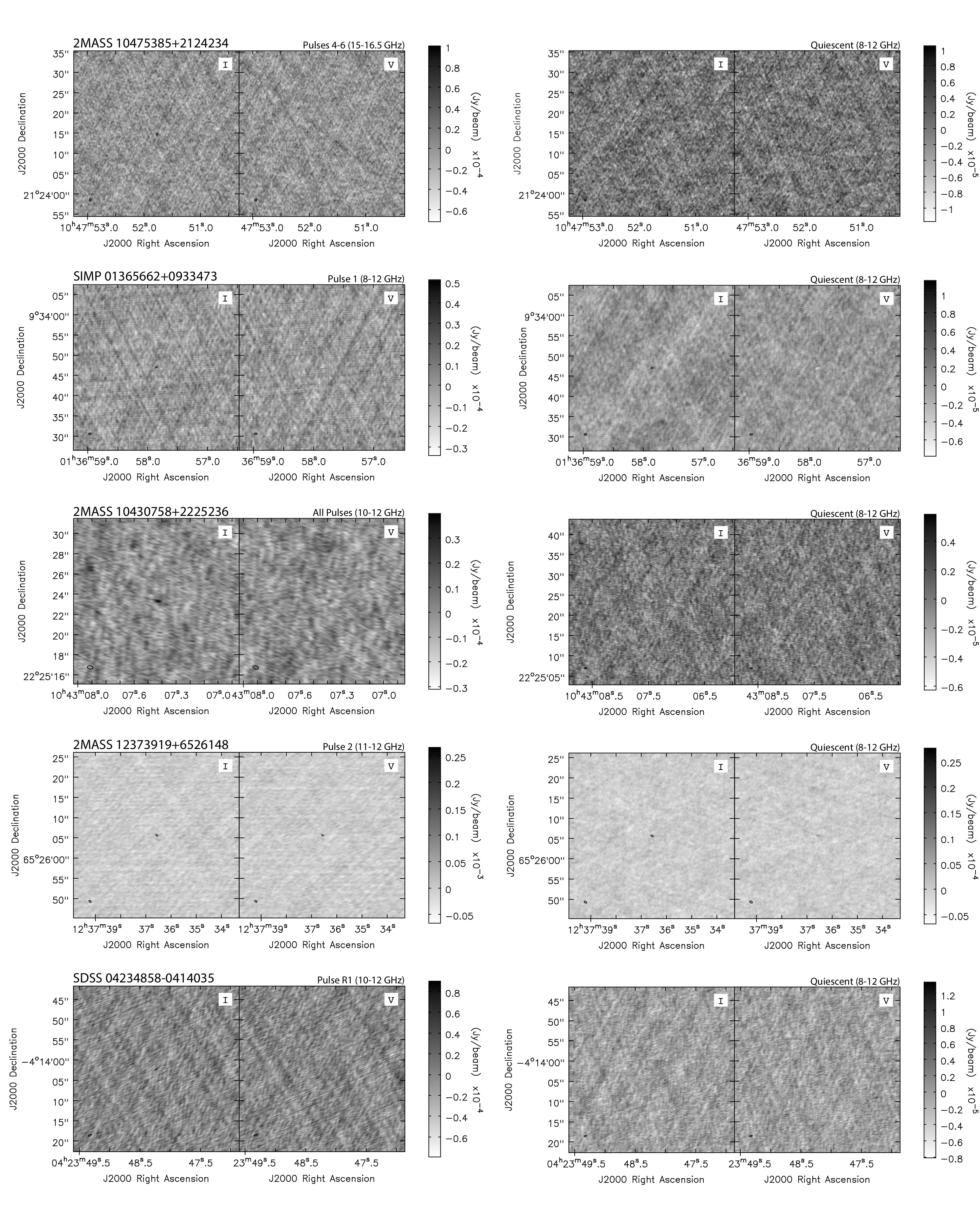}
\caption{Stokes I and Stokes V images of pulsed emission (left) and quiescent emission (right).  Images are centered over measured target coordinates and ellipses in bottom left corners depict synthesized beam dimensions.   No quiescent emission is detectable from 2M1043 or SDSS0423.  A measurement of the flux density at the expected coordinates for 2M1047 yields a tentative detection, but a point source is not clearly distinguishable by eye.  \label{fig:15a374quiescentEmission} }
\end{figure*}

We additionally measure quiescent emission by removing the full width of each pulse across the entire 4- or 6-GHz bandwidth from our data and imaging the remaining emission, shown in Figure \ref{fig:15a374quiescentEmission}.  We report the characteristics of the pulsed and quiescent emission in Tables \ref{table:2m1047}, \ref{table:simp0136},  \ref{table:2m1237}, and \ref{table:sdss0423}.

\subsection{Measuring Rotation Periods} \label{sec.Rotation_15a374} 

Our data are well-sampled with respect to pulse widths but very noisy and may contain low-amplitude or wide duty cycle peaks.  Previous attempts have benefited from fitting the time series of relatively bright $\sim$mJy pulses \citep{hallinan2007, hallinan2008, williams2015,routeWolszczan2016}, an order of magnitude brighter than the pulses in our targets.  In contrast, for our data, some pulses do not become apparent until the data have been averaged to 60s or 120s resolutions, further introducing uncertainty when attempting to accurately identify pulses and their arrival times.  For these reasons, we elected not to pursue a Levenberg-Marquardt or Monte Carlo time-of-arrival fitting \citep{williams2015, routeWolszczan2016} and instead employ three independent algorithms widely used in exoplanet transit and radial velocity searches.  Using these algorithms has the added benefit of independently verifying the pulses that we identified in \S\ref{sec.Timeseries_15a374}.  The first is the classic Lomb-Scargle (L-S) periodogram, which relies on decomposing time series into Fourier components and is optimized to identify sinusoidally-shaped periodic signals in time-series data, making this algorithm most appropriate for testing periodicity in broader pulses such as those observed in the SDSS0423 and SIMP0136 time series or even our targets' quiescent emission.  The second method is the Plavchan periodogram, a brute force method that derives periodicities in a method similar to that employed by phase dispersion minimization \citep{stellingwerf1978}, but circumvents period aliasing because it is binless \citep{plavchan2008, parksPlavchan2014}.  The Plavchan algorithm is not dependent on pulse shape and thus is sensitive to both sinusoid-dominated variability and other pulse profiles.  Finally, the shapes of some of the pulses bear resemblance to inverse light curves of planet transits, for which the Box-fitting Least Squares (BLS) algorithm is optimized \citep{kovacs2002}. 

We generate periodograms for all of our objects using the 10s time-averaged time series for the full bandwidth data and at all sub-bands using the MATLAB Lomb-Scargle function \texttt{plomb} and the NASA Exoplanet Archive Periodogram Service\footnote{https://exoplanetarchive.ipac.caltech.edu/cgi-bin/Pgram/nph-pgram} for Plavchan and BLS periodograms.  The Plavchan algorithm depends on two input parameters: number of outliers and fractional phase smoothing width, which we vary between 10\%--30\% of total data points and 0.025 - 0.1, respectively.  BLS depends on three input parameters: number of points per bin, minimum fractional period coverage by pulse, and maximum fraction period coverage.  For BLS, we hold the minimum fractional period coverage constant at 0.01, and we vary the number of points per bin and maximum fractional period coverage between 10--100 and 0.1--0.3, respectively. In most cases, the recovered periodicities do not depend significantly on these parameters and we discuss exceptions in \S \ref{sec.Results_15a374}.

\begin{figure*}
\epsscale{1.1}
\plotone{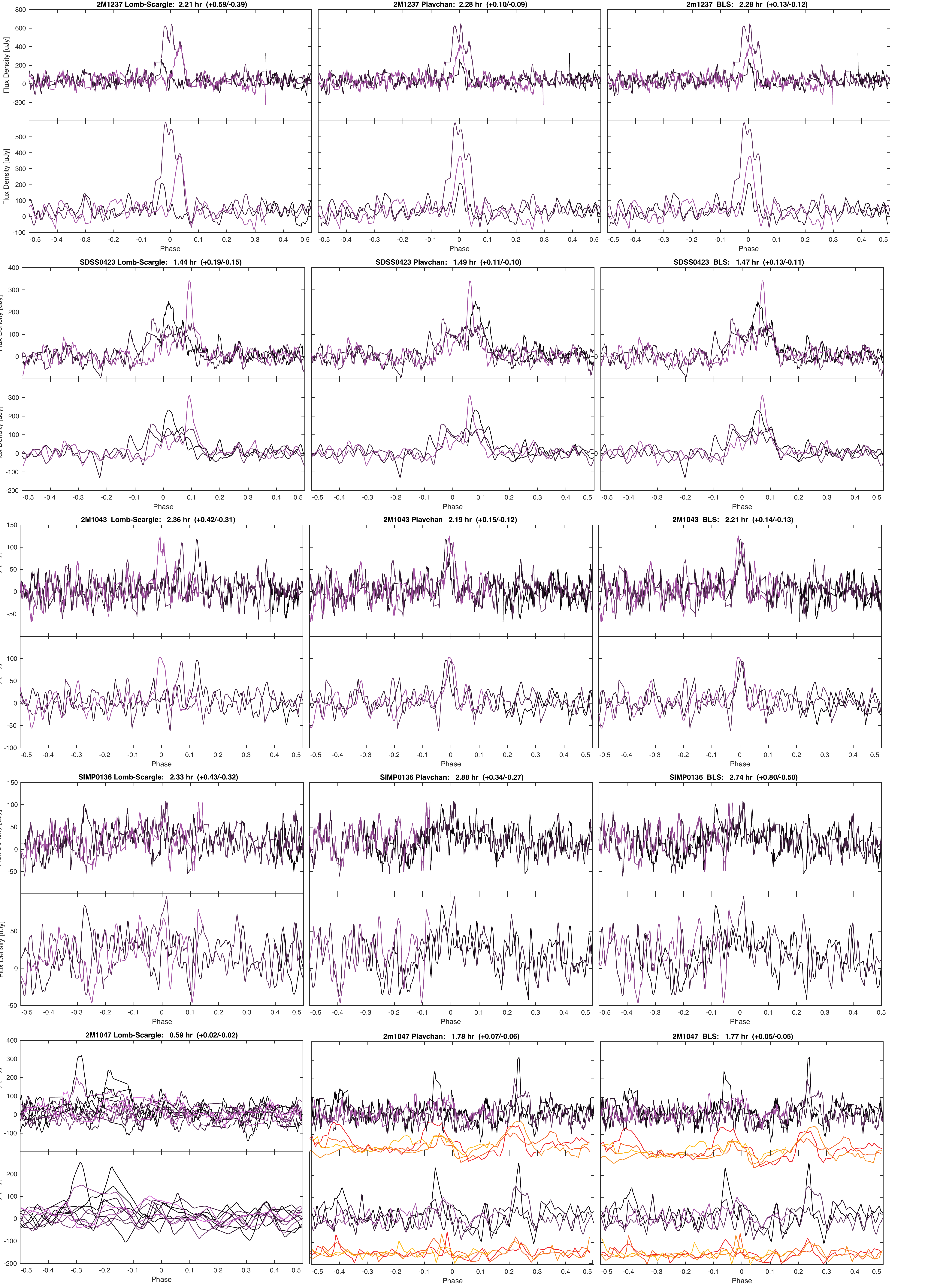}
\caption{From left to right: Phase-folded 10s time series using periods from Lomb-Scargle (L-S), Plavchan, and Box-fitting Least Squares (BLS) periodograms. Top panels are raw data, bottom panels are smoothed data. 60s time series are overplotted in orange.  \label{fig:phaseFolded}}
\end{figure*}

We compare peaks with false alarm probability less than 10\% returned by the the Lomb-Scargle algorithm to the most significant periods returned by the other algorithms in Figure~\ref{fig:periodograms} and visually inspect periods by phase-folding the time series in Figure~\ref{fig:phaseFolded} with the most significant period returned by each algorithm.  We estimate uncertainties as the inverse of the FWHM of the frequency power peaks.
 We list periods returned by each algorithm in Table~\ref{table:periods} and adopt the periods that result in the folded time series with the most visual agreement in pulse overlaps.

\section{Results}\label{sec.Results_15a374}

\begin{figure*}[hp!]
\epsscale{1.15}
\plotone{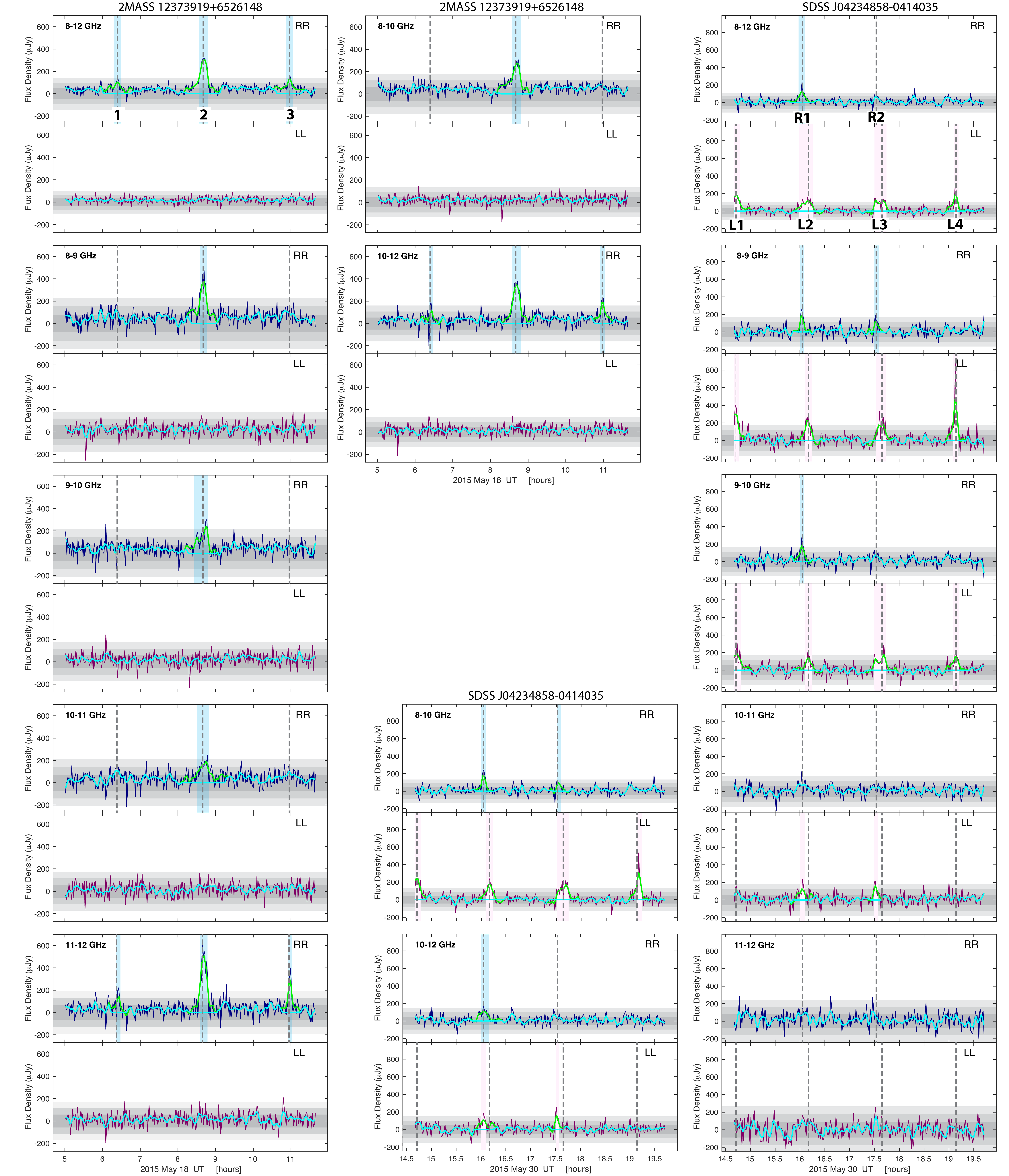}
\caption{\label{fig:2m1237_sdss0423timeseries} 60s time series of rr- and ll-correlated (blue and maroon, respectively) flux densities for 2M1237 and SDSS0423.  Green lines are smoothed time series used for identifying pulse candidates and overlaid cyan lines show removed pulse candidates for calculating rms noise and imaging quiescent emission. Light blue and pink bars highlight pulses identified by algorithm.  Grey dashed lines are aligned to pulse peaks.  Grey regions indicates 1$\sigma$, 2$\sigma$, and 3$\sigma$ rms noise. }
\end{figure*}

\textbf{2MASS~J12373919+6526148}.
We detect 2M1237 in initial Stokes I and Stokes V imaging with signal-to-noise ratios (SNR) 35.0 and 14.1, respectively.  Table~\ref{table:imageDetections} gives the measured mean flux density and rms noise. These strong detections are due to weakly circularly polarized ($\sim$35\%) quiescent emission ($27.8\pm1.3$ $\mu$Jy mean flux density) present throughout the entire 8--12 GHz band, as well as Pulse 2, a very bright ($159.7\pm5.3$ $\mu$Jy mean flux density) and highly circularly polarized ($\sim$80\%) pulse occurring near the center of the observation time window.   Pulse 2 is observable at all subbands within the full 8--12~GHz band, though its flux density varies from band to band by a factor of nearly 3 (see \S \ref{sec.Cutoff_15a374} for discussion about such frequency-dependent variability).  Two substantially weaker pulses with mean flux densities $41.3\pm5.4$ $\mu$Jy and $61.0\pm5.7$ $\mu$Jy additionally occur before and after Pulse 2. Figure~\ref{fig:2m1237_sdss0423timeseries} shows the time series for 2M1237 and  we report the characteristics of the pulsed and quiescent emission in Table \ref{table:2m1237}.

\begin{figure*}
\epsscale{1.1}
\plotone{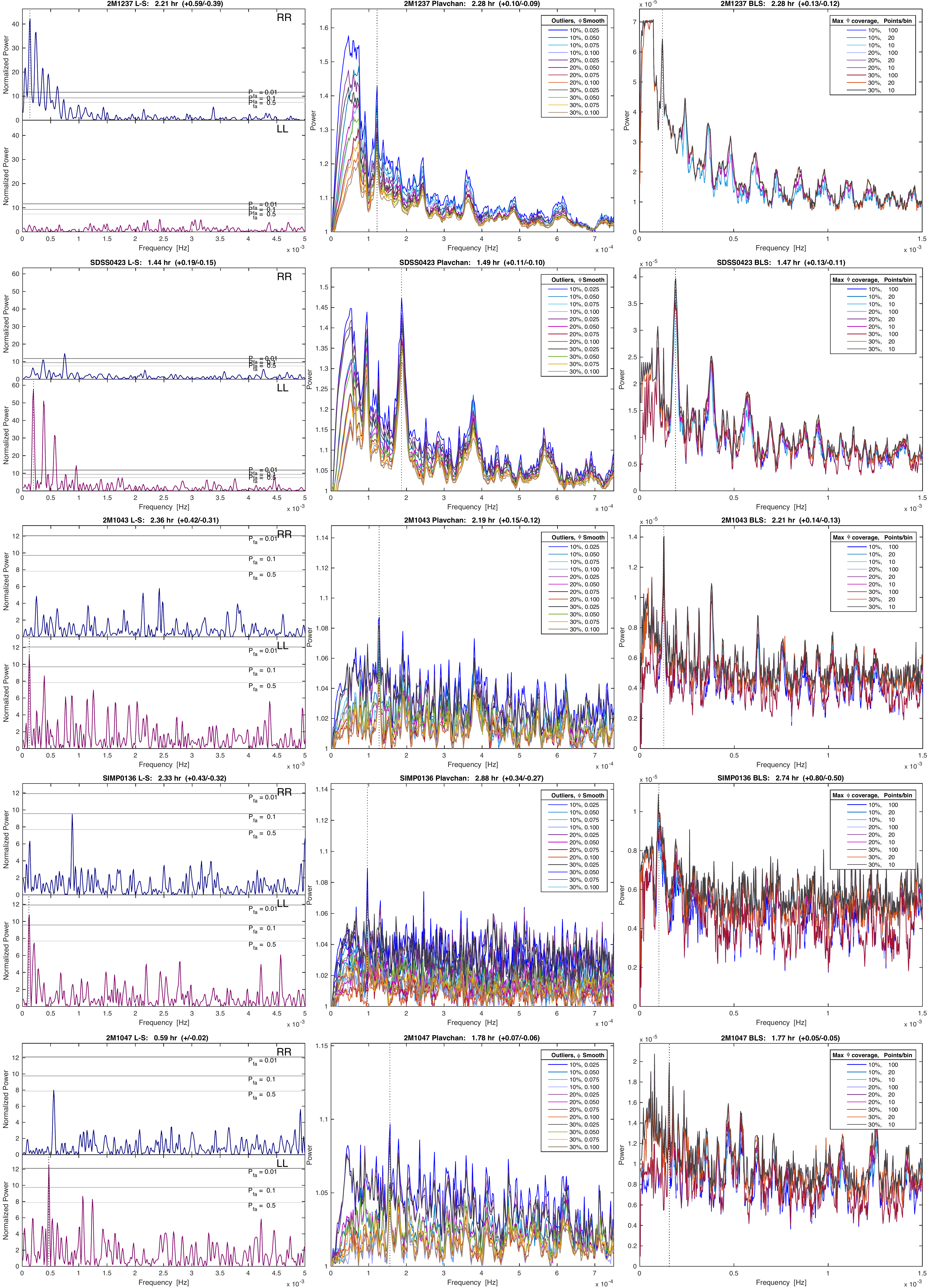}
\caption{From left to right: Lomb-Scargle (L-S), Plavchan, and Box-fitting Least Squares (BLS) periodograms.  RR and LL periodograms are shown for the L-S periodogram to show relative powers of peaks between time series with apparent periodic variation and ones without.  Periodograms for Plavchan and BLS algorithms are for correlations with strongest L-S peaks.   \label{fig:periodograms} }
\end{figure*}

Such strong pulses suggested a straightforward period analysis, and indeed, the periods returned by the L-S, Plavchan, and BLS periodogram algorithms are consistime seriestent within uncertainties (see Table~\ref{table:periods}). However, the data for 2M1237 do not appear to provide enough phase coverage to adequately sample periods longer than $\sim$3.77 hours.  Plavchan peak power locations at and longer than this $\sim$3.77-hour period change dramatically depending on input variables and especially on the fractional amount of outliers (Figure~\ref{fig:periodograms}).  Specifically, Plavchan periodograms with a lower fraction of allowed outliers are biased in favor of a period that is approximately two times longer than the periods favored when allowed outlier fractions are higher. This occurs because the flux density of Pulse 2 deviates strongly from the mean amplitude of the smaller pulses before and after it. When the algorithm is not allowed to ignore datapoints from this strong pulse, it will favor a rotation period that generates a time series akin to one with a main transit and a secondary eclipse.  Additional phase coverage to characterize the variable behavior of the pulse profile is necessary to resolve the ambiguity between period harmonics.

\textbf{2MASS~10475385+2124234}.
We detect 2M1047 in initial Stokes I imaging with SNR 16.8.  In contrast, there is no clear Stokes V detection, with a SNR of only 2.6.   Table~\ref{table:imageDetections} gives the measured mean flux density and rms noise. 
Highly circularly polarized pulses are clearly evident in the 10s, 60s, and 120s sub-band time series for 2M1047, with two large-amplitude pulses occurring near the beginning of the observation time window (Pulse 1 and Pulse 2).  Pulse 1 occurred during a time range when strong RFI caused all 12--12.8~GHz data to be flagged, affecting noise properties and especially so for the 12-13.5~GHz subband. To check if Pulse 1 could be attributed to this additional noise, we created time series for a nearby object at 10$^{\mathrm{h}}$47$^{\mathrm{m}}$54\fs95 +21\degr24\arcmin13\farcs40s and searched for variability that correlates with Pulse 1.  We include this comparison time series in the 2M1047 time series figures for 120s resolution.  This comparison object does not exhibit any evidence of highly circularly polarized pulses at any of the frequencies or timestamps associated with the pulses detected for 2M1047. 

\begin{figure*}[hp!]
\epsscale{1.15}
\plotone{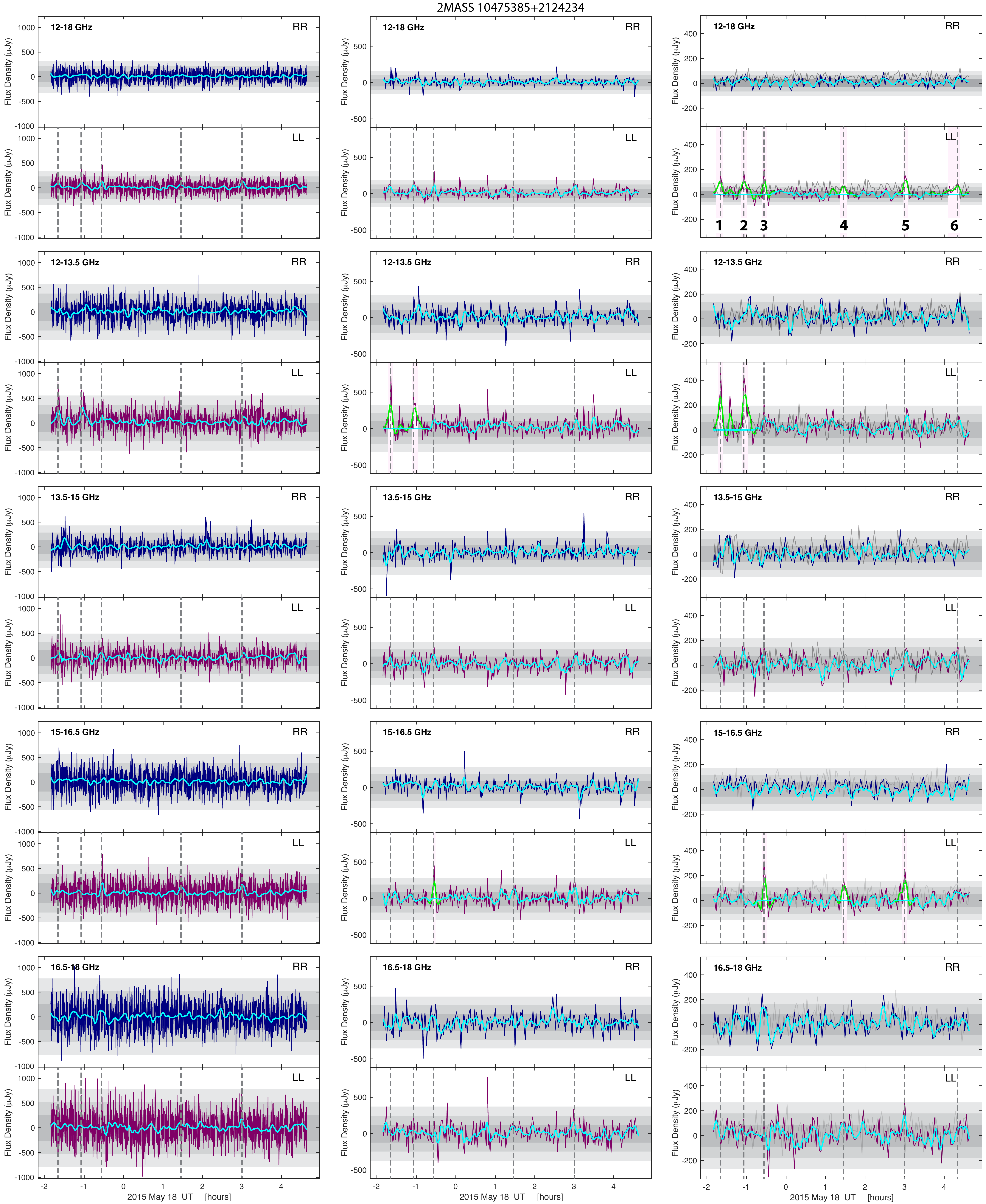}
\caption{\label{fig:2m1047timeseries}  10s, 60s, and 120s time series of rr- and ll-correlated (blue and maroon, respectively) flux densities for 2M1047 showing the emergence of apparent pulses at 12--13.5~GHz and 15--16.5~GHz.  Green lines are smoothed time series used for identifying pulse candidates and overlaid cyan lines show removed pulse candidates for calculating rms noise and imaging quiescent emission. Light blue and pink bars highlight pulses identified by algorithm.  Grey dashed lines are aligned to 12--13.5~GHz and 15--16.5~GHz pulse peaks.  Grey regions indicates 1$\sigma$, 2$\sigma$, and 3$\sigma$ rms noise. Comparison time series of a nearby object are plotted in dark grey in the 120s column. }
\end{figure*}

When checking each pulse individually with imaging, Pulses 3--5 were very faint and were difficult to individually distinguish by eye in the imaging (see Figure~\ref{fig:2m1047timeseries}).  To further check these pulses, we averaged them together to reduce rms noise and report measured flux densities for this averaged image in Table \ref{table:2m1047}.  Pulses 3--5 were clearly detectable by eye in the 12--18 GHz and 15--16.5 GHz images. 

Pulse 5 may extend into the 16.5--18 GHz time series. We measured Stokes I and Stokes V flux densities of $91.5 \pm 28.7$ $\mu$Jy and $-94.9 \pm 24.9$ $\mu$Jy, respectively, where negative values indicate left-circular
 polarization.  The percent circular polarization is expected to lie between [-100\%, -58.0\%] with 68.27\% confidence and [-100\%, -14.3\%] with 99.73\% confidence.  However, there is not a clear point source in the associated images.  The bootstrapped Stokes I significance is 99.29\%.  The significance increases to 99.63\% and 99.99\% when we constrain the acceptable percent circular polarization to lie within the 99.73\% and 68.27\% confidence intervals, respectively.  We classify the 16.5--18 GHz detection as a tentative detection.  We report the characteristics of the pulsed and quiescent emission in Table \ref{table:2m1047}.

When applying the periodogram analyses, 2M1047 stood out as the sole object whose periods returned by the L-S, Plavchan, and BLS algorithms were inconsistent with each other (see Table~\ref{table:periods} and Figure~\ref{fig:periodograms}).  
The Lomb-Scargle periodogram returns a $\sim$0.59~hr period, while Plavchan returns $\sim$1.77~hr, and BLS returns either $\sim$3.54~hr or $\sim$1.77~hr depending on the maximum allowed rotation pulse phase coverage and phase binning.  Happily, these periods are all harmonics, suggesting a non-spurious origin.  Similar to 2M1237, the longest period is favored by the BLS algorithm for the cases with the least number of data points per bin, emphasizing the significance of the strongest peaks.  The Plavchan periodogram also reflects this behavior, although its most significant period is consistently $\sim$1.77~hr irrespective of input parameters.  For ground-based transit surveys, a typical number of points per bin is of order a few tens to a hundred, which would correspond to a $\sim$1.77~hr period.  

Owing to the observed intermittency of the pulses, the periodogram results are tantalizing but inconclusive.   However, the periodogram detects periodicity consistent with the expected period as measured by \citet{williams2015} using 10-hr C-band (4--6~GHz) observations, suggesting that our detected periodicity may be due to the pulsed emission and/or the quiescent emission. Given the ambiguities arising from the periodogram analysis of 2M1047 and the lack of clear pulse periodicity in the phase-folded lightcurves, we treat the periodogram analysis as a confirmation of the period measured by \citet{williams2015}.

\begin{figure*}[hp!]
\epsscale{1.15}
\plotone{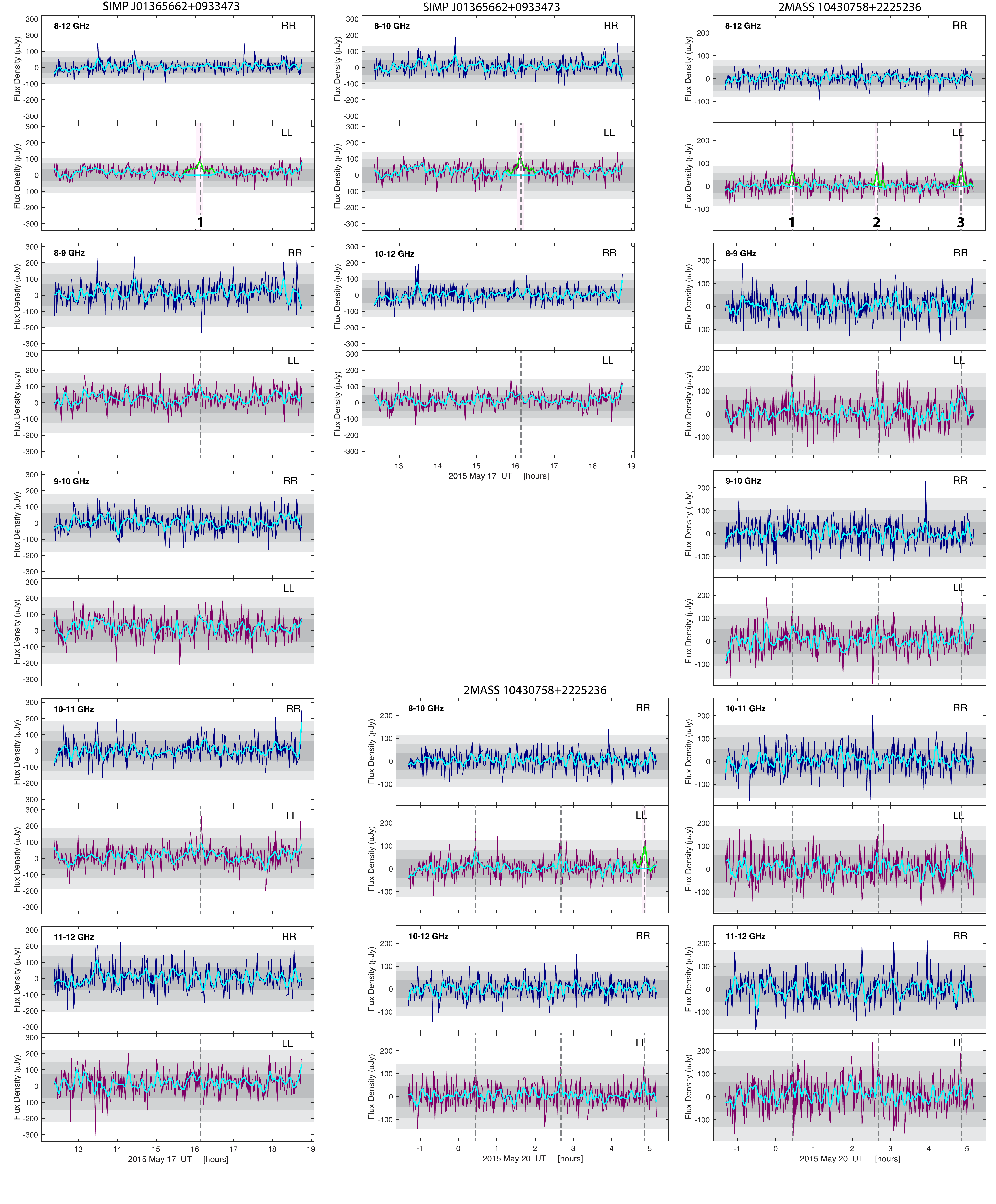}
\caption{\label{fig:simp0136timeseries} 60s time series of rr- and ll-correlated (blue and maroon, respectively) flux densities for SIMP0136 and 2M1043.  Green lines are smoothed time series used for identifying pulse candidates and overlaid cyan lines show removed pulse candidates for calculating rms noise and imaging quiescent emission. Light blue and pink bars highlight pulses identified by algorithm.  Grey dashed lines are aligned to pulse peaks.  Grey regions indicates 1$\sigma$, 2$\sigma$, and 3$\sigma$ rms noise.  }
\end{figure*}

\textbf{SIMP~J01365662+0933473}.
We detect SIMP0136 in initial Stokes I and Stokes V imaging with SNR 65.9 and 21.6, respectively.  Table~\ref{table:imageDetections} gives the measured mean flux density and rms noise.
SIMP0136 appears to have broadly variable quasi-quiescent radio emission with a single broad peak (Pulse 1) that is persistent across 60s and 120s sub-band time series (see Figure~\ref{fig:simp0136timeseries}).  We confirm Pulse 1 with imaging and report the characteristics of the pulsed and quiescent emission in Table \ref{table:simp0136}.   

At first glance, the 8--12 GHz time-averaged quasi-quiescent emission from SIMP0136 is similarly circularly polarized as for Pulse 1 ($\sim$60\%).  Upon closer examination, Pulse 1 is more strongly circularly polarized than the quasi-quiescent emission at 8--10 GHz ($\sim$60\% vs. $\sim$40\%).  At the 10--12 GHz subband, any Stokes V emission that may be present cannot be distinguished from the rms noise for either Pulse 1 or the quasi-quiescent emission.  Although the 10--12 GHz Pulse 1 detection is tentative ($40.5 \pm 8.5$ $\mu$Jy with 99.67\% bootstrapped significance), it is important to note that the quasi-quiescent emission is undetectable in Stokes I down to a $3\sigma_{\mathrm{rms}}$ level of 6.2 $\mu$Jy.  The significantly lower rms noise results from the longer time coverage of the quasi-quiescent emission as compared to the narrower time-width of Pulse 1.   When we further examine the SIMP0136 time series at 1 GHz bandwidths, the Stokes I detection remains clear for Pulse 1 at 8--9 GHz ($69.9\pm12.9$ $\mu$Jy) and becomes more tentative at 9--10 GHz and 10--11 GHz ($44.3\pm12.2$ $\mu$Jy with 98.78\% bootstrapped significance and $41.5\pm12.0$ $\mu$Jy with 98.80\% bootstrapped significance, respectively),  finally becoming indistinguishable from rms noise at 11-12 GHz.  These tentative detections are further bolstered by measured flux densities that are consistent with those measured for the 8--10 GHz and 10--12 GHz subbands.  In contrast to the persistence of Pulse 1 emission up through 11 GHz, the Stokes I quasi-quiescent emission becomes undetectable above 10 GHz, at $3\sigma_{\mathrm{rms}}$ rms noise levels of 9.0 $\mu$Jy and 10.5 $\mu$Jy.  Given these comparisons, we are confident of the 8--9 GHz Pulse 1 detection and classify the 9--10 GHz and 10--11 GHz detections as tentative.

Infrared cloud variability studies of SIMP0136 suggest that its rotation period is $P = 2.3895\pm 0.0005$ hr.  This \textit{a priori} knowledge of the expected pulse periodicity allows us to search for pulses at expected occurrence times in our observing block.   A pulse occurring before the above-noted time series peak would have directly coincided with a phase calibrator observation and thus possibly prevented its detection.  A pulse occurring after would have taken place near the middle of the target integration block, when phase errors would be greatest and may possibly smear out flux from a pulse.  To check for the effects of phase errors on flux densities, we imaged a bright nearby object at 01$^{\mathrm{h}}$36$^{\mathrm{m}}$47\fs63s +09\degr34\arcmin04\farcs25 and well within the $\sim$4.5\arcmin primary beam during `edge' and `middle' observing scans.  `Edge' scans are directly adjacent to a phase calibration scan whereas `middle' scans are sandwiched by the edge scans and therefore likely suffer from the worst phase calibration errors.  We measured only a $3.2\pm1.8$\% decrease in flux, suggesting that phase calibration errors cannot account for a possible missing pulse. We conclude that either another pulse exists but is not detectable, or there is not another pulse.  See \S \ref{sec.Cutoff_15a374} for an in-depth discussion.

Despite the single pulse, we include SIMP0136 in the periodogram analysis for the sake of completeness. The period returned by the L-S, Plavchan, and BLS algorithms are consistent with each other within uncertainties, and appear to be based on the variability occurring in the quasi-quiescent emission. 
We adopt a period of $2.88^{+0.34}_{-0.27}$ hr for the quasi-quiescent emission at X band.  
We analyzed the 4--8~GHz data from \citet{kao2016} and find that the C-band period appears nominally consistent with $\sim$2.88~hr, but the data are inconclusive because the total C-band observing block was only 4 hours long. In contrast to the X-band period, the photometric period is $2.3895\pm 0.0005$ hr.  These periods are not statistically distinct.

 With only one visually apparent  ECM pulse, we cannot confirm a cloud-independent rotation period for SIMP0136.  Since ECM emission is more clearly discerned at 4--8 GHz for SIMP0136, we recommend a future rotation study using long-duration observations at 4--8 GHz to determine the cloud-independent rotation period of SIMP0136. Because the mechanism generating the non-pulsed but varying quiescent emission and its location within the brown dwarf system remain unknown, while the infrared variability is expected to occur within the brown dwarf atmosphere, we adopt the rotation period measured by photometric studies for our discussion in \S\ref{sec.Discussion_15a374}.

\textbf{2MASS~J10430758+2225236}.
We detect 2M1043 in initial Stokes I imaging with SNR 9.5.  The Stokes V detection is very faint, with a SNR of 4.7.   Table~\ref{table:imageDetections} gives the measured mean flux density and rms noise. 
In its time series, 2M1043 has three very faint pulses that become clearly evident when the data are averaged across the full 8--12 GHz bandwidth (see Figure~\ref{fig:simp0136timeseries}).   At the full 4 GHz bandwidth, the pulses have flux densities that range from $40.8 \pm 8.0$ $\mu$Jy through $60.5 \pm 7.4$ $\mu$Jy.  When imaged individually, these pulses are difficult to distinguish by eye in the imaging.   To reduce the rms noise, we averaged the three pulses together to check for them in subband imaging.  We include measured flux densities for these averaged images in the ``All Pulses" column in Table \ref{table:simp0136}.

In the time series, the pulses are most clearly visually evident at the 8--12 GHz and 8--10 GHz bands. In the imaging, the pulses remain evident through the 9--10 GHz subband for both Stokes I and Stokes V.  At 10--11 GHz, the Stokes I component of the averaged-together pulses remains clear with flux density $40.1 \pm 7.8$ $\mu$Jy, but the Stokes V component is undetectable up to a $3\sigma_{\mathrm{rms}}$ flux density of 25.8 $\mu$Jy when all three pulses are averaged together.  When judging if these pulses are truly present or not, we compared the ``All Pulses" flux density measurements in each 1 GHz subband to the flux density measurements for quiescent emission.  In contrast to clear Stokes I pulsed emission up through 10--11 GHz, 2M1043 does not appear to have any detectable quiescent emission $\geq$6.9 $\mu$Jy (3$\sigma_{\mathrm{rms}}$) in that subband, or $\geq$3.6 $\mu$Jy (3$\sigma_{\mathrm{rms}}$) for the full 8--12~GHz bandwidth. We therefore conclude that the pulses are present through the 10--11 GHz subband.

The period returned by the L-S, Plavchan, and BLS periodogram algorithms are consistent within uncertainties. Given the sharpness of the pulses, we rule out the period returned by the L-S algorithm as our adopted period.  This is because the L-S algorithm relies on Fourier analysis and therefore is not well-suited to time series with sharp pulses, which require many high-order sinusoids to reproduce.  Following our methodology outlined in \S \ref{sec.Rotation_15a374}, we adopt the period returned by BLS, which results in a folded time series with the most visual agreement in pulse overlaps.

\textbf{SDSS~J04234858-0414035}.
We detect SDSS0423 in initial Stokes I imaging with SNR 12.9 and no Stokes V detection.   Table~\ref{table:imageDetections} gives the measured mean flux density and rms noise. 
SDSS0423 has four left-circularly polarized pulses that are clearly evident through 10 GHz.  At 8--9 GHz and 9--10 GHz, the peak flux density ranges from $60.6 \pm12.0$ $\mu$Jy for the faintest pulse to $218.1\pm21.0$ $\mu$Jy for the brightest pulse. At these frequency ranges, the pulses are strongly circularly polarized, with highest-likelihood percent polarizations between $-54.8$\% and $-95.4$\%. At 10--11 GHz and 11--12 GHz, these pulses fade and become undetectable up to Stokes I $3\sigma_{\mathrm{rms}}$ limits between 31.5 $\mu$Jy and 88.2 $\mu$Jy.  However, when the left-circularly polarized pulses are averaged over 2 GHz bandwidths, Pulses L1 and L3 remain clearly detectable in Stokes I with flux densities $67.2 \pm12.9$ $\mu$Jy and $53.1 \pm8.8$ $\mu$Jy, respectively.   

In addition to the left-circularly polarized pulses, there are two fainter right circularly polarized pulses, with peak Stokes I flux densities between $73.8 \pm 18.5$ $\mu$Jy and $111.6 \pm 14.0$ $\mu$Jy throughout the 8--9 GHz and 9--10 GHz bands.  Except for Pulse R1 at 8--9 GHz, these right-circularly polarized pulses are less polarized than the left circularly polarized pulses.  They are undetectable in Stokes V up to $3\sigma_{\mathrm{rms}}$ limits between 44.1 $\mu$Jy and 54.6 $\mu$Jy, with corresponding upper limits on the highest-likelihood percent circular polarization between 38.9\% and 48.8\%.  Pulse R1 at 8--9 GHz is strongly polarized, with a Stokes V flux density of $65.7 \pm 16.0$ $\mu$Jy and a highest-likelihood percent circular polarization of 83.9\%.  At 10--11 GHz, only Pulse R1 remains detectable in Stokes I, with a flux density of $82.7 \pm 17.9$ $\mu$Jy.  However, its Stokes V flux density fades and cannot be detected above a $3\sigma_{\mathrm{rms}}$ limit of 57.0 $\mu$Jy.  At 11--12 GHz, both right circularly polarized pulses become undetectable above a Stokes I $3\sigma_{\mathrm{rms}}$ limit between 66.0 $\mu$Jy and 71.4 $\mu$Jy.   

With stronger left-circularly polarized pulses than right-circularly polarized pulses, these X-band observations directly contrast with the C-band observations for SDSS0423, in which the right-circularly polarized pulses are stronger than the left-circularly polarized ones \citep{kao2016}.  Also in contrast to its C-band behavior, SDSS0423 does not appear to have any detectable quiescent emission above a Stokes I $3\sigma_{\mathrm{rms}}$ limit of 5.1 $\mu$Jy for the full 4 GHz bandwidth (see \S \ref{sec.Quiescent_15a374} for discussion).  

The multiple pulses in the SDSS0423 time series allows for a straightforward periodogram analysis. The periods returned by the L-S, Plavchan, and BLS periodogram algorithms are consistent within uncertainties (see Table~\ref{table:periods}).   Following our methodology outlined in \S \ref{sec.Rotation_15a374}, we adopt the $\sim$1.47 hr period returned by BLS, which results in a folded time series with the most visual agreement in pulse overlaps. 
This period is consistent with the $2.0 \pm 0.4$ hr J-band variability period reported by \citet{clarke2008}.  Additionally, with a $v \sin i = 60 \pm 10$ km s$^{-1}$ \citep{prato2015}, the corresponding lower bound radius $R \sin i = 0.71 \pm 0.13$ R$_{\mathrm{J}}$.  This lower bound radius is consistent with the $\sim$0.9--1.0 R$_{\mathrm{J}}$ radii inferred from dynamical masses measured by \citet{dupuy2017}.


\setlength{\tabcolsep}{0.03in}
\begin{deluxetable*}{llccccc}[ht!]
\tablecaption{2M1237: Pulsed and Quiescent Emission \label{table:2m1237}}
\tablehead{
	\colhead{}								&
	\colhead{}								&
	\colhead{}								&
	\colhead{Pulse 1}						&
	\colhead{Pulse 2}						&
	\colhead{Pulse 3}						&
	\colhead{Quiescent}				            }             
\startdata
\textbf{8--12GHz}           &            &&                         &                           &                           		&                            \\  
Stokes $I$\tablenotemark{a} & ($\mu$Jy)  && 41.3$\pm$5.4            &   159.7$\pm$5.3           &  61.0$\pm$5.7             		& 27.8$\pm$1.3               \\  
Stokes $V$\tablenotemark{a} & ($\mu$Jy)  && 26.5$\pm$6.4            &   127.3$\pm$5.5           &  34.2$\pm$4.6             		& 9.7$\pm$1.4               \\  
S/N                         & ($I$, $V$) && 7.6, 4.1	            &   30.1, 23.1              &  10.7, 7.4                		& 21.4, 6.9                  \\  
Circ. Poln\tablenotemark{b} & (\%)       && 63.1$_{-15.4}^{+18.5}$  &   79.6$_{-4.1}^{+4.6}$    &  55.6$_{-7.9}^{+10.8}$    		& 34.8$_{-5.1}^{+5.5}$      \\  
\textbf{8--10GHz}           &            &&                         &                           &                           		&                            \\  
Stokes $I$\tablenotemark{a} & ($\mu$Jy)  && 30.0$\pm$9.1            &   151.5$\pm$9.0           &  52.4$\pm$7.3             		& 32.5$\pm$1.9               \\  
Stokes $V$\tablenotemark{a} & ($\mu$Jy)  && $<$24.9		            &   122.6$\pm$7.8           &  $<$20.7		            		& 9.2$\pm$1.9                \\  
S/N                         & ($I$, $V$) && 3.3, \nodata            &   16.8, 15.7              & 7.2, \nodata	            		& 17.1, 4.8                  \\  
Circ. Poln\tablenotemark{b} & (\%)       && (76.2$_{-30.5}^{+13.4}$)& 80.6$_{-6.3}^{+7.7}$      & (38.8$_{-12.4}^{+17.1}$)			& 28.2$_{-5.8}^{+6.4}$       \\  
\textbf{10--12GHz}          &            &&                         &                           &                           		&                            \\  
Stokes $I$\tablenotemark{a} & ($\mu$Jy)  && 44.4$\pm$7.6            &   174.3$\pm$9.1           &  71.3$\pm$8.6             		& 22.8$\pm$1.8               \\  
Stokes $V$\tablenotemark{a} & ($\mu$Jy)  && $<$24.0                 &   144.1$\pm$9.0           &  57.6$\pm$7.5  		           	& 10.2$\pm$1.8               \\  
S/N                         & ($I$, $V$) && 5.8, \nodata	        &   19.2, 16.0              &  8.3, 7.7             	    	& 12.7, 5.7                  \\  
Circ. Poln\tablenotemark{b} & (\%)       && (52.5$_{-17.0}^{+23.0}$)& 82.4$_{-6.2}^{+7.2}$      & 79.6$_{-12.4}^{+11.8}$    		& 44.5$_{-8.0}^{+9.6}$       \\  
\textbf{8--9GHz}            &            &&                         &                           &                           		&                            \\  
Stokes $I$\tablenotemark{a} & ($\mu$Jy)  &&  $<$36.9                & 197.7$\pm$11.5            &  57.0$\pm$11.4         		   	& 33.8$\pm$2.7               \\  
Stokes $V$\tablenotemark{a} & ($\mu$Jy)  &&  $<$36.6                & 145.6$\pm$10.7            &  $<$30.0                  		& 11.5$\pm$2.2\,\tablenotemark{d}               \\  
S/N                         & ($I$, $V$) && \nodata	 			    & 17.2, 13.6                &  5.0, \nodata             		& 12.5, 5.2                 \\  
Circ. Poln\tablenotemark{b} & (\%)       && \nodata	 			    & 73.4$_{-6.2}^{+7.7}$      & (50.6$_{-16.1}^{+24.4}$)			& 33.8$_{-6.5}^{+7.8}$       \\  
\textbf{9--10GHz}           &            &&                         &                           &                           		&                            \\  
Stokes $I$\tablenotemark{a} & ($\mu$Jy)  && $<$36.0		            &  97.1$\pm$10.9            &  $<$34.2                  		& 30.1$\pm$2.2               \\  
Stokes $V$\tablenotemark{a} & ($\mu$Jy)  && $<$35.1	             	&  94.5$\pm$10.4            &  $<$36.9                  		& $<$7.2	                 \\  
S/N                         & ($I$, $V$) && \nodata	 			    &  8.9, 9.1                 &   \nodata                 		&   13.7, \nodata            \\  
Circ. Poln\tablenotemark{b} & (\%)       && \nodata	 			    &  96.1$_{-17.0}^{+0.6}$    &   \nodata                 		& (23.8$_{-7.9}^{+8.6}$)     \\  
\textbf{10--11GHz}          &            &&                         &                           &                           		&                            \\  
Stokes $I$\tablenotemark{a} & ($\mu$Jy)  && 54.4$\pm$12.6           &  96.7$\pm$12.2            & 45.0$\pm$11.7\,\tablenotemark{c}	& 21.5$\pm$2.5               \\  
Stokes $V$\tablenotemark{a} & ($\mu$Jy)  && $<$30.9	   	 	        &  76.3$\pm$11.7            &  $<$35.7	             		    & 11.7$\pm$2.5               \\  
S/N                         & ($I$, $V$) && 4.3, \nodata		    &  7.9, 6.5                 &   3.8, \nodata           		    & 8.6, 4.7                   \\  
Circ. Poln\tablenotemark{b} & (\%)       && (54.0$_{-17.4}^{+25.5}$)&  77.7$_{-13.7}^{+12.9}$   &  (74.4$_{-28.8}^{+14.8}$)		    & 53.7$_{-11.4}^{+15.7}$     \\  
\textbf{11-12GHz}           &            &&                         &                           &                       		    &                            \\  
Stokes $I$\tablenotemark{a} & ($\mu$Jy)  &&  $<$36.3	            &  269.8$\pm$13.6           & 99.2$\pm$10.9        		  		& 22.6$\pm$2.7               \\  
Stokes $V$\tablenotemark{a} & ($\mu$Jy)  &&  $<$36.0		        &  222.4$\pm$12.6           & 86.4$\pm$11.2         		    & 9.6$\pm$2.7\,\tablenotemark{d}               \\  
S/N                         & ($I$, $V$) && \nodata	 			    &  19.8, 17.7               & 9.1, 7.7              		    & 8.4, 3.6                   \\  
Circ. Poln\tablenotemark{b} & (\%)       && \nodata	 			    &  82.2$_{-5.7}^{+6.8}$     & 86.1$_{-14.2}^{+8.0}$ 		    & 41.9$_{-11.5}^{+15.2}$     \\  
\enddata 
\tablenotetext{a}{Reported flux densities are integrated over the FWHM of the full-bandwidth 60 s resolution data.  Fixing fit parameters can result in overestimated uncertainties on the integrated and peak flux densities, so we report the rms image noise as the uncertainty $\sigma_\mathrm{rms}$.   For targets with a clear visual non-detection, we list  3$\sigma_\mathrm{rms}$. }
\tablenotetext{b}{Reported polarization fractions are highest-likelihood values, given the measured Stokes $I$ and Stokes $V$ flux densities.  Uncertainties reflect upper and lower bounds of 68.27\% confidence intervals. Negative and positive values indicate left- and right- circular polarizations, respectively.  Lower-bound 68.27\% and 99.73\% confidence intervals are given for sub-bands with 100\% circular polarization.  Upper bounds are given in parentheses for objects without detectable levels of Stokes $V$ emission, assuming a 3$\sigma_\mathrm{rms}$ flux density and right circular polarization.}
\tablenotetext{c}{Tentative image detection (no clearly visually distinguishable ($I$, $V$) point source). Bootstrapped significance is 99.20\%.}
\tablenotetext{d}{Tentative image detection (no clearly visually distinguishable Stokes $V$ point source). Possible Stokes $I$ point sources are apparent at the expected location of 2M1237 but are not clearly distinguishable by eye from the noise in the image.   Bootstrapped significance is 99.93\% (8--9 GHz) and 99.39\% (11--12~GHz). }
\end{deluxetable*}

\setlength{\tabcolsep}{0.03in}
\begin{deluxetable*}{llccccccccc}[ht!]
\tablecaption{2M1047: Pulsed and Quiescent Emission \label{table:2m1047}}
\tablehead{
	\colhead{}								&
	\colhead{}								&
	\colhead{}								&
	\colhead{Pulse 1}						&
	\colhead{Pulse 2}						&
	\colhead{Pulse 3}						&
	\colhead{Pulse 4}						&
	\colhead{Pulse 5}						&
	\colhead{Pulse 6}						&
	\colhead{Pulses 3--5}						&
	\colhead{Quiescent}				            }             
\startdata
\textbf{12--18GHz} 	        & 			&&							& 							&							&							& 						&		        			&		        		&							\\	
Stokes $I$\tablenotemark{a}	& ($\mu$Jy)	&& 47.0$\pm$14.8\,\tablenotemark{c} & 50.7$\pm$13.3		& 63.8$\pm$12.9				& $<$46.8					& 71$\pm$11.6			& 31.0$\pm$7.0  			&  54.0$\pm$7.1 		& 7.4$\pm$2.2\,\tablenotemark{e}\\	
Stokes $V$\tablenotemark{a}   & ($\mu$Jy)	&& -46.4$\pm$14.3\,\tablenotemark{c}& $<$36.3			& $<$44.7					& $<$50.1 					& -56$\pm$10.6			& $<$19.5				& -33.3$\pm$8.3 		& $<$5.4				\\	
S/N					        & ($I$, $V$)	&& 3.2, 	3.2				& 3.8, \nodata				& 4.9, \nodata				&	 	\nodata				& 6.1, 5.3				& 4.4, \nodata  			&	7.6, 4.0    		&  3.4, \nodata				\\	
Circ. Poln\tablenotemark{b} & (\%)		&&-90.2$_{-2.0}^{+38.3}$	& (-67.1$_{-20.0}^{+24.2}$)	&(-67.3$_{-19.4}^{+24.0}$)	&	 	\nodata				&-76.8$_{-13.8}^{+16.8}$&(-59.9$_{-23.4}^{+20.1}$)&-60.6$_{-39.0}^{+45.4}$& 		\nodata				\\
\textbf{12--13.5GHz}        & 			&&							&							&							&							&						&		      			&		        		&							\\	
Stokes $I$\tablenotemark{a}   & ($\mu$Jy)	&& $<$91.5\,\tablenotemark{c}		& 143.4$\pm$17.6	& $<$84.6					& $<$78.9					& $<$81.0				& $<$38.4 	 			&	 $<$48.3			& 20.4$\pm$4.1\,\tablenotemark{e}\\	
Stokes $V$\tablenotemark{a}   & ($\mu$Jy)	&&-129.6$\pm$24.6\,\tablenotemark{c}& -78.9$\pm$21.7	& $<$81.3					& $<$77.7					& $<$78.6				& $<$38.3				&	 $<$45.6	   	 	& $<$12.3				\\	
S/N					        & ($I$, $V$)	&&	\nodata, 5.3			& 8.1, 3.6					& \nodata					& \nodata					& \nodata				& \nodata				& 	\nodata			& 5.0, \nodata			\\	
Circ. Poln\tablenotemark{b} & (\%)		&&	(-72.8, -38.0)		 	&-54.2$_{-18.8}^{+14.6}$	& \nodata					& \nodata					& \nodata				& \nodata				& 	\nodata			& 		\nodata			\\
\textbf{13.5--15GHz}        & 			&&							&							&							&							&						&		      		 	&		        		&						\\	
Stokes $I$\tablenotemark{a}   & ($\mu$Jy)	&&	$<$105.0				& $<$71.7					& $<$80.4					& $<$72.0					& $<$72.3				& $<$41.7				&	$<$44.1     		& $<$10.5				\\	
Stokes $V$\tablenotemark{a}   & ($\mu$Jy)	&&	$<$110.4				& $<$68.4					& $<$81.6					& $<$75.9					& $<$71.1				& $<$40.8				&	$<$43.5	    		& $<$11.1				\\	
S/N					        & ($I$, $V$)	&&	\nodata					& \nodata					& \nodata					& \nodata					& \nodata				& \nodata  				& 	\nodata			& \nodata               \\	
Circ. Poln\tablenotemark{b} & (\%)		&&	\nodata					& \nodata					& \nodata					& \nodata					& \nodata				& \nodata  				& 	\nodata			& \nodata               \\	
\textbf{15--16.5GHz}        & 			&&							&							&							&							& 						&       					&		        		& 						\\	
Stokes $I$\tablenotemark{a}   & ($\mu$Jy)	&&	$<$77.4					& $<$66.3					& 125.4$\pm$25.8				& 93.3$\pm$19.9				& 93.7$\pm$24.0			& $<$38.7  				&  1($I$, $V$)05.2$\pm$13.7 	& $<$12.3				\\	
Stokes $V$\tablenotemark{a}   & ($\mu$Jy)	&&	$<$77.9					& $<$67.8					& $<$84.6					& $<$69.6					& $<$63.9				& $<$41.1  				&  -46.7$\pm$12.8 	& $<$12.0				\\	
S/N					        & ($I$, $V$)	&&	\nodata					& \nodata					& 4.9,  \nodata				& 9.4, \nodata				& 3.9,  \nodata			& \nodata  				&	7.7, 3.6	  		& \nodata               \\	
Circ. Poln\tablenotemark{b} & (\%)		&&	\nodata					& \nodata					& (-64.8$_{-20.8}^{+19.2}$)	& (-71.4$_{-16.8}^{+26.5}$)	&(-64.1$_{-21.8}^{+22.4}$)& \nodata  			&-43.6$_{-50.9}^{+35.7}$& \nodata               \\	
\textbf{16.5--18GHz}        & 			&&							&							&							&							& 						&		       		 	&		        		&						\\	
Stokes $I$\tablenotemark{a}   & ($\mu$Jy)	&&	$<$99.3					& $<$90.3					& $<$102.9					& $<$91.2					& 91.5$\pm$28.7\,\tablenotemark{d}	& $<$54.0	&	$<$57.3		  	& $<$15.6				\\	
Stokes $V$\tablenotemark{a}   & ($\mu$Jy)	&&	$<$108.6				& $<$88.8					& $<$95.1					& $<$99.9					& -94.9$\pm$24.9\,\tablenotemark{d}	& $<$52.8	&	$<$56.4	        	& $<$15.6			\\	
S/N					        & ($I$, $V$)	&&	\nodata					& \nodata					& \nodata					& \nodata					& 3.2, 3.8				& \nodata       			& 	\nodata			& \nodata 				\\	
Circ. Poln\tablenotemark{b} & (\%)		&&	\nodata					& \nodata					& \nodata					& \nodata					& -58.0, -14.3			& \nodata       			& 	\nodata			& \nodata               \\	
\enddata
\tablenotetext{a}{Reported flux densities are integrated over the FWHM of the full-bandwidth 60 s resolution data.  Fixing fit parameters can result in overestimated uncertainties on the integrated and peak flux densities, so we report the rms image noise as the uncertainty $\sigma_\mathrm{rms}$.  For targets with a clear visual non-detection, we list  3$\sigma_\mathrm{rms}$. }
\tablenotetext{b}{Reported polarization fractions are highest-likelihood values, given the measured Stokes $I$ and Stokes $V$ flux densities.  Uncertainties reflect upper and lower bounds of 68.27\% confidence intervals. Negative and positive values indicate left and right circular polarizations, respectively.  Upper-bound 68.27\% and 99.73\% confidence intervals are given for sub-bands with -100\% circular polarization.  Lower bounds are given in parentheses for objects without detectable levels of Stokes $V$ emission, assuming a 3$\sigma_\mathrm{rms}$ flux density and left circular polarization.}
\tablenotetext{c}{Possible Stokes $I$ point sources at the expected location of 2M1047 for 12--18~GHz and 12--13.5~GHz are not clearly distinguishable by eye from the noise in the image.  For 12--18~GHz, the significance of the measured Stokes $I$ and Stokes $V$ flux densities bootstrapped from 10,000 trials in a 2.7\arcmin$\times$2.7\arcmin image are 99.24\% and 99.32\%, respectively. For 12--13.5~GHz, the measured flux density at the expected location for 2M1047 is 104.4$\pm$30.5 $\mu$Jy, with a bootstrapped significance of 99.92\%.  However, the Stokes $V$ flux density may be statistically significant with a bootstrapped significance of $\geq$99.99\%.  Although the Stokes $V$ flux is higher than the measured flux for Stokes $I$, the discrepancy is within the rms noise.   We classify these detections as tentative. }
\tablenotetext{d}{Tentative detection. Bootstrapped significance is 99.29\% (Stokes $I$ only), 99.63\% (Stokes $I$ with acceptable percent circular polarization constrained to 99.73\% confidence interval), and 99.99\% (Stokes $I$ with acceptable percent circular polarization constrained to 68.27\%  confidence interval).  For additional discussion, see \S\ref{sec.Timeseries_15a374}.}
\tablenotetext{e}{Tentative detections. Possible Stokes $I$ point sources are apparent at the expected location of 2M1047 for 12--18~GHz and 12--13.5~GHz, but they are not clearly distinguishable by eye from the rms noise image.  Bootstrapped significance levels are 99.59\%  and 99.98\%, respectively. }
\tablenotetext{\,}{   \hspace{3cm}}
\tablenotetext{\,}{   \hspace{3cm}}
\end{deluxetable*}

\setlength{\tabcolsep}{0.03in}
\begin{deluxetable*}{llccccccccc}[htp]
\tablecaption{SIMP0136 \& 2M1043: Pulsed and Quiescent Emission \label{table:simp0136}}
\tablehead{
	\colhead{}								&
	\colhead{}								&
	\colhead{\hspace*{.5cm}}				&
	\multicolumn{2}{l}{SIMP0136}		    &
	\colhead{\hspace*{.5cm}}				&
	\multicolumn{5}{l}{2M1043}			     \\
	\colhead{}								&
	\colhead{}								&
	\colhead{}								&
	\colhead{Pulse 1}						&
	\colhead{Quiescent}						&
	\colhead{}								&
	\colhead{Pulse 1}						&
	\colhead{Pulse 2}						&
	\colhead{Pulse 3}						&
	\colhead{All Pulses}					&
	\colhead{Quiescent}						
			}                              
\startdata              	       
\textbf{8--12~GHz} 	        & 			&&								&							&&							&							&							&							&							\\		
Stokes $I$\tablenotemark{a} & ($\mu$Jy)	&&	51.5$\pm$5.7					&	11.5$\pm$1.2			&&	40.8$\pm$8.0			&	60.5$\pm$7.4			&	51.5$\pm$5.6			&	49.3$\pm$4.2			&	$<$3.6					\\		
Stokes $V$\tablenotemark{a}	& ($\mu$Jy)	&&	-33.3$\pm$5.9 				&	-7.1$\pm$1.2			&&	-34.7$\pm$8.3			&	$<$24.6					&	-36.5$\pm$6.6			&	-30.3$\pm$4.3			&	$<$3.6					\\
S/N					        & ($I$, $V$)&&	9.0, 5.6						&	9.6, 5.9				&&	5.1, 4.2				&	8.2, \nodata			&	9.2, 5.5				&	11.7, 7.0				&		 \nodata			\\			
Circ. Poln\tablenotemark{b} & (\%)		&& -63.9$_{-15.5}^{+11.5}$		&	-61.1$_{-14.4}^{+10.5}$ &&	-82.0$_{-10.0}^{+24.1}$	&(-40.1$_{-16.7}^{+13.0}$)	&	-70.0$_{-15.2}^{+13.2}$	&	-61.0$_{-11.7}^{+8.9}$	&		 \nodata			\\		
\textbf{8--10~GHz}  	   	& 			&&								&							&&							&							&							&							&							\\		
Stokes $I$\tablenotemark{a} & ($\mu$Jy)	&&	57.2$\pm$8.6					&	20.9$\pm$1.8 			&&	50.1$\pm$11.2			&	54.8$\pm$9.3			&	55.1$\pm$8.6			&	55.9$\pm$5.8			&	$<$4.8					\\	
Stokes $V$\tablenotemark{a} & ($\mu$Jy)	&&-34.9$\pm$8.1\,\tablenotemark{c}&	-8.1$\pm$1.8			&&	-48.7$\pm$10.9			&	$<$33.6					&	-48.7$\pm$9.0			&	-44.3$\pm$5.9			&	$<$4.8					\\
S/N					        & ($I$, $V$)&&	6.7, 4.3						&	11.6, 4.5				&&	4.5, 4.5				&	5.9, \nodata			&	6.4, 5.4				&	9.6, 7.5				&		 \nodata			\\			
Circ. Poln\tablenotemark{b} & (\%)		&&	-59.7$_{-19.4}^{+13.8}$		&	-38.5$_{-10.2}^{+8.5}$	&&	-92.7$_{-1.3}^{+29.8}$	& (-59.6$_{-22.1}^{+19.9}$)	& -86.3$_{-7.3}^{+20.6}$	& -78.4$_{-12.0}^{+11.8}$	&		 \nodata			\\			
\textbf{10--12~GHz} 	    	& 			&&								&							&&							&							&							&							&							\\			
Stokes $I$\tablenotemark{a} & ($\mu$Jy)	&&40.5$\pm$8.5\,\tablenotemark{d}&	$<$6.3					&&	$<$32.7					&	58.9$\pm$11.6			&	44.0$\pm$8.5			&	42.1$\pm$5.7			&	$<$5.1					\\		
Stokes $V$\tablenotemark{a} & ($\mu$Jy)	&&	$<$29.7						&	$<$4.8					&&	$<$33.3					&	$<$35.7					&	$<$30.3		       		&	$<$18.0					&	$<$4.8					\\			
S/N					        & ($I$, $V$)&&	4.7, \nodata 				&   	\nodata			    &&	\nodata		 			&	5.1, \nodata			&	5.2, \nodata			&	7.4, \nodata			&		 \nodata			\\
Circ. Poln\tablenotemark{b} & (\%)		&&	(-70.3$_{-17.6}^{+25.8}$)	&   	\nodata			    &&	\nodata					& (-58.4$_{-23.2}^{+19.4}$)	& (-66.4$_{-19.8}^{+23.4}$)	& (-42.0$_{-18.2}^{+13.5}$)	&		 \nodata			\\	
\textbf{8--9~GHz}   	    	& 			&&								&							&&							&							&							&							&							\\			
Stokes $I$\tablenotemark{a} & ($\mu$Jy)	&&	69.9$\pm$12.9				&	20.2$\pm$2.1			&&	43.4$\pm$17.5			&	$<$47.7					&	59.3$\pm$12.6			&	53.0$\pm$9.0			&	$<$7.2					\\		
Stokes $V$\tablenotemark{a} & ($\mu$Jy)	&&	$<$38.1						&	-7.5$\pm$2.0			&&	-67.1$\pm$15.8			&	$<$49.2					&	-51.1$\pm$11.8			&	-51.5$\pm$8.1			&	$<$10.2					\\	
S/N				    	    	& ($I$, $V$)&&  5.4, \nodata           			&	9.6, 3.8			    &&		4.1, 5.0            & 		 \nodata			&	4.7, 4.3		   		&	5.9, 6.4				&		 \nodata			\\
Circ. Poln\tablenotemark{b} & (\%)		&&  (-52.7$_{-23.7}^{+17.0}$)	& -36.7$_{-12.1}^{+9.6}$	&&	-73.0, -35.6 			& 		 \nodata			&	-82.5$_{-9.7}^{+24.0}$	&	-94.5$_{-1.0}^{+22.9}$	&		 \nodata			\\	
\textbf{9--10~GHz}	        & 			&&								&							&&							&							&							&							&							\\			
Stokes $I$\tablenotemark{a} & ($\mu$Jy)	&&44.3$\pm$12.2\,\tablenotemark{d}&	13.2$\pm$2.4			&&	$<$45.0					&57.7$\pm$13.3\,\tablenotemark{e}&	56.1$\pm$12.0		&	59.7$\pm$7.8			&	$<$6.9					\\	
Stokes $V$\tablenotemark{a} & ($\mu$Jy)	&&	$<$39.6						&	-9.1$\pm$2.1			&&	$<$42.0					&	$<$45.9					&	-48.0$\pm$13.5			&	-36.3$\pm$8.5			&	$<$6.9					\\	
S/N				    	    	& ($I$, $V$)&&   3.6, \nodata		    			&	5.5, 4.3				&&  		\nodata			& 	4.3, \nodata			&	4.7, 3.6				&	7.7, 4.3				&		 \nodata			\\		
Circ. Poln\tablenotemark{b} & (\%)		&&  (-83.2$_{-7.7}^{+35.3}$)  	& -66.8$_{-19.2}^{+16.1}$  	&&  		\nodata			& (-75.6$_{-13.9}^{+29.4}$)	&	-81.9$_{-9.6}^{+28.3}$	& -59.8$_{-18.4}^{+13.9}$	&		 \nodata			\\	
\textbf{10--11~GHz}         & 			&&								&							&&							&							&							&							&							\\			
Stokes $I$\tablenotemark{a} & ($\mu$Jy)	&&41.5$\pm$12.0\,\tablenotemark{d}&		$<$9.0				&&	$<$48.3					&	$<$49.8					&	$<$41.4					&	40.1$\pm$7.8			&	$<$6.9					\\	
Stokes $V$\tablenotemark{a} & ($\mu$Jy)	&&	$<$33.9						&	    $<$6.9				&&	$<$44.7					&	$<$50.1					&	$<$42.9					&	$<$25.8					&	$<$6.9					\\
S/N					        & ($I$, $V$)&&	 3.5, \nodata				&  		\nodata			   	&&  		\nodata		   	& 		 \nodata			&	\nodata					&	5.1, \nodata			&		 \nodata			\\	
Circ. Poln\tablenotemark{b} & (\%)		&&  (-75.6$_{-14.0}^{+29.9}$)	&   	\nodata			    &&  		\nodata			& 		 \nodata			&	\nodata					&(-62.0$_{-21.9}^{+21.1}$)	&		 \nodata			\\	
\textbf{11--12~GHz} 	    	& 			&&								&							&&							&							&							&							&							\\	
Stokes $V$\tablenotemark{a} & ($\mu$Jy)	&&	$<$43.8 		    				&		$<$10.5				&&	$<$48.0					&	$<$51.0					&	$<$46.2					&	$<$27.6					&	$<$7.5					\\	
Stokes $V$\tablenotemark{a} & ($\mu$Jy)	&&	$<$44.1	        				&		$<$9.0				&&	$<$50.7					&	$<$52.2					&	$<$43.8					&	$<$27.9					&	$<$7.2					\\			
S/N				    	    		& Stokes $I$	&&	\nodata						&   	\nodata			    &&  		\nodata			& 		 \nodata			&	\nodata					&		 \nodata			&		 \nodata			\\	
Circ. Poln\tablenotemark{b} & (\%)		&&	\nodata						&   	\nodata			    &&  		\nodata			& 		 \nodata			&	\nodata					&		 \nodata			&		 \nodata			\\ \hline
\enddata
\tablenotetext{a}{Reported flux densities are integrated over the FWHM of the full-bandwidth 60 s resolution data.  Fixing fit parameters can result in overestimated uncertainties on the integrated and peak flux densities, so we report the rms image noise as the uncertainty $\sigma_\mathrm{rms}$.   For targets with a clear visual non-detection, we list 3$\sigma_\mathrm{rms}$. }
\tablenotetext{b}{Reported polarization fractions are highest-likelihood values, given the measured Stokes $I$ and Stokes $V$ flux densities.  Uncertainties reflect upper and lower bounds of 68.27\% confidence intervals. Negative and positive values indicate left- and right- circular polarizations, respectively.  Upper-bound 68.27\% and 99.73\% confidence intervals are given for sub-bands with -100\% circular polarization.  Lower bounds are given in parentheses for objects without detectable levels of Stokes $V$ emission, assuming a 3$\sigma_\mathrm{rms}$ flux density  and left circular polarization.}
\tablenotetext{c}{Tentative image detection (no clearly visually distinguishable Stokes $V$ point source). Bootstrapped significance is 99.67\%.}
\tablenotetext{d}{Tentative image detection (no clearly visually distinguishable Stokes $I$ point source). Bootstrapped significance is 99.66\% (10--12~GHz), 98.78\% (9--10~GHZ), 98.80\% (10--11~GHZ).  }
\tablenotetext{e}{Tentative image detection (no clearly visually distinguishable Stokes $I$ point source). Bootstrapped significance is 99.54\%.}
\end{deluxetable*}

\setlength{\tabcolsep}{0.03in}
\begin{deluxetable*}{llcccccccc}[h!]
\tablecaption{SDSS0423: Pulsed and Quiescent Emission \label{table:sdss0423}}
\tablehead{
	\colhead{}								&
	\colhead{}								&
	\colhead{}								&
	\colhead{Pulse R1}						&
	\colhead{Pulse R2}						&
	\colhead{Pulse L1}						&
	\colhead{Pulse L2}						&
	\colhead{Pulse L3}						&
	\colhead{Pulse L4}						&
	\colhead{Quiescent}				            }             
\startdata
\textbf{8--12GHz} 	       	& 			&& 		 					& 						& 		 		        	& 						& 		 				& 				    & 		 				\\		
Stokes $I$\tablenotemark{a} & ($\mu$Jy)	&& 86.9$\pm$9.6				& 82.0$\pm$9.5			& 	99.2$\pm$8.2	    		& 58.0$\pm$6.6			& 64.6$\pm$5.0 	 		& 101.0$\pm$9.1     & 	$<$5.1 				\\	
Stokes $V$\tablenotemark{a} & ($\mu$Jy)	&& $<$29.7					& $<$24.0				& 	-94.2$\pm$6.7	    	& -37.0$\pm$7.0			&-34.3$\pm$4.6	 		& -99.3$\pm$10.1    & 	$<$5.7				\\	
S/N					        & ($I$, $V$)&& 	9.1, \nodata 			& 	8.6, \nodata 		& 	12.1, 14.1	 	    	& 10.1,	7.6 				& 12.9, 7.5				& 11.1, 9.8 	    	& 	\nodata 			\\	
Circ. Poln\tablenotemark{b} & (\%)		&&(33.8$_{-11.0}^{+13.5}$)	&(28.9$_{-9.4}^{+11.8}$)&-94.3$_{-2.8}^{+10.9}$ 	&-63.0$_{-16.0}^{+12.1}$& -52.8$_{-9.3}^{+7.4}$	& -81.8,-62.2	    & 	\nodata 			\\	
\textbf{8--10GHz} 	        & 			&& 		 					& 						& 		 		    	    & 						& 		 				& 				    & 		 				\\	
Stokes $I$\tablenotemark{a} & ($\mu$Jy)	&& 	90.2$\pm$11.4			& 96.5$\pm$10.6	 		& 121.4$\pm$11.7    	  	& 69.3$\pm$8.7			& 82.6$\pm$6.0			& 152.6$\pm$13.3    & 	$<$6.6				\\	
Stokes $V$\tablenotemark{a} & ($\mu$Jy)	&& 	51.9$\pm$10.9			& $<$34.5				&-132.3$\pm$12.5        	& -67.1$\pm$9.8			& -49.6$\pm$6.0			& -151.9$\pm$15.8	& 	$<$6.6 				\\	
S/N				    	    & ($I$, $V$)&& 	7.9, 4.8 				& 	9.1, \nodata 		& 10.4, 10.6	      	  	& 8.0, 6.8				& 	13.8, 8.3			& 11.5, 9.6		    & 	\nodata 			\\	
Circ. Poln\tablenotemark{b} & (\%)		&&56.6$_{-11.8}^{+16.8}$		&(35.3$_{-11.5}^{+14.1}$)& 	-86.4, -68.0 	    	&-95.3$_{-0.7}^{+20.6}$	&-59.7$_{-9.6}^{+7.6}$  & -82.3, -62.5		& 	\nodata 			\\
\textbf{10--12GHz} 	        & 			&& 		 					& 						& 		 	        		& 						& 		 				& 				    & 		 				\\	
Stokes $I$\tablenotemark{a} & ($\mu$Jy)	&& 	83.7$\pm$14.5			& 56.5$\pm$13.8			& 67.2$\pm$12.9\,\tablenotemark{d}&  $<$10.5		& 53.1$\pm$8.8			& $<$41.7		    & 	$<$7.2 				\\	
Stokes $V$\tablenotemark{a} & ($\mu$Jy)	&& $<$39.9					& $<$39.0				& $<$37.5		        	&  $<$30.3				& $<$23.4				& $<$45.9		    & 	$<$7.5 				\\	
S/N					        & ($I$, $V$)&& 	5.8, \nodata 			&  4.1, \nodata 			& 5.2, \nodata			& 	\nodata 				& 	6.0, \nodata 		& \nodata		    & 	\nodata 			\\	
Circ. Poln\tablenotemark{b} & (\%)		&&(46.3$_{-14.7}^{+22.2}$)	&(65.2$_{-23.0}^{+21.0}$)&(-53.8$_{-24.0}^{+17.4}$)& 	\nodata 			&(-42.9$_{-55.4}^{+42.9}$)& \nodata		    & 	\nodata 			\\
\textbf{8--9GHz}    	    & 			&& 		 					& 						& 		 		        	& 						& 		 				& 				    & 		 				\\	
Stokes $I$\tablenotemark{a} & ($\mu$Jy)	&& 73.8$\pm$18.5				& 111.6$\pm$14.0			& 133.5$\pm$16.3	   	 	& 72.2$\pm$12.9 			& 95.5$\pm$8.9			& 218.1$\pm$21.0		& 	$<$8.4				\\	
Stokes $V$\tablenotemark{a} & ($\mu$Jy)	&& 65.7$\pm$16.0\,\tablenotemark{c}& $<$44.1			&-166.7$\pm$16.1	    		& -78.5$\pm$13.7			& -52.8$\pm$9.1			& -209.9$\pm$21.0	& 	$<$8.4   			\\	
S/N				          	& ($I$, $V$)&& 	4.0, 4.1					& 8.0,  \nodata 			& 8.2, 10.4		        	& 5.6, 5.7				& 10.7, 5.8				& 10.4, 10.0	 		& 	\nodata 			\\	
Circ. Poln\tablenotemark{b} & (\%)		&& 83.9$_{-26.6}^{+8.3}$		&(38.9$_{-12.6}^{+16.4}$)& 	-88.4, -70.2			& -73.4 ,-42.1 			&-54.8$_{-12.5}^{+9.5}$	&-95.4$_{-1.5}^{+14.8}$& 	\nodata 		\\	
\textbf{9--10GHz} 	        & 			&& 		 					& 						& 		 		        	& 						& 		 				& 				    & 		 				\\	
Stokes $I$\tablenotemark{a} & ($\mu$Jy)	&& 110.2$\pm$19.0			& 93.6$\pm$14.8			& 102.3$\pm$15.9 	    	& 60.6$\pm$12.0			& 69.6$\pm$8.8 			& 86.5$\pm$18.2	    & 	$<$8.7  			\\	
Stokes $V$\tablenotemark{a} & ($\mu$Jy)	&& $<$54.6					& $<$46.8				& -103.3$\pm$15.3       	& -56.5$\pm$12.4			& -49.8$\pm$8.5			& -107.0$\pm$21.0   & 	$<$8.7 				\\	
S/N				    	    & ($I$, $V$)&& 	5.8, \nodata 			& 6.3,  \nodata 			& 6.4, 6.8  		    		& 5.0, 4.6				& 7.9, 5.9				& 4.8, 5.1		    & 	\nodata 			\\	
Circ. Poln\tablenotemark{b} & (\%)		&&(48.1$_{-15.3}^{+22.6}$)	&(48.8$_{-15.7}^{+21.8}$)& -74.7, -47.8	    		&-89.8$_{-10.2}^{+62.8}$&-70.4$_{-15.6}^{+12.8}$& -72.9, -36.8		& 	\nodata 			\\	
\textbf{10--11GHz}          & 			&& 		 					& 						& 		 		        	& 						& 		 				& 				    & 		 				\\	
Stokes $I$\tablenotemark{a} & ($\mu$Jy)	&& 	82.7$\pm$17.9			& $<$52.8				& $<$49.5		        	& $<$39.0				& $<$31.5				& $<65.7$		    & 	$<$8.4 				\\	
Stokes $V$\tablenotemark{a} & ($\mu$Jy)	&& 	$<$57.0					& $<$48.9				& $<$48.0		        	& $<$38.7				& $<$30.3				& $<60.3$		    &  	$<$8.4 				\\	
S/N				    	    & ($I$, $V$)&& 	4.6, \nodata 			&   \nodata 				& 	\nodata 				& 	\nodata 				&  \nodata		    	& \nodata		    & 	\nodata 			\\	
Circ. Poln\tablenotemark{b} & (\%)		&&(65.9$_{-23.3}^{+20.3}$)	& 	\nodata 				& 	\nodata 				& 	\nodata 				&  \nodata		    	& \nodata		    & 	\nodata 			\\	
\textbf{11-12GHz}   	    & 			&& 		 					& 						& 		 		        	& 						& 		 				& 				    & 		 				\\	
Stokes $I$\tablenotemark{a} & ($\mu$Jy)	&& 	$<$71.4 					& $<$66.0				& 	$<$88.2		        	& $<$51.3				& $<$39.9	 			& $<$75.3		    & 	$<$12.9 			\\	
Stokes $V$\tablenotemark{a} & ($\mu$Jy)	&& 	$<$72.6					& $<$65.7				& 	$<$97.2 	       	 	& $<$51.3				& $<$38.1				& $<$76.5		    & 	$<$16.2				\\	
S/N			    		    & ($I$, $V$)&& 	\nodata 					& \nodata				& 	\nodata 		    		& \nodata				& 	\nodata 				& \nodata		    & 	\nodata 			\\	
Circ. Poln\tablenotemark{b} & (\%)		&& 	\nodata 					& \nodata				& 	\nodata 		    		& \nodata				& 	\nodata 				& \nodata		    & 	\nodata 			\\	
\enddata
\tablenotetext{a}{Reported flux densities are integrated over the FWHM of the full-bandwidth 60 s resolution data.  Fixing fit parameters can result in overestimated uncertainties on the integrated and peak flux densities, so we report the rms image noise as the uncertainty $\sigma_\mathrm{rms}$. For targets with a clear visual non-detection, we list  3$\sigma_\mathrm{rms}$. }
\tablenotetext{b}{Reported polarization fractions are highest-likelihood values, given the measured Stokes $I$ and Stokes $V$ flux densities.  Uncertainties reflect the upper and lower bounds of the 68.27\% confidence intervals. Negative and positive values indicate left and right circular polarizations, respectively.  Lower-bound (upper-bound) 68.27\% and 99.73\% confidence intervals are given for objects with 100\% right (left) circular polarization.  For pulses without detectable levels of Stokes $V$ emission, we give upper bounds on the percent circular polarization in parentheses by assuming a 3$\sigma_\mathrm{rms}$ flux density  and right circular polarization (for R1 and R2) or left circular polarization (for L1 - L4).}
\tablenotetext{c}{Tentative image detection (no clearly visually distinguishable Stokes $V$ point source).  Bootstrapped significance is 99.27\%.}
\tablenotetext{d}{Tentative Stokes $I$ image detection is difficult to distinguish from image noise.  Bootstrapped significance is  $\geq$99.99\%.}
\end{deluxetable*}

\setlength{\tabcolsep}{0.05in}
\begin{deluxetable}{lcccc}[htp]
\tabletypesize{\scriptsize}
\tablecaption{Periodogram Results \label{table:periods}}
\tablehead{
	\colhead{}						  &
	\colhead{L-S}					  &
	\colhead{Plavchan}			      &
	\colhead{BLS}					  &
	\colhead{Adopted}										            				
		    \\
	\colhead{Object}					 &
	\colhead{(hr)}				     &
	\colhead{(hr)}		 			 &
	\colhead{(hr)}			         &	
	\colhead{(hr)}				}                              
\startdata
2M1047\tablenotemark{a}  	& 0.59$^{+0.02}_{-0.02}$ & 1.78$^{+0.07}_{-0.06}$ & 1.77$^{+0.05}_{-0.05}$ & 1.78$^{+0.07}_{-0.06}$\\
SIMP0136\tablenotemark{b}	& 2.33$^{+0.43}_{-0.32}$ & 2.88$^{+0.34}_{-0.27}$ & 2.74$^{+0.80}_{-0.50}$ & 2.88$^{+0.34}_{-0.27}$\\
2M1043						& 2.36$^{+0.42}_{-0.31}$ & 2.19$^{+0.15}_{-0.12}$ & 2.21$^{+0.14}_{-0.13}$ & 2.21$^{+0.14}_{-0.13}$\\
2M1237						& 2.21$^{+0.59}_{-0.39}$ & 2.28$^{+0.10}_{-0.09}$ & 2.28$^{+0.13}_{-0.12}$ & 2.28$^{+0.10}_{-0.09}$\\
SDSS0423					& 1.44$^{+0.19}_{-0.15}$ & 1.49$^{+0.11}_{-0.10}$ & 1.47$^{+0.13}_{-0.11}$ & 1.47$^{+0.13}_{-0.11}$\\
\enddata
\tablenotetext{a}{No periodicity is clearly observable in the pulsed emission from 2M1047.  The detected periodicity is consistent with the $\sim$1.77 hr C-band pulse period measured by \citet{williams2015}, suggesting that our detected periodicity may be due to the pulsed emission and/or the quiescent emission. For our discussion in \S\ref{sec.c09}, we adopt the rotation period measured by \citet{williams2015}.}
\tablenotetext{b}{No periodicity is observable in the pulsed emission from SIMP0136.  The periods listed here correspond to the non-pulsed quasi-quiescent emission. For our discussion in \S\ref{sec.c09}, we adopt the photometric rotation period $P= 2.3895\pm 0.0005$ hr measured by \citet{croll2016}.}
\end{deluxetable}

\newpage
\section{Discussion}\label{sec.Discussion_15a374}

\subsection{The Curious Case of Highly Circularly Polarized and/or Disappearing Quiescent Emission \label{sec.Quiescent_15a374}}

\citet{kao2016} noted that all known radio brown dwarfs exhibited detectable levels of quiescent emission, and \cite{pineda2017} showed that the quiescent radio luminosities correlated with H$\alpha$ luminosities for confirmed auroral emitters (i.e. with clear rotational modulation in the highly circularly polarized emission component).  This suggested a possible connection between pulsed and quiescent radio processes.  

In contrast, we do not observe detectable levels of quiescent emission from SDSS0423 and 2M1043 for 8--12 GHz or individual 1 or 2 GHz sub-bands, down to Stokes $I$ $3\sigma_{\mathrm{rms}}$ noise levels of 5.1--12.9~$\mu$Jy and 3.6--7.5~$\mu$Jy, respectively.  We also do not observe detectable quiescent emission from 2M1047 at frequencies $\gtrsim$13.5~GHz down to rms noise levels of 10.5--15.6~$\mu$Jy.  For SDSS0423, \citet{kao2016} measured a 4--8~GHz mean quiescent flux density of 26.7$\pm$3.1~$\mu$Jy.  Assuming an upper $3\sigma_{\mathrm{rms}}$ detection limit of 5.1~$\mu$Jy for flux density averaged over 8--12~GHz, the upper limit spectral index is $\alpha \lesssim -3.2 \pm 0.7$ and the corresponding mildly relativistic power-law electron distribution index is $\delta \gtrsim 5.0$. For 2M1043, \citet{kao2016} measured a 4--8~GHz mean quiescent flux density of 16.3$\pm$2.5~$\mu$Jy, which leads to $\alpha \lesssim -3.0 \pm 0.7$  and $\delta \gtrsim 4.7$. 

In the stellar case, typical spectral indices for quiescent radio emission from active M dwarfs are much flatter at $\alpha \sim-0.3$ \citep[e.g.][and references therein]{gudel1993, gudel1994}, although there may be fundamental differences for the brown dwarf case.  While evidence exists that much of the quiescent emission from ultracool dwarfs exhibits behavior consistent with incoherent synchrotron or gyrosynchrotron emisssion \citep[e.g.][]{ravi2011, williams2015b}, there have been some objects that depart from this model.

At least some component of the `quiescent' non-pulsed emission may be coherent.  The steep spectral index implied by the drop-off in quiescent emission is atypical (but not impossible) for nonthermal gyrosynchrotron or synchrotron emission \citep{dulk1985, melrose2006} and may be more indicative of an emission cutoff.  Such a model has been proposed for solar quiescent emission with electron power-law indices $\delta \approx 2-4$ and weak $\sim$100~G fields \citep{pallavicini1985, white1989, whiteFranciosini1995, umana1998}, including for both plasma and gyrosynchrotron emission.  

Evidence for a coherent mechanism at play in the quiescent component precedes the data presented here.  For instance, the L3.5 dwarf 2MASS~J00361617+1821104 exhibits broadly varying emission with duty cycles $\sim$30\% of the rotational period \citep{berger2002, hallinan2008}.  This emission can be decomposed into two components: (1) a periodic and highly circularly polarized component, which \citet{hallinan2008} attributed to ECM, and (2) a component that was largely unpolarized for two out of three of the observed rotation periods.  In the third rotation period, this second component emitted two narrower peaks with up to $\sim$75\% right- and left-circular polarization, respectively.  This same feature was observed in data separated by 18 months, which demonstrated the longevity of this high degree of circular polarization and ruled out incoherent gyrosynchrotron as a mechanism.  To explain the observed short-term variability in the degree of polarization, \citet{hallinan2008} argued that local conditions in the emitting region could plausibly depolarize the emission, a phenomenon that commonly occurs in the strongly circularly polarized millisecond spikes of solar radio emission, such that polarization fractions can range from 0\% to 10\% \citep{benz1986}.

Similar dual-component varying emission has been observed in the T6 dwarf WISEP~J112254.73+255021.5.  The first component comprises clear bursts in left-circular polarization \citep{routeWolszczan2016}.  The second component is broadly varying in both the right- and left-circularly polarized flux density, with spectral index $\alpha = -1.5\pm 0.3$ and a high degree of circular polarization ($>$50\%) that is present for nearly the entire duration of a 162-minute observation \citep{williams2017}.  This second component is similar to what we observed in SIMP0136 and 2M1237. These two objects have flatter spectra than SDSS0423 and 2M1043 if no variability is assumed, with spectral indices $\alpha \approx -2.1 \pm 0.4$ and $\alpha \approx -0.9 \pm 0.3$, respectively.

In the case that the non-pulsed emission is coherent, plasma emission is unlikely because the plasma density in a cool brown dwarf such as SDSS0423 is expected to be tenuous in comparison to the solar corona, and the plasma frequency scales with the electron density as $\nu_p \propto n_e^{1/2}$.  For a gas to exhibit plasma-like behaviors, electron-electron interactions should dominate over electron-neutral interactions.  In models of thermal ionization for temperatures characteristic of M--T dwarfs, \citet{rodriguez-barrera2015} find that while M dwarfs can expect $\sim$10$^{-1}$ fraction of ionization in their atmospheres, this rapidly drops to $\sim$10$^{-4}-10^{-3}$ for 1000~K objects.  Additionally, the presence of plasma would correlate with X-ray emission, but L and later brown dwarfs remain underluminous in X-ray compared to their warmer counterparts \citep{williams2014}. 
 The other plausible coherent mechanism would be ECM emission in the form of superposed flares, as observed for 2MASS~J00361617+1821104 \citep{hallinan2008}.  However, if the mechanism generating this quiescent emission is indeed related to the pulsed emission, the presence of the pulses observed in the same frequency bands would preclude the observed cutoff, unless the emitting regions traced different magnetic field strengths.  This scenario could account for the strong circular polarization of the non-flaring emission from SIMP0136, 2M1237, and WISEP~J112254.73+255021.5.  

Another possible explanation is that the quiescent emission may exhibit long-term variability.  Such variability has been previously reported in other brown dwarfs.  For instance, \citet{antonova2007} did not detect any radio emission from a 9~hr observation (with 3$\sigma$ upper limit $\sim$45~$\mu$Jy) of 2MASS~J05233822-1403022 (L2.5) on 2006 September 23, which \citet{berger2010} also reported for observations on 2008 December 30. Archival data analyzed by \citet{antonova2007} revealed that this same object was also not detected on 2004 May 03 with a 3$\sigma$ upper limit of 42~$\mu$Jy, yet it was detected without the flare on 2004 May 17 with flux density $95\pm 19$~$\mu$Jy and also on 2004 June 18 with flux density $230\pm 17$~$\mu$Jy, the latter of which was previously reported by \citet{berger2006}.  Similarly, \citet{berger2010} reported no detectable emission from BRI~0021 (M9.5) with $3\sigma$ upper limits of 54~$\mu$Jy and 48~$\mu$Jy for 4.9~GHz and 8.5~GHz, despite a previous marginal detection of its quiescent emission at $40\pm13$~$\mu$Jy as well as a flare with a peak flux density of $360 \pm 70$~$\mu$Jy.  In the case that the quiescent emission is variable over longer timescales, long-term monitoring of radio brown dwarfs would be necessary to quantify how much the current detection rate underestimates the true detection rate and may warrant revisiting previously undetected objects with H$\alpha$ or infrared variability such as SDSS~J12545393-0122474 \citep{kao2016}. 

The radio emission from 2M1047 differs from both the strongly polarized two-component behavior observed from SIMP0136 and 2M1237 and the single-component (pulsing only) behavior of SDSS0423 and 2M1043.  
Like SIMP0136 and 2M1237, the non-pulsing radio emission from 2M1047 is also relatively flat. \citet{kao2016} measured a 4--8~GHz mean quiescent flux density of 17.5$\pm$3.6~$\mu$Jy, implying $\alpha \approx -0.9 \pm 0.4$ and $\delta \approx 2.4$, when we take the 12--18~GHz mean quiescent flux density.  This confirms the spectral indices measured by \citet{williams2015} at 4--8 GHz ($\alpha = 0.0 \pm 0.3$) and 8--12 GHz ($\alpha = -0.7 \pm 0.7$). However, unlike SIMP0136 and 2M1237, the non-pulsing 12--18 GHz emission from 2M1047 is not circularly polarized, and \cite{williams2015} reported ``quasi-quiescent" emission from 2M1047 at 4--8~GHz that was not circularly polarized.
\subsection{Intermittent Pulses: Implications for ECM Emission Frequency Cutoff}\label{sec.Cutoff_15a374}

At these high frequencies, pulses appear to be more intermittent compared to previous 4--8~GHz observations, with short-duration variability in both time and frequency. For instance, while the central pulse in 2M1237 is present at all bandwidths, the right-most peak is clearly apparent only at 11--12~GHz.  SDSS0423 emits two faint right-circularly polarized pulses at 8--9~GHz, but the right pulse appears to drop out at higher frequencies.  For 2M1047, the multi-peaked and/or long-lived left-circularly polarized pulse at 12.8--13.5~GHz early in the observing block drops out at higher frequencies, while three fainter left-circularly polarized pulses emerge at 15-16 GHz. In contrast, these objects' C-band (4--8~GHz) pulses are present at all sub-bands \citep{kao2016}. 

This pulse variability suggests that the conditions for current systems driving these auroral emissions  may be much less stable or more variable close to the surface of the star, where fields are expected to be stronger and emitting frequencies are higher. One possibility for variable conditions is magnetic flux.  While large-scale fields appear necessary to drive solar system auroral currents and the same may occur in isolated brown dwarfs like our targets, evolving and complex small-scale fields may also begin to emerge near the object surface. As radiating electrons traverse the large-scale field lines inward, they will radiate at higher frequencies corresponding to the increased magnetic fluxes that they see. Some fully convective dynamo models capable of generating kilogauss fields suggest that these small-scale fields may be driven by convection near the surface, where convective turnover times are shorter and small-scale intermittent features begin to appear in convective flows.  In contrast, more stable large-scale fields may be tied to slowly overturning convection in the deep interiors  \citep{browning2008}.

Other examples of intermittent auroral pulse structures exist in the literature.  As an example, the dynamic spectrum of LSR~J1835+3259 shows one pulse per rotation extending through $\sim$4--8~GHz, one extending through $\sim$4--6~GHz, and one only extending through $\sim$4.5~GHz, with emission from each pulse appearing to fade away or renew again at different frequencies \citep{hallinan2015}.  Narrowband and intermittent pulses have also been observed in terrestrial, Jovian, and Saturnian auroral kilometric radiation (AKR).  High-resolution dynamic spectra reveal that rather than one continuous pulse through frequency, AKR actually consists of many small-scale micropulses from individually radiating sources that are highly time variable and narrowly-spaced in frequency, with widths of order $\sim$10--1000~Hz corresponding to bunched groups of these local AKR sources traveling very rapidly through space.  The origin of this fine structure remains unknown, but it is speculated that they may reflect a number of physical processes including propagation and absorption effects or small-scale field parallel current structures \citep[][and references therein]{gurnett1981, pottelette1999, treumann2006}.

While we do observe what appears to be the disappearance of highly circularly polarized pulsed auroral emission in SIMP0136, 2M1043, and SDSS0423 at 11--12~GHz, in light of the observed behavior in 2M1237 and 2M1047 and the above-discussed cases, we classify these dropoffs only as very tentative evidence of ECM emission cutoff.  The known intermittent behavior of AKR suggests that observations through a much wider bandwidth of high frequencies are necessary to confirm a true emission cutoff.

\subsection{Comparison to Luminosity-Driven Model}\label{sec.c09}
\begin{figure*}
\epsscale{.9}
\plotone{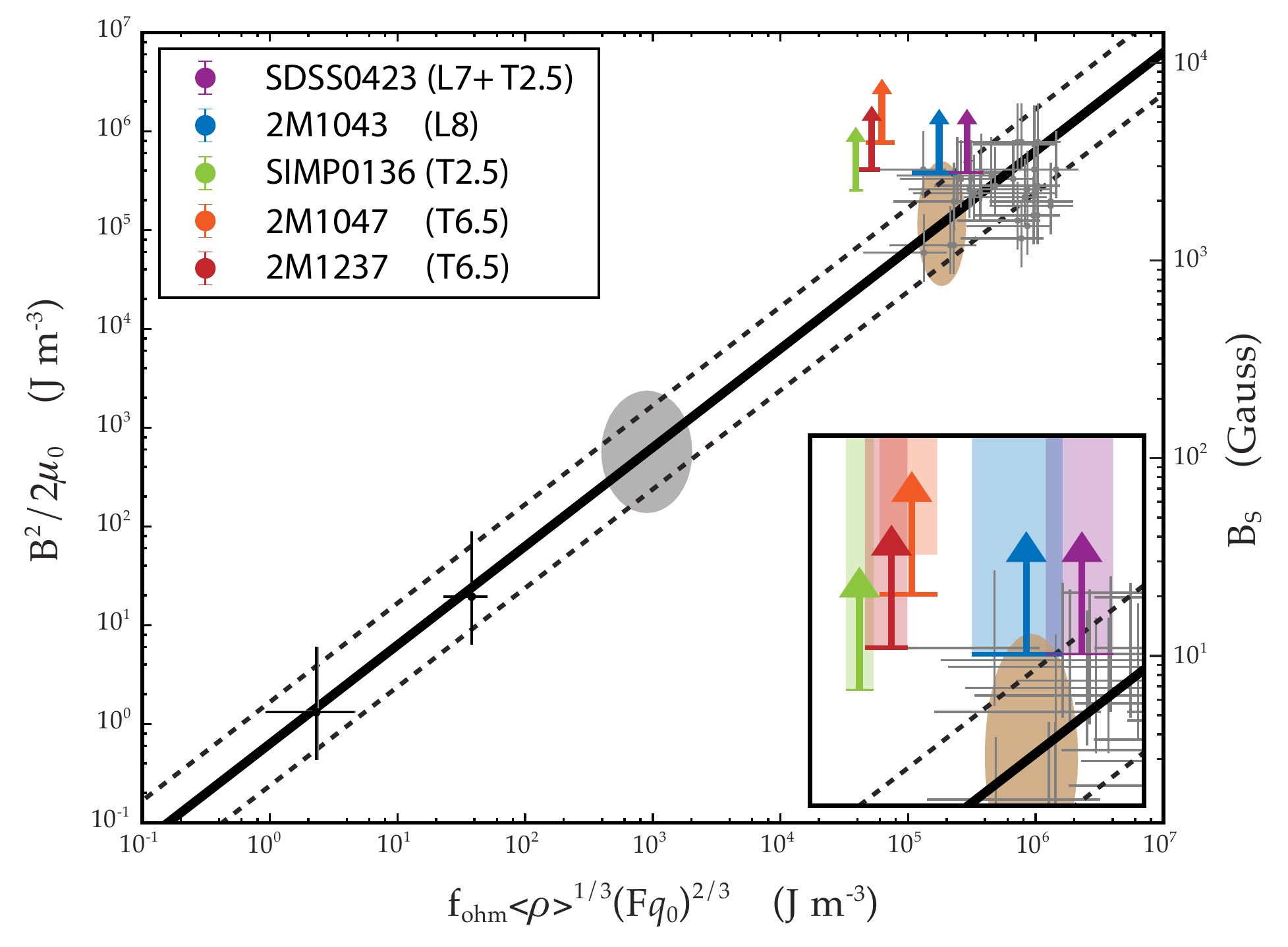}
\caption{\label{fig:christensenFig2}
A comparison of estimated lower-bound magnetic field energy densities for our targets (overplotted arrows) to values predicted by the \citet{christensen2009} scaling relation (black solid line) between convected energy density (x-axis, $q_0$) and magnetic energy density (left y-axis) for fully convective dipole-dominated rapid rotators.  Black dashed lines are 3$\sigma$ uncertainties on the model and horizontal bars on arrows are our estimated uncertainties. Previous constraints were T~Tauri stars and old M dwarfs (gray crosses). Black points represent Earth and Jupiter.  Brown and grey ellipses are predicted positions for a 1500~K brown dwarf and a 7 M$_{\mathrm{J}}$ exoplanet, respectively.  Right y-axis values are predicted surface-averaged fields $B_s$.}
\end{figure*}

Previously, \citet{kao2016} found tentative evidence of a T dwarf departure from a predominantly luminosity-driven dynamo for rapid rotators (P$<$4 days).  This model extended planetary dynamo models to stellar-mass objects including T~Tauri stars and old M-dwarfs, whose Zeeman broadening and Zeeman Doppler imaging measurements were empirically consistent with a scaling relationship linking internal magnetic energy density to convected energy flux and dynamo region density, while being largely independent of both magnetic diffusivity and rotation rate \citep[][hereafter C09]{christensen2009}. The broad span through planetary and stellar parameter spaces suggested that the scaling law may in fact present a unifying principle governing the magnetic field generation in all rapidly rotating, dipole-dominated fully convective objects -- namely, that the bolometric flux $q_0$ sets the magnetic field strength averaged over the whole volume of the dynamo region $\langle B^2\rangle$, with a weak dependence on the mean density of the dynamo region $\langle\rho\rangle$:
\begin{equation}
\langle B^2\rangle \propto \langle\rho\rangle^{1/3} q_0^{2/3} \; .
\end{equation}

Because the C09 model is specific to dipole-dominated fields ($>$35\% of field strength in the dipole component) in rapid rotators, possible explanations for the observed tentative inconsistency between late L and T dwarf magnetic fields with the C09 model included: (1) higher order non-dipole fields may dominate our objects or (2) several of our targets may be slower rotators.

Regarding the possibility that our objects may not have dipole-dominated field topologies, auroral radio emission by itself is currently insufficient for confirming magnetic field topologies.  This is because the frequency of the emission corresponds only to localized emitting regions in the magnetospheres of our targets.  Therefore, in this work we make no attempt to assume a particular magnetic field topology and instead follow the formalism presented in \citet{kao2016} to convert the local magnetic fields measured with ECM emission $B_{\mathrm{ECM}}$ to lower bound mean surface field magnitudes $B_{\mathrm{s, dip}}$, which we list in Table~\ref{table:magneticFields}.  For this conversion, we conservatively adopt ECM emission cutoff frequencies corresponding to the middle of the last sub-band with imaging detections of auroral pulses in Stokes I and V, since evidence of cutoffs in the ECM emission frequency is inconclusive (\S\ref{sec.Cutoff_15a374}). As described in \citet{kao2016}, $B_{\mathrm{s, dip}}$ is equivalent to a lower bound Zeeman broadening measurement of a surface-averaged field strength $B_s$, and the presence of any higher-order fields would raise this estimate.   We convert $B_{\mathrm{s, dip}}$  to a mean internal field strength $\langle B \rangle$ for comparison to the C09 relation by following the conversions outlined in C09 and summarized in \citet{kao2016}. 

\setlength{\tabcolsep}{0.05in}
\begin{deluxetable}{lccc}[htp]
\tabletypesize{\small}
\tablecaption{Adopted Magnetic Fields \label{table:magneticFields}}
\tablehead{
	\colhead{}						                &
	\colhead{Tentative }		&
	\colhead{ Local field }			    &
	\colhead{ Min avg field}	   	   								            				
		    \\
	\colhead{Object}						                &
	\colhead{$\nu_{\mathrm{cutoff}}$\,\tablenotemark{a}}		&
	\colhead{ $B_{\mathrm{ECM}}$\,\tablenotemark{b}}			    &
	\colhead{$B_{\mathrm{s, dip}}$\,\tablenotemark{c}}	   	   								            				
		    \\
	\colhead{}					 &
	\colhead{(GHz)}				     &
	\colhead{(kG)}		 			 &
	\colhead{(kG)}			         }                              
\startdata
2M1047   	& 15.75 & 5.6  & 4.0  \\
SIMP0136	& 9.0   & 3.2  & 2.3  \\
2M1043		& 11.0  & 3.9  & 2.8  \\
2M1237		& 11.5  & 4.1  & 2.9  \\
SDSS0423	& 11.0  & 3.9  & 2.8   \\
\enddata
\tablenotetext{a}{Center of highest subband with non-tentative imaging detection of ECM pulse. }
\tablenotetext{b}{$B_{\mathrm{ECM~[kG]}}= \nu_{\mathrm{ECM~[GHz]}}$ / 2.8 \, \citep{treumann2006}}
\tablenotetext{c}{$\langle B_{\mathrm{s, dip}}^2 \rangle = \frac{1}{2} B^2_{\mathrm{ECM}}$ \hspace{1.6cm} \citep{kao2016}}
\end{deluxetable}

It is important to note that though the \citet{kao2016} formalism assumes that a dipole field powers the ECM emission, the lower bound mean surface magnetic field calculated from this assumption accommodates multipolar field topologies. Therefore, in adopting this formalism to interpret our measured magnetic fields, we are not making a concrete statement on brown dwarf field topologies.  However, the \citet{kao2016} formalism depends on the observed and modeled magnetic behavior of fully convective M dwarfs extending to late L and T dwarfs. We believe that an analogy to fully convective M dwarfs is appropriate here, as the dynamo action occurring in fully convective M dwarfs is likely similar to what occurs in very cold brown dwarfs.  This is because the dynamo regions of $\sim$M4 and later dwarfs, including L and T dwarfs, are expected to be fully convective, with an important exception that we discuss below.  In contrast, higher mass stars have dynamo region structures where both convection and differential rotation are important to the fluid dynamics driving the dynamos. 

The differing fluid dynamics in these dynamo regions lead to different magnetic field behaviors. Strong differential rotation in the dynamo region tends to destroy the dominant dipolar component generated by convection \citep{jones2014, gastine2012}, leading to toroidal magnetic fields \citep[e.g.][]{browning2008, gastine2012, yadav2016} and magnetic cycles \citep{yadav2016}.  While many late-type M dwarfs and brown dwarfs are expected to have fully convective dynamo regions, differential rotation may be able to arise in some.  Observational evidence suggests that low-mass M dwarfs may be able to generate magnetic fields that undergo cycles, pointing to dynamo mechanisms that may be solar-like \citep{wrightDrake2016}.  However, the onset of differential rotation suggested by such magnetic cycles seems to occur only in slowly rotating objects \citep{browning2008, yadav2016}.  Hence, for our assumption that the dynamo mechanism in fully convective M dwarfs is analogous to those in late L and T dwarfs to hold, our objects must be rapid rotators.

Regarding the rapid rotation requirement for both the C09 model and our use of the \citet{kao2016} formalism, the periodicities that we recover in \S\ref{sec.Rotation_15a374} together with cloud variability studies for SIMP0136 and C-band observations for 2M1047 by \citet{williams2015} unambiguously confirm that our targets are indeed rapid rotators, with rotation periods between $\sim$1.44--2.88 hours.  While SIMP0136 does not have any clearly periodic pulse structure, infrared cloud variability studies suggest a rotation period of $2.3895\pm0.0005$~hr \citep{artigau2009, croll2016}.  This rotation period is  not inconsistent with the recovered periodicity in its quasi-quiescent emission, which we measure to be 2.88$^{+0.34}_{-0.27}$~hr. Our data confirm that to date, all pulsing radio brown dwarfs with rotation period measurements have reported rotational periods less than 4 hours \citep[][and references therein]{pineda2017}.  These rotation periods likely fall well within the limit of rapid rotation (Rossby number $\mathrm{Ro} <  0.1$), with measured rotation periods on the order of just a few hours compared to convective turnover times that may be in the tens to hundreds of days \citep[e.g.][]{noyes1984, pizzolato2003, mclean2012, landin2010}.

The previous statement comes with some important caveats.  First, empirical estimations and numerical calculations of convective turnover times with observable properties such as X-ray luminosity do not extend to L and T dwarfs.  Second, dynamo regions can span a wide range of fluid densities, with density stratification ranging from $\sim$20\% in incompressible fluids such as in the geodynamo to at least $\sim$10$^6$--10$^{10}$ in stars and likely also cool brown dwarfs \citep{saumon1995}.  In highly stratified regimes, fluids in the most diffuse regions become less efficient at transporting heat and small-scale motions with accompanying shorter convective turnover times may become increasingly important.  Defining an appropriate Rossby number is not straightforward, since it is unclear where in the dynamo region is most important for generating fields that auroral radio emission probes.  

We present our resulting field constraints on a reproduction of the C09 scaling law in Figure~\ref{fig:christensenFig2}, with x-axis values determined from the physical parameters of our targets summarized in \S\ref{sec.Targets_15a374}.  The T dwarfs 2M1047, 2M1237, and SIMP0136 clearly depart by an order of magnitude from C09 magnetic energy predictions.  While the late L dwarfs lie near the outer bounds of the 3$\sigma$ error on the scaling relationship, these are in fact conservative constraints; no emission frequency cutoff has been conclusively detected, pointing to the possibility of yet stronger fields.  This tantalizingly hints at a possible ultracool brown dwarf locus that may not age along the predicted luminosity-magnetic field sequence \citep{reinersChristensen2010}.  Additional studies identifying aurorally pulsing radio brown dwarfs and characterizing their physical parameters could reveal such a locus.

As we previously pointed out, this emerging departure from the C09 predictions may be due to magnetic topologies that are not dominated by dipoles.   In such a case, a comparison to the C09 predictions would be inappropriate, since the models employed by C09 are specific to objects with dipole-dominated fields. Here, we examine that possibility. While several attempts to model brown dwarf ECM emission with dipole fields have been successful \citep{nichols2012, kuznetsov2012, leto2016}, \citet{lynch2015} found that a dipole field was unable to reproduce observed dynamic spectra for the M9 dwarf TVLM~513-46546 and L0+L1.5 binary 2MASS~J0746425+200032.  Instead, ECM sources at the footprints of coronal magnetic loops with radial extents of only 1.2--2.7 R$_s$, where R$_s$ is the dwarf radius, are more able to reproduce the observed emission, suggesting that a global multipolar field may be responsible for powering ECM emission in some cool dwarfs. It is plausible that the magnetic fields of these cool dwarfs may be dominated by small-scale multipolar fields, since dynamo models suggest that objects with very low Rossby numbers ($<$0.1) can have magnetic topologies with most of the magnetic energy either in the dipole component or in multipolar components \citep{gastine2013}, and ZDI studies show that dwarfs with kilogauss dipoles have order-of-magnitude weaker multipole fields and vice versa \citep{morin2010}. 

The radio pulsing of some dwarfs, including TVLM~513-46546, the M8.5 dwarf LSR J1835+3259, and our targets here, is stable for many months and even years.  The long-term persistence of these ECM pulses requires continuous quasi-stable particle acceleration. In the coronal loop model explored by \citet{lynch2015}, this would require reconnection in an active region that persists for many months, such as a planetary field continuously interacting with the dwarf field \citep{lanza2013}.  In this scenario, the radio emission would be seen near the footprints of a sequence of coronal loops.

Quasi-stable particle acceleration can also be explained by an auroral model \citep{hallinan2015}.  If the emission that we observe from brown dwarfs is indeed auroral in nature, evidence points to a strong dipole field rather than a strong multipolar field.  Auroral emissions rely on coupling energy from locations where there is a large $v \times B$ into the magnetosphere \citep{nichols2012}. This is best achieved by having strong magnetic fields far away from the planet (e.g. in the middle or outer magnetosphere), where rotational speed $v$ will be large as well.  Dipoles drop off much more slowly than higher order fields and are more likely to dominate auroral power for this reason, suggesting that ECM emission of auroral origin likely probes the dipole components of our objects. Indeed, all of the examples of planetary aurorae in our Solar System demonstrate that the ECM emission can originate from a dipole component of our targets' magnetic fields.  Similarly, models of the co-rotation breakdown mechanism that occurs in the Jovian auroral system assuming dipolar magnetic fields show close agreement between modeled and observed auroral radio luminosities for TVLM~513-46546 (M9), LSR~J1835+3259 (M8.5), and 2MASS~J00361617+1821104 (L3.5) \citep{nichols2012, turnpenney2017}.  This model also predicted rotation periods between $\sim$2.1--2.8~hr for 2M1047, which is not inconsistent with the rotation period measured by \citet{williams2015}.

\subsection{Consideration of Age-Related Models}
\begin{figure*}
\epsscale{0.8}
\plotone{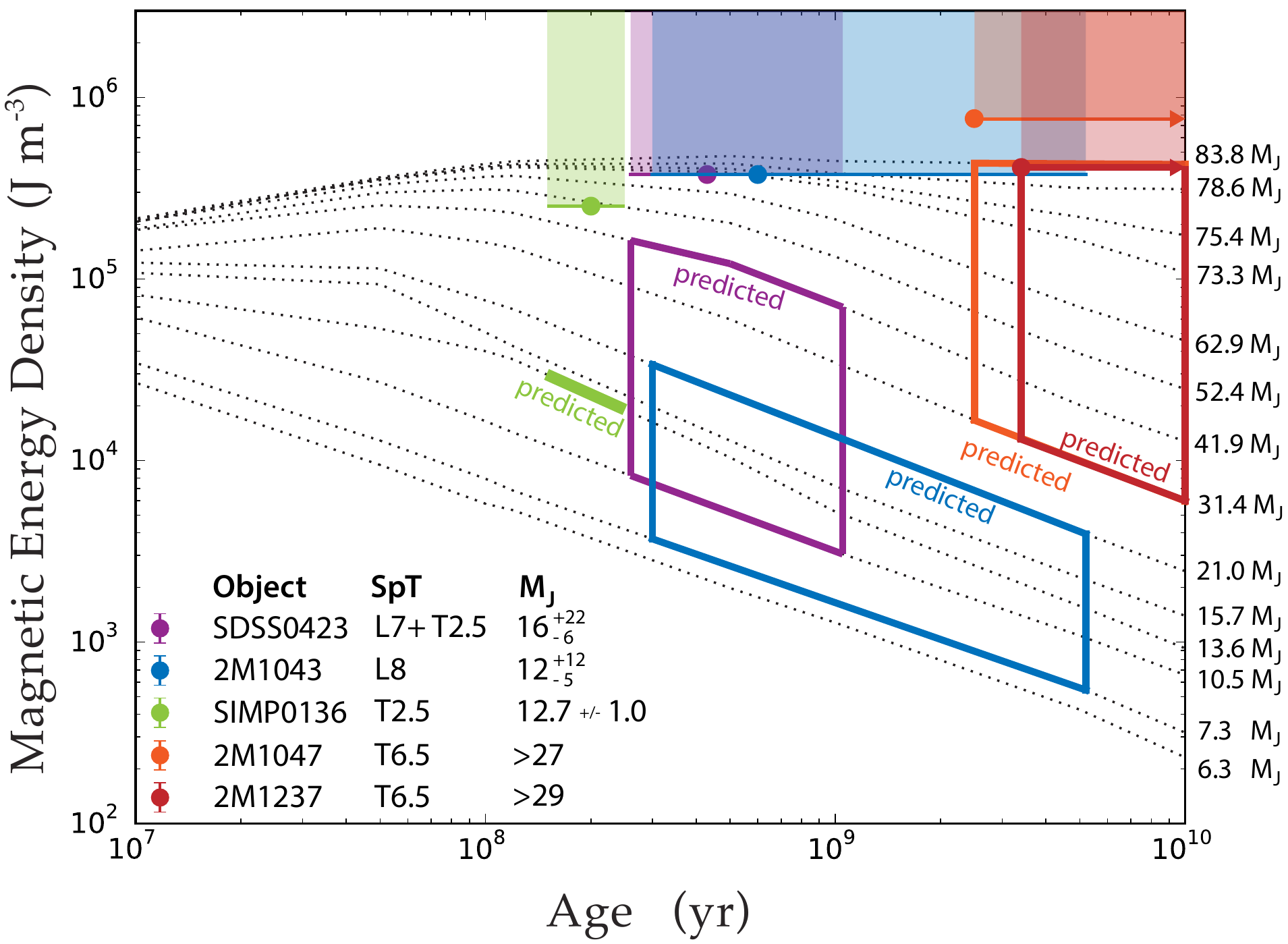}
\caption{\label{fig:reinersChristensen2010_15a374}
When effective temperature, age, and mass are accounted for by applying the \citet{christensen2009} scaling relationship to \citet{baraffe2003} brown dwarf evolutionary tracks, they are not sufficient to explain the strong magnetic fields generated by the dynamo mechanisms in our targets.  Here, we compare inferred lower-bound magnetic field energy densities (colored circles) to predictions from dynamo evolutions tracks (outlined colored regions).   Masses are adopted from \citet{kao2016} and \citet{gagne2017} and are provided in the bottom left corner.  Outlined colored regions include mass tracks falling within the mass uncertainties or those nearest the edge cases. For objects with only lower bound constraints on masses, we adopt 0.08~M$_{\odot}$ as the nominal hydrogen burning limit.  Shaded colored regions depict age uncertainties and extend upward to indicate that magnetic energy densities inferred from auroral radio magnetic field measurements are lower bounds. Age constraints for 2M1047 and 2M1237 give only lower bounds.}

\end{figure*}

The possibility that magnetic energy may scale with luminosity in rapidly rotating convective objects supports a picture in which brown dwarf magnetic fields are expected to decay with age as they cool through the L/T/Y spectral sequence and become increasingly less luminous.  Indeed, field strengths can wane by a factor of 10\% over the lifetime of a brown dwarf when evolutionary tracks are applied to the C09 model \citep{reinersChristensen2010}.

The luminosity of a brown dwarf depends both on its age and its mass, and these factors may account for some of the possible emerging disagreement between the C09 relationship and our targets.  Using the \citep{baraffe2003} brown dwarf evolutionary tracks and the C09 relationship, we calculated predicted age-evolving magnetic energy densities for each mass grid point and overplotted our objects in Figure~\ref{fig:reinersChristensen2010_15a374}. Given the disagreement between our objects and the C09 relation, it is no surprise that our objects also depart from these age-related predictions.  However, while our T dwarf data appeared to disagree somewhat with the C09 model in \S \ref{sec.c09}, a departure was less clear for our L dwarfs.  Accounting for the effects of age and mass on luminosity hints at a stronger departure from the C09 scaling law for our warmer but less massive and younger L dwarfs than was initially evident when mass and age were folded into luminosity.  Regardless, a much larger sample is needed before any concrete conclusions can be drawn about how age affects convective dynamos, and the simplest prediction to test is whether objects with similar masses have stronger fields when younger.

In the event that luminosity (T$_{\mathrm{eff}}$) does not play a dominant role in brown dwarf dynamos, it is worth noting that magnetic field strengths do not appear to vary much by age across an order of magnitude between $\sim$0.2--3.4~Gyr. Of course, no definitive ECM emission cutoff frequency has been observed for any brown dwarfs yet, including our targets, so the plotted mean surface field strengths are merely lower bounds and the future addition of constraints from higher frequencies and a broader range of ages, masses, and temperatures may yet reveal a correlation between age and field strength.  

Presenting our data within the context of age has an important implication for ongoing efforts to detect exoplanet radio emission.  While such efforts have focused on hot Jupiters (which see high flux from host stars thus increasing the luminosity of solar-wind generated aurorae) and hot young exoplanets \citep{lazioFarrell2007,lazio2010, hallinan2013, murphy2015, lynch2017}, old objects appear to also be capable of generating strong fields along with the associated radio emission, and broader searches may be warranted.

\subsection{First Radio Detection of a Planetary-Mass Object?}\label{planetaryMass}
Recently, \citet{gagne2017} reported that SIMP0136 may be a member of the $\sim$200 Myr old Carina-Near moving group based on its kinematics, with a field interloper probability of only 0.0001\%.  Using an empirical measurement of its bolometric luminosity and the the \cite{saumonMarley2008} models, they inferred $R = 1.22\pm0.01$ $R_{\mathrm{J}}$, which together predicted T$_{\mathrm{eff}} = 1098\pm6$ K and $M=12.7\pm1.0$ $ M_{\mathrm{J}}$. 

This low mass is further supported by new $v \sin i$ measurements that, in combination with its photometric periodicity, constrains its inclination angle at $i=55.9^{+1.6\degree}_{-1.5\degree}$.  This inclination angle leads to a lower-bound radius and upper bounds on age and mass of $R>1.01\pm0.02$ $R_{\mathrm{J}}$, $\tau < 910^{+26}_{-110}$ Myr and $M<42.6^{+2.5}_{-2.4}$ $M_{\mathrm{J}}$.  Finally, models of the photometric variability assuming a single spot are also in agreement, constraining its inclination at $i<60\degree$, which would increase the lower bound radius to $R>1.17\pm0.02$ $R_{\mathrm{J}}$ and further support the young age and low mass derived for SIMP0136 if it is indeed a member of the Carina-Near moving group.

This low mass of SIMP0136 is notable for its proximity to the $\sim$12--13 $M_{\mathrm{J}}$ deuterium burning limit, or the mass above which compact gaseous objects are expected to burn deuterium \citep{spiegel2011, molliere2012, bodenheimer2013}.  The deuterium burning limit is one way to distinguish between gas giant planets and brown dwarfs.

\section{Conclusions}\label{sec.Conclusion_15a374}

We detected auroral radio emission from four L7--T6.5 dwarfs up through 10--12~GHz, and one T6.5 object up through 15--16.5~GHz, corresponding to 3.2--5.6~kG local magnetic field strengths and 2.3--4.0~kG minimum surface averaged fields.   Additionally, we reported a tentative 16.5--18~GHz auroral pulse detection for the T6.5 dwarf 2M1047, corresponding to 6.2~kG local magnetic field strengths and 4.4~kG minimum surface averaged fields.  Pulses appear to be more intermittent in frequency at higher frequencies compared to previous observations of lower frequency counterparts, which can be interpreted as evidence of a higher degree of variability in the conditions necessary to generate auroral radio emission near the surfaces of brown dwarfs.  While we observe the fading out of auroral pulses at 11--12~GHz for some targets, observations at higher frequencies are necessary to affirm definitive cut-offs in the auroral radio emission.  We additionally observe no detectable quiescent emission for SDSS0423 but do observe highly circularly polarized non-pulsed emission from SIMP0136 and in some sub-bands also for 2M1237.  The behavior of SDSS0423 may point to long term variability in the quiescent emission mechanism, while SIMP0136 and 2M1237 are more suggestive of coherent processes.  

The presented detections are strong direct constraints on dynamo theory at the substellar-planetary boundary.  We presented data suggesting that a scaling relation between convected energy flux and magnetic energy density \citep{christensen2009} may not fit.  We also show that age, mass, and temperature together cannot account for the strong magnetic fields produced by our targets. Using the rotational modulation of auroral radio emission, we measured rotational periods between 1.47--2.28~hr for SDSS0423, 2M1043, and 2M1237.  These short rotation periods are consistent with periods measured for earlier-type brown dwarfs using auroral radio emission and reiterates that rapid rotators can host strong large-scale fields.  Finally, we find that our oldest targets ($>$2.5~Gyr) can generate fields that are as strong as those measured in our youngest targets ($\sim$200--600~Myr), suggesting that old exoplanets may also host fields with strengths comparable to their younger siblings and serving as preliminary and very tentative evidence that age dependence in dynamo mechanisms may be weak.  The absence of an emission frequency cut-off means that we have not broken any degeneracies in our analyses and a larger, more well-characterized sample is required.

Included in our sample was the archetypal cloud variable SIMP0136, which was recently found to be a member of a nearby $\sim$200~Myr moving group.  This new age constraint reduces its estimated mass to a mere $12.7\pm 1.0$~M$_\mathrm{J}$, possibly making SIMP0136 the first known planetary mass object detected in the radio.  If SIMP0136 is indeed a field exoplanet, its detection demonstrates that auroral radio emission can open a new avenue to detecting exoplanets, including elusive rogue planets.


\acknowledgments
\section{Acknowledgements}\label{sec.Acknowledgements}
MMK thanks Jackie Villadsen for helping to troubleshoot calibrations and Rakesh Yadav for thoughtful and instructive discussions about dynamo modeling.  MMK additionally thanks the enthusiastically supportive staff at the National Radio Astronomy Observatory for their technical and scientific mentorship. 

Support for this work was provided by the NSF through the Grote Reber Fellowship Program administered by Associated Universities, Inc./National Radio Astronomy Observatory.  The National Radio Astronomy Observatory is a facility of the National Science Foundation operated under cooperative agreement by Associated Universities, Inc. 

This material is based in part upon work supported by the National Science Foundation under Grant AST-1654815 and the NASA Solar System Exploration Virtual Institute cooperative agreement 80ARC017M0006. GH acknowledges the support of the Alfred P. Sloan Foundation and the Research Corporation for Science Advancement.

JSP was supported by a grant from the National Science Foundation Graduate Research Fellowship under grant no. DGE-1144469.

This publication makes use of data products from the Two Micron All Sky Survey, which is a joint project of the University of Massachusetts and the Infrared Processing and Analysis Center/California Institute of Technology, funded by the National Aeronautics and Space Administration and the National Science Foundation.

\facility{VLA}
\software{CASA \citep{mcmullin2007}} 
\software{MATLAB \citep{matlabSPTB}}

\bibliography{mkaoHighFreq}

\begin{thebibliography}{}
\expandafter\ifx\csname natexlab\endcsname\relax\def\natexlab#1{#1}\fi

\bibitem[{{Ackerman} \& {Marley}(2001)}]{ackermanMarley2001}
{Ackerman}, A.~S., \& {Marley}, M.~S. 2001, \apj, 556, 872

\bibitem[{{Antonova} {et~al.}(2007){Antonova}, {Doyle}, {Hallinan}, {Golden},
  \& {Koen}}]{antonova2007}
{Antonova}, A., {Doyle}, J.~G., {Hallinan}, G., {Golden}, A., \& {Koen}, C.
  2007, \aap, 472, 257

\bibitem[{{Antonova} {et~al.}(2013){Antonova}, {Hallinan}, {Doyle}, {Yu},
  {Kuznetsov}, {Metodieva}, {Golden}, \& {Cruz}}]{antonova2013}
{Antonova}, A., {Hallinan}, G., {Doyle}, J.~G., {et~al.} 2013, \aap, 549, A131

\bibitem[{{Apai} {et~al.}(2013){Apai}, {Radigan}, {Buenzli}, {Burrows}, {Reid},
  \& {Jayawardhana}}]{apai2013}
{Apai}, D., {Radigan}, J., {Buenzli}, E., {et~al.} 2013, \apj, 768, 121

\bibitem[{{Artigau} {et~al.}(2009){Artigau}, {Bouchard}, {Doyon}, \&
  {Lafreni{\`e}re}}]{artigau2009}
{Artigau}, {\'E}., {Bouchard}, S., {Doyon}, R., \& {Lafreni{\`e}re}, D. 2009,
  \apj, 701, 1534

\bibitem[{{Artigau} {et~al.}(2006){Artigau}, {Doyon}, {Lafreni{\`e}re},
  {Nadeau}, {Robert}, \& {Albert}}]{artigau2006}
{Artigau}, {\'E}., {Doyon}, R., {Lafreni{\`e}re}, D., {et~al.} 2006, \apjl,
  651, L57

\bibitem[{{Artigau} {et~al.}(2003){Artigau}, {Nadeau}, \&
  {Doyon}}]{artigau2003}
{Artigau}, E., {Nadeau}, D., \& {Doyon}, R. 2003, in IAU Symposium, Vol. 211,
  Brown Dwarfs, ed. E.~{Martin}, 451

\bibitem[{{Badman} {et~al.}(2015){Badman}, {Branduardi-Raymont}, {Galand},
  {Hess}, {Krupp}, {Lamy}, {Melin}, \& {Tao}}]{badman2015}
{Badman}, S.~V., {Branduardi-Raymont}, G., {Galand}, M., {et~al.} 2015, \ssr,
  187, 99

\bibitem[{{Baraffe} {et~al.}(2003){Baraffe}, {Chabrier}, {Barman}, {Allard}, \&
  {Hauschildt}}]{baraffe2003}
{Baraffe}, I., {Chabrier}, G., {Barman}, T.~S., {Allard}, F., \& {Hauschildt},
  P.~H. 2003, \aap, 402, 701

\bibitem[{{Batygin} \& {Stevenson}(2010)}]{batygin2010}
{Batygin}, K., \& {Stevenson}, D.~J. 2010, \apjl, 714, L238

\bibitem[{{Benz}(1986)}]{benz1986}
{Benz}, A.~O. 1986, \solphys, 104, 99

\bibitem[{{Berdyugina} \& {Solanki}(2002)}]{berdyugina2002}
{Berdyugina}, S.~V., \& {Solanki}, S.~K. 2002, \aap, 385, 701

\bibitem[{{Berger}(2002)}]{berger2002}
{Berger}, E. 2002, \apj, 572, 503

\bibitem[{{Berger}(2006)}]{berger2006}
---. 2006, \apj, 648, 629

\bibitem[{{Berger} {et~al.}(2001){Berger}, {Ball}, {Becker}, {Clarke}, {Frail},
  {Fukuda}, {Hoffman}, {Mellon}, {Momjian}, {Murphy}, {Teng}, {Woodruff},
  {Zauderer}, \& {Zavala}}]{berger2001}
{Berger}, E., {Ball}, S., {Becker}, K.~M., {et~al.} 2001, \nat, 410, 338

\bibitem[{{Berger} {et~al.}(2005){Berger}, {Rutledge}, {Reid}, {Bildsten},
  {Gizis}, {Liebert}, {Martin}, {Basri}, {Jayawardhana}, {Brandeker},
  {Fleming}, {Johns-Krull}, {Giampapa}, {Hawley}, \& {Schmitt}}]{berger2005}
{Berger}, E., {Rutledge}, R.~E., {Reid}, I.~N., {et~al.} 2005, \apj, 627, 960

\bibitem[{{Berger} {et~al.}(2009){Berger}, {Rutledge}, {Phan-Bao}, {Basri},
  {Giampapa}, {Gizis}, {Liebert}, {Martin}, \& {Fleming}}]{berger2009}
{Berger}, E., {Rutledge}, R.~E., {Phan-Bao}, N., {et~al.} 2009, \apj, 695, 310

\bibitem[{{Berger} {et~al.}(2010){Berger}, {Basri}, {Fleming}, {Giampapa},
  {Gizis}, {Liebert}, {Martin}, {Phan-Bao}, \& {Rutledge}}]{berger2010}
{Berger}, E., {Basri}, G., {Fleming}, T.~A., {et~al.} 2010, \apj, 709, 332

\bibitem[{{Bodenheimer} {et~al.}(2013){Bodenheimer}, {D'Angelo}, {Lissauer},
  {Fortney}, \& {Saumon}}]{bodenheimer2013}
{Bodenheimer}, P., {D'Angelo}, G., {Lissauer}, J.~J., {Fortney}, J.~J., \&
  {Saumon}, D. 2013, \apj, 770, 120

\bibitem[{{Brain} {et~al.}(2015){Brain}, {McFadden}, {Halekas}, {Connerney},
  {Bougher}, {Curry}, {Dong}, {Dong}, {Eparvier}, {Fang}, {Fortier}, {Hara},
  {Harada}, {Jakosky}, {Lillis}, {Livi}, {Luhmann}, {Ma}, {Modolo}, \&
  {Seki}}]{brain2015}
{Brain}, D.~A., {McFadden}, J.~P., {Halekas}, J.~S., {et~al.} 2015, \grl, 42,
  9142

\bibitem[{{Browning}(2008)}]{browning2008}
{Browning}, M.~K. 2008, \apj, 676, 1262

\bibitem[{{Burgasser}(2007)}]{burgasser2007}
{Burgasser}, A.~J. 2007, \apj, 659, 655

\bibitem[{{Burgasser} {et~al.}(2006{\natexlab{a}}){Burgasser}, {Burrows}, \&
  {Kirkpatrick}}]{burgasser2006b}
{Burgasser}, A.~J., {Burrows}, A., \& {Kirkpatrick}, J.~D. 2006{\natexlab{a}},
  \apj, 639, 1095

\bibitem[{{Burgasser} {et~al.}(2006{\natexlab{b}}){Burgasser}, {Geballe},
  {Leggett}, {Kirkpatrick}, \& {Golimowski}}]{burgasser2006a}
{Burgasser}, A.~J., {Geballe}, T.~R., {Leggett}, S.~K., {Kirkpatrick}, J.~D.,
  \& {Golimowski}, D.~A. 2006{\natexlab{b}}, \apj, 637, 1067

\bibitem[{{Burgasser} {et~al.}(2003){Burgasser}, {Kirkpatrick}, {Liebert}, \&
  {Burrows}}]{burgasser2003_redOpticalData}
{Burgasser}, A.~J., {Kirkpatrick}, J.~D., {Liebert}, J., \& {Burrows}, A. 2003,
  \apj, 594, 510

\bibitem[{{Burgasser} {et~al.}(2000){Burgasser}, {Kirkpatrick}, {Reid},
  {Liebert}, {Gizis}, \& {Brown}}]{burgasser2000b}
{Burgasser}, A.~J., {Kirkpatrick}, J.~D., {Reid}, I.~N., {et~al.} 2000, \aj,
  120, 473

\bibitem[{{Burgasser} {et~al.}(2002{\natexlab{a}}){Burgasser}, {Liebert},
  {Kirkpatrick}, \& {Gizis}}]{burgasser2002b}
{Burgasser}, A.~J., {Liebert}, J., {Kirkpatrick}, J.~D., \& {Gizis}, J.~E.
  2002{\natexlab{a}}, \aj, 123, 2744

\bibitem[{{Burgasser} {et~al.}(2002{\natexlab{b}}){Burgasser}, {Marley},
  {Ackerman}, {Saumon}, {Lodders}, {Dahn}, {Harris}, \&
  {Kirkpatrick}}]{burgasser2002clouds}
{Burgasser}, A.~J., {Marley}, M.~S., {Ackerman}, A.~S., {et~al.}
  2002{\natexlab{b}}, \apjl, 571, L151

\bibitem[{{Burgasser} {et~al.}(2005){Burgasser}, {Reid}, {Leggett},
  {Kirkpatrick}, {Liebert}, \& {Burrows}}]{burgasser2005b}
{Burgasser}, A.~J., {Reid}, I.~N., {Leggett}, S.~K., {et~al.} 2005, \apjl, 634,
  L177

\bibitem[{{Burgasser} {et~al.}(1999){Burgasser}, {Kirkpatrick}, {Brown},
  {Reid}, {Gizis}, {Dahn}, {Monet}, {Beichman}, {Liebert}, {Cutri}, \&
  {Skrutskie}}]{burgasser1999}
{Burgasser}, A.~J., {Kirkpatrick}, J.~D., {Brown}, M.~E., {et~al.} 1999, \apjl,
  522, L65

\bibitem[{{Carson} {et~al.}(2011){Carson}, {Marengo}, {Patten}, {Luhman},
  {Sonnett}, {Hora}, {Schuster}, {Allen}, {Fazio}, {Stauffer}, \&
  {Schnupp}}]{carson2011}
{Carson}, J.~C., {Marengo}, M., {Patten}, B.~M., {et~al.} 2011, \apj, 743, 141

\bibitem[{{Chabrier} \& {K{\"u}ker}(2006)}]{chabrierKuker2006}
{Chabrier}, G., \& {K{\"u}ker}, M. 2006, \aap, 446, 1027

\bibitem[{{Christensen} {et~al.}(2009){Christensen}, {Holzwarth}, \&
  {Reiners}}]{christensen2009}
{Christensen}, U.~R., {Holzwarth}, V., \& {Reiners}, A. 2009, \nat, 457, 167

\bibitem[{{Clarke} {et~al.}(2008){Clarke}, {Hodgkin}, {Oppenheimer},
  {Robertson}, \& {Haubois}}]{clarke2008}
{Clarke}, F.~J., {Hodgkin}, S.~T., {Oppenheimer}, B.~R., {Robertson}, J., \&
  {Haubois}, X. 2008, \mnras, 386, 2009

\bibitem[{{Croll} {et~al.}(2016){Croll}, {Muirhead}, {Lichtman}, {Han},
  {Dalba}, \& {Radigan}}]{croll2016}
{Croll}, B., {Muirhead}, P.~S., {Lichtman}, J., {et~al.} 2016, ArXiv e-prints,
  arXiv:1609.03587

\bibitem[{{Cruz} {et~al.}(2003){Cruz}, {Reid}, {Liebert}, {Kirkpatrick}, \&
  {Lowrance}}]{cruz2003}
{Cruz}, K.~L., {Reid}, I.~N., {Liebert}, J., {Kirkpatrick}, J.~D., \&
  {Lowrance}, P.~J. 2003, \aj, 126, 2421

\bibitem[{{Cruz} {et~al.}(2007){Cruz}, {Reid}, {Kirkpatrick}, {Burgasser},
  {Liebert}, {Solomon}, {Schmidt}, {Allen}, {Hawley}, \& {Covey}}]{cruz2007}
{Cruz}, K.~L., {Reid}, I.~N., {Kirkpatrick}, J.~D., {et~al.} 2007, \aj, 133,
  439

\bibitem[{{Dulk}(1985)}]{dulk1985}
{Dulk}, G.~A. 1985, \araa, 23, 169

\bibitem[{{Dupuy} \& {Liu}(2017)}]{dupuy2017}
{Dupuy}, T.~J., \& {Liu}, M.~C. 2017, \apjs, 231, 15

\bibitem[{{Enoch} {et~al.}(2003){Enoch}, {Brown}, \& {Burgasser}}]{enoch2003}
{Enoch}, M.~L., {Brown}, M.~E., \& {Burgasser}, A.~J. 2003, \aj, 126, 1006

\bibitem[{{Gagn{\'e}} {et~al.}(2017){Gagn{\'e}}, {Faherty}, {Burgasser},
  {Artigau}, {Bouchard}, {Albert}, {Lafreni{\`e}re}, {Doyon}, \& {Bardalez
  Gagliuffi}}]{gagne2017}
{Gagn{\'e}}, J., {Faherty}, J.~K., {Burgasser}, A.~J., {et~al.} 2017, \apjl,
  841, L1

\bibitem[{{Gastine} {et~al.}(2012){Gastine}, {Duarte}, \&
  {Wicht}}]{gastine2012}
{Gastine}, T., {Duarte}, L., \& {Wicht}, J. 2012, \aap, 546, A19

\bibitem[{{Gastine} {et~al.}(2013){Gastine}, {Morin}, {Duarte}, {Reiners},
  {Christensen}, \& {Wicht}}]{gastine2013}
{Gastine}, T., {Morin}, J., {Duarte}, L., {et~al.} 2013, \aap, 549, L5

\bibitem[{{Geballe} {et~al.}(2002){Geballe}, {Knapp}, {Leggett}, {Fan},
  {Golimowski}, {Anderson}, {Brinkmann}, {Csabai}, {Gunn}, {Hawley},
  {Hennessy}, {Henry}, {Hill}, {Hindsley}, {Ivezi{\'c}}, {Lupton}, {McDaniel},
  {Munn}, {Narayanan}, {Peng}, {Pier}, {Rockosi}, {Schneider}, {Smith},
  {Strauss}, {Tsvetanov}, {Uomoto}, {York}, \& {Zheng}}]{geballe2002}
{Geballe}, T.~R., {Knapp}, G.~R., {Leggett}, S.~K., {et~al.} 2002, \apj, 564,
  466

\bibitem[{{G\"udel}(1994)}]{gudel1994}
{G\"udel}, M. 1994, \apjs, 90, 743

\bibitem[{{G\"udel} {et~al.}(1993){G\"udel}, {Schmitt}, {Bookbinder}, \&
  {Fleming}}]{gudel1993}
{G\"udel}, M., {Schmitt}, J.~H.~M.~M., {Bookbinder}, J.~A., \& {Fleming}, T.~A.
  1993, \apj, 415, 236

\bibitem[{{Gurnett} {et~al.}(1981){Gurnett}, {Kurth}, \& {Scarf}}]{gurnett1981}
{Gurnett}, D.~A., {Kurth}, W.~S., \& {Scarf}, F.~L. 1981, \nat, 292, 733

\bibitem[{{Hallinan} {et~al.}(2006){Hallinan}, {Antonova}, {Doyle}, {Bourke},
  {Brisken}, \& {Golden}}]{hallinan2006}
{Hallinan}, G., {Antonova}, A., {Doyle}, J.~G., {et~al.} 2006, \apj, 653, 690

\bibitem[{{Hallinan} {et~al.}(2008){Hallinan}, {Antonova}, {Doyle}, {Bourke},
  {Lane}, \& {Golden}}]{hallinan2008}
---. 2008, \apj, 684, 644

\bibitem[{{Hallinan} {et~al.}(2013){Hallinan}, {Sirothia}, {Antonova},
  {Ishwara-Chandra}, {Bourke}, {Doyle}, {Hartman}, \& {Golden}}]{hallinan2013}
{Hallinan}, G., {Sirothia}, S.~K., {Antonova}, A., {et~al.} 2013, \apj, 762, 34

\bibitem[{{Hallinan} {et~al.}(2007){Hallinan}, {Bourke}, {Lane}, {Antonova},
  {Zavala}, {Brisken}, {Boyle}, {Vrba}, {Doyle}, \& {Golden}}]{hallinan2007}
{Hallinan}, G., {Bourke}, S., {Lane}, C., {et~al.} 2007, \apjl, 663, L25

\bibitem[{{Hallinan} {et~al.}(2015){Hallinan}, {Littlefair}, {Cotter},
  {Bourke}, {Harding}, {Pineda}, {Butler}, {Golden}, {Basri}, {Doyle}, {Kao},
  {Berdyugina}, {Kuznetsov}, {Rupen}, \& {Antonova}}]{hallinan2015}
{Hallinan}, G., {Littlefair}, S.~P., {Cotter}, G., {et~al.} 2015, \nat, 523,
  568

\bibitem[{{Hartmann} {et~al.}(2016){Hartmann}, {Herczeg}, \&
  {Calvet}}]{hartmann2016}
{Hartmann}, L., {Herczeg}, G., \& {Calvet}, N. 2016, \araa, 54, 135

\bibitem[{{Johns-Krull} \& {Valenti}(1996)}]{johnskrullValenti1996}
{Johns-Krull}, C.~M., \& {Valenti}, J.~A. 1996, \apjl, 459, L95

\bibitem[{{Johns-Krull} \& {Valenti}(2000)}]{johnskrull2000ASPC}
{Johns-Krull}, C.~M., \& {Valenti}, J.~A. 2000, in Astronomical Society of the
  Pacific Conference Series, Vol. 198, Stellar Clusters and Associations:
  Convection, Rotation, and Dynamos, ed. R.~{Pallavicini}, G.~{Micela}, \&
  S.~{Sciortino}, 371

\bibitem[{{Jones}(2014)}]{jones2014}
{Jones}, C.~A. 2014, \icarus, 241, 148

\bibitem[{{Kao} {et~al.}(2016){Kao}, {Hallinan}, {Pineda}, {Escala},
  {Burgasser}, {Bourke}, \& {Stevenson}}]{kao2016}
{Kao}, M.~M., {Hallinan}, G., {Pineda}, J.~S., {et~al.} 2016, \apj, 818, 24

\bibitem[{{Kervella} {et~al.}(2016){Kervella}, {M{\'e}rand}, {Ledoux},
  {Demory}, \& {Le Bouquin}}]{kervella2016}
{Kervella}, P., {M{\'e}rand}, A., {Ledoux}, C., {Demory}, B.-O., \& {Le
  Bouquin}, J.-B. 2016, \aap, 593, A127

\bibitem[{{Kirkpatrick} {et~al.}(2008){Kirkpatrick}, {Cruz}, {Barman},
  {Burgasser}, {Looper}, {Tinney}, {Gelino}, {Lowrance}, {Liebert},
  {Carpenter}, {Hillenbrand}, \& {Stauffer}}]{kirkpatrick2008}
{Kirkpatrick}, J.~D., {Cruz}, K.~L., {Barman}, T.~S., {et~al.} 2008, \apj, 689,
  1295

\bibitem[{{Kov{\'a}cs} {et~al.}(2002){Kov{\'a}cs}, {Zucker}, \&
  {Mazeh}}]{kovacs2002}
{Kov{\'a}cs}, G., {Zucker}, S., \& {Mazeh}, T. 2002, \aap, 391, 369

\bibitem[{{Kuznetsov} {et~al.}(2012){Kuznetsov}, {Doyle}, {Yu}, {Hallinan},
  {Antonova}, \& {Golden}}]{kuznetsov2012}
{Kuznetsov}, A.~A., {Doyle}, J.~G., {Yu}, S., {et~al.} 2012, \apj, 746, 99

\bibitem[{{Landin} {et~al.}(2010){Landin}, {Mendes}, \& {Vaz}}]{landin2010}
{Landin}, N.~R., {Mendes}, L.~T.~S., \& {Vaz}, L.~P.~R. 2010, \aap, 510, A46

\bibitem[{{Lanza}(2013)}]{lanza2013}
{Lanza}, A.~F. 2013, \aap, 557, A31

\bibitem[{{Lazio} {et~al.}(2010){Lazio}, {Carmichael}, {Clark}, {Elkins},
  {Gudmundsen}, {Mott}, {Szwajkowski}, \& {Hennig}}]{lazio2010}
{Lazio}, T.~J., {Carmichael}, S., {Clark}, J., {et~al.} 2010, \aj, 139, 96

\bibitem[{{Lazio} \& {Farrell}(2007)}]{lazioFarrell2007}
{Lazio}, T.~J., \& {Farrell}, W.~M. 2007, \apj, 668, 1182

\bibitem[{{Leblanc} {et~al.}(2015){Leblanc}, {Modolo}, {Curry}, {Luhmann},
  {Lillis}, {Chaufray}, {Hara}, {McFadden}, {Halekas}, {Eparvier}, {Larson},
  {Connerney}, \& {Jakosky}}]{leblanc2015}
{Leblanc}, F., {Modolo}, R., {Curry}, S., {et~al.} 2015, \grl, 42, 9135

\bibitem[{{Leto} {et~al.}(2016){Leto}, {Trigilio}, {Buemi}, {Umana},
  {Ingallinera}, \& {Cerrigone}}]{leto2016}
{Leto}, P., {Trigilio}, C., {Buemi}, C.~S., {et~al.} 2016, \mnras, 459, 1159

\bibitem[{{Lynch} {et~al.}(2015){Lynch}, {Mutel}, \& {G{\"u}del}}]{lynch2015}
{Lynch}, C., {Mutel}, R.~L., \& {G{\"u}del}, M. 2015, \apj, 802, 106

\bibitem[{{Lynch} {et~al.}(2017){Lynch}, {Murphy}, {Kaplan}, {Ireland}, \&
  {Bell}}]{lynch2017}
{Lynch}, C.~R., {Murphy}, T., {Kaplan}, D.~L., {Ireland}, M., \& {Bell}, M.~E.
  2017, \mnras, 467, 3447

\bibitem[{{Marley} {et~al.}(2010){Marley}, {Saumon}, \&
  {Goldblatt}}]{marley2010}
{Marley}, M.~S., {Saumon}, D., \& {Goldblatt}, C. 2010, \apjl, 723, L117

\bibitem[{MATLAB(R2016a, version 9.0.0.341360)}]{matlabSPTB}
MATLAB. R2016a, version 9.0.0.341360, Signal Processing Toolbox,  Natick, MA,
  USA: The MathWorks Inc.

\bibitem[{{McLean} {et~al.}(2012){McLean}, {Berger}, \& {Reiners}}]{mclean2012}
{McLean}, M., {Berger}, E., \& {Reiners}, A. 2012, \apj, 746, 23

\bibitem[{{McMullin} {et~al.}(2007){McMullin}, {Waters}, {Schiebel}, {Young},
  \& {Golap}}]{mcmullin2007}
{McMullin}, J.~P., {Waters}, B., {Schiebel}, D., {Young}, W., \& {Golap}, K.
  2007, in Astronomical Society of the Pacific Conference Series, Vol. 376,
  Astronomical Data Analysis Software and Systems XVI, ed. R.~A. {Shaw},
  F.~{Hill}, \& D.~J. {Bell}, 127

\bibitem[{{Melrose}(2006)}]{melrose2006}
{Melrose}, D.~B. 2006, \apj, 637, 1113

\bibitem[{{Miles-P{\'a}ez} {et~al.}(2017){Miles-P{\'a}ez}, {Metchev}, {Heinze},
  \& {Apai}}]{miles2017}
{Miles-P{\'a}ez}, P.~A., {Metchev}, S.~A., {Heinze}, A., \& {Apai}, D. 2017,
  \apj, 840, 83

\bibitem[{{Mohanty} \& {Basri}(2003)}]{mohantyBasri2003}
{Mohanty}, S., \& {Basri}, G. 2003, \apj, 583, 451

\bibitem[{{Mohanty} {et~al.}(2002){Mohanty}, {Basri}, {Shu}, {Allard}, \&
  {Chabrier}}]{mohanty2002}
{Mohanty}, S., {Basri}, G., {Shu}, F., {Allard}, F., \& {Chabrier}, G. 2002,
  \apj, 571, 469

\bibitem[{{Molli{\`e}re} \& {Mordasini}(2012)}]{molliere2012}
{Molli{\`e}re}, P., \& {Mordasini}, C. 2012, \aap, 547, A105

\bibitem[{{Morin} {et~al.}(2010){Morin}, {Donati}, {Petit}, {Delfosse},
  {Forveille}, \& {Jardine}}]{morin2010}
{Morin}, J., {Donati}, J.-F., {Petit}, P., {et~al.} 2010, \mnras, 407, 2269

\bibitem[{{Murphy} {et~al.}(2015){Murphy}, {Bell}, {Kaplan}, {Gaensler},
  {Offringa}, {Lenc}, {Hurley-Walker}, {Bernardi}, {Bowman}, {Briggs},
  {Cappallo}, {Corey}, {Deshpande}, {Emrich}, {Goeke}, {Greenhill}, {Hazelton},
  {Hewitt}, {Johnston-Hollitt}, {Kasper}, {Kratzenberg}, {Lonsdale}, {Lynch},
  {McWhirter}, {Mitchell}, {Morales}, {Morgan}, {Oberoi}, {Ord}, {Prabu},
  {Rogers}, {Roshi}, {Shankar}, {Srivani}, {Subrahmanyan}, {Tingay},
  {Waterson}, {Wayth}, {Webster}, {Whitney}, {Williams}, \&
  {Williams}}]{murphy2015}
{Murphy}, T., {Bell}, M.~E., {Kaplan}, D.~L., {et~al.} 2015, \mnras, 446, 2560

\bibitem[{{Nichols} {et~al.}(2012){Nichols}, {Burleigh}, {Casewell}, {Cowley},
  {Wynn}, {Clarke}, \& {West}}]{nichols2012}
{Nichols}, J.~D., {Burleigh}, M.~R., {Casewell}, S.~L., {et~al.} 2012, \apj,
  760, 59

\bibitem[{{Noyes} {et~al.}(1984){Noyes}, {Hartmann}, {Baliunas}, {Duncan}, \&
  {Vaughan}}]{noyes1984}
{Noyes}, R.~W., {Hartmann}, L.~W., {Baliunas}, S.~L., {Duncan}, D.~K., \&
  {Vaughan}, A.~H. 1984, \apj, 279, 763

\bibitem[{{Pallavicini} {et~al.}(1985){Pallavicini}, {Willson}, \&
  {Lang}}]{pallavicini1985}
{Pallavicini}, R., {Willson}, R.~F., \& {Lang}, K.~R. 1985, \aap, 149, 95

\bibitem[{{Parks} {et~al.}(2014){Parks}, {Plavchan}, {White}, \&
  {Gee}}]{parksPlavchan2014}
{Parks}, J.~R., {Plavchan}, P., {White}, R.~J., \& {Gee}, A.~H. 2014, \apjs,
  211, 3

\bibitem[{{Perley} \& {Butler}(2013)}]{perleyButler2013}
{Perley}, R.~A., \& {Butler}, B.~J. 2013, \apjs, 204, 19

\bibitem[{{Pineda} {et~al.}(2017){Pineda}, {Hallinan}, \& {Kao}}]{pineda2017}
{Pineda}, J.~S., {Hallinan}, G., \& {Kao}, M.~M. 2017, \apj, 846, 75

\bibitem[{{Pineda} {et~al.}(2016){Pineda}, {Hallinan}, {Kirkpatrick}, {Cotter},
  {Kao}, \& {Mooley}}]{pineda2016}
{Pineda}, J.~S., {Hallinan}, G., {Kirkpatrick}, J.~D., {et~al.} 2016, \apj,
  826, 73

\bibitem[{{Pizzolato} {et~al.}(2003){Pizzolato}, {Maggio}, {Micela},
  {Sciortino}, \& {Ventura}}]{pizzolato2003}
{Pizzolato}, N., {Maggio}, A., {Micela}, G., {Sciortino}, S., \& {Ventura}, P.
  2003, \aap, 397, 147

\bibitem[{{Plavchan} {et~al.}(2008){Plavchan}, {Jura}, {Kirkpatrick}, {Cutri},
  \& {Gallagher}}]{plavchan2008}
{Plavchan}, P., {Jura}, M., {Kirkpatrick}, J.~D., {Cutri}, R.~M., \&
  {Gallagher}, S.~C. 2008, \apjs, 175, 191

\bibitem[{{Pottelette} {et~al.}(1999){Pottelette}, {Ergun}, {Treumann},
  {Berthomier}, {Carlson}, {McFadden}, \& {Roth}}]{pottelette1999}
{Pottelette}, R., {Ergun}, R.~E., {Treumann}, R.~A., {et~al.} 1999, \grl, 26,
  2629

\bibitem[{{Prato} {et~al.}(2015){Prato}, {Mace}, {Rice}, {McLean},
  {Kirkpatrick}, {Burgasser}, \& {Kim}}]{prato2015}
{Prato}, L., {Mace}, G.~N., {Rice}, E.~L., {et~al.} 2015, \apj, 808, 12

\bibitem[{{Radigan}(2014)}]{radigan2014b}
{Radigan}, J. 2014, \apj, 797, 120

\bibitem[{{Radigan} {et~al.}(2014){Radigan}, {Lafreni{\`e}re}, {Jayawardhana},
  \& {Artigau}}]{radigan2014a}
{Radigan}, J., {Lafreni{\`e}re}, D., {Jayawardhana}, R., \& {Artigau}, E. 2014,
  \apj, 793, 75

\bibitem[{{Ravi} {et~al.}(2011){Ravi}, {Hallinan}, {Hobbs}, \&
  {Champion}}]{ravi2011}
{Ravi}, V., {Hallinan}, G., {Hobbs}, G., \& {Champion}, D.~J. 2011, \apjl, 735,
  L2

\bibitem[{{Reiners}(2012)}]{reiners2012}
{Reiners}, A. 2012, Living Reviews in Solar Physics, 9, 1

\bibitem[{{Reiners} \& {Basri}(2007)}]{reinersBasri2007}
{Reiners}, A., \& {Basri}, G. 2007, \apj, 656, 1121

\bibitem[{{Reiners} \& {Basri}(2008)}]{reinersBasri2008}
---. 2008, \apj, 684, 1390

\bibitem[{{Reiners} \& {Basri}(2009)}]{reinersBasri2009}
---. 2009, \aap, 496, 787

\bibitem[{{Reiners} \& {Basri}(2010)}]{reinersBasri2010}
---. 2010, \apj, 710, 924

\bibitem[{{Reiners} \& {Christensen}(2010)}]{reinersChristensen2010}
{Reiners}, A., \& {Christensen}, U.~R. 2010, \aap, 522, A13

\bibitem[{{Rodriguez-Barrera} {et~al.}(2015){Rodriguez-Barrera}, {Helling},
  {Stark}, \& {Rice}}]{rodriguez-barrera2015}
{Rodriguez-Barrera}, M.~I., {Helling}, C., {Stark}, C.~R., \& {Rice}, A.~M.
  2015, \mnras, 454, 3977

\bibitem[{{Ros{\'e}n} {et~al.}(2015){Ros{\'e}n}, {Kochukhov}, \&
  {Wade}}]{rosen2015}
{Ros{\'e}n}, L., {Kochukhov}, O., \& {Wade}, G.~A. 2015, \apj, 805, 169

\bibitem[{{Route}(2016)}]{route2016}
{Route}, M. 2016, \apjl, 830, L27

\bibitem[{{Route} \& {Wolszczan}(2012)}]{route2012}
{Route}, M., \& {Wolszczan}, A. 2012, \apjl, 747, L22

\bibitem[{{Route} \& {Wolszczan}(2016)}]{routeWolszczan2016}
---. 2016, \apjl, 821, L21

\bibitem[{{Saumon} {et~al.}(1995){Saumon}, {Chabrier}, \& {van
  Horn}}]{saumon1995}
{Saumon}, D., {Chabrier}, G., \& {van Horn}, H.~M. 1995, \apjs, 99, 713

\bibitem[{{Saumon} \& {Marley}(2008)}]{saumonMarley2008}
{Saumon}, D., \& {Marley}, M.~S. 2008, \apj, 689, 1327

\bibitem[{{Schmidt} {et~al.}(2015){Schmidt}, {Hawley}, {West}, {Bochanski},
  {Davenport}, {Ge}, \& {Schneider}}]{schmidt2015}
{Schmidt}, S.~J., {Hawley}, S.~L., {West}, A.~A., {et~al.} 2015, \aj, 149, 158

\bibitem[{{Schmidt} {et~al.}(2010){Schmidt}, {West}, {Hawley}, \&
  {Pineda}}]{schmidt2010}
{Schmidt}, S.~J., {West}, A.~A., {Hawley}, S.~L., \& {Pineda}, J.~S. 2010, \aj,
  139, 1808

\bibitem[{{Schmitt} \& {Rosso}(1988)}]{schmittRosso1988}
{Schmitt}, J.~H.~M.~M., \& {Rosso}, C. 1988, \aap, 191, 99

\bibitem[{{Shulyak} {et~al.}(2017){Shulyak}, {Reiners}, {Engeln}, {Malo},
  {Yadav}, {Morin}, \& {Kochukhov}}]{shulyak2017}
{Shulyak}, D., {Reiners}, A., {Engeln}, A., {et~al.} 2017, Nature Astronomy, 1,
  0184

\bibitem[{{Shulyak} {et~al.}(2010){Shulyak}, {Reiners}, {Wende}, {Kochukhov},
  {Piskunov}, \& {Seifahrt}}]{shulyak2010}
{Shulyak}, D., {Reiners}, A., {Wende}, S., {et~al.} 2010, \aap, 523, A37

\bibitem[{{Skrutskie} {et~al.}(2006){Skrutskie}, {Cutri}, {Stiening},
  {Weinberg}, {Schneider}, {Carpenter}, {Beichman}, {Capps}, {Chester},
  {Elias}, {Huchra}, {Liebert}, {Lonsdale}, {Monet}, {Price}, {Seitzer},
  {Jarrett}, {Kirkpatrick}, {Gizis}, {Howard}, {Evans}, {Fowler}, {Fullmer},
  {Hurt}, {Light}, {Kopan}, {Marsh}, {McCallon}, {Tam}, {Van Dyk}, \&
  {Wheelock}}]{2mass2006}
{Skrutskie}, M.~F., {Cutri}, R.~M., {Stiening}, R., {et~al.} 2006, \aj, 131,
  1163

\bibitem[{{Spiegel} {et~al.}(2011){Spiegel}, {Burrows}, \&
  {Milsom}}]{spiegel2011}
{Spiegel}, D.~S., {Burrows}, A., \& {Milsom}, J.~A. 2011, \apj, 727, 57

\bibitem[{{Stellingwerf}(1978)}]{stellingwerf1978}
{Stellingwerf}, R.~F. 1978, \apj, 224, 953

\bibitem[{{Treumann}(2006)}]{treumann2006}
{Treumann}, R.~A. 2006, \aapr, 13, 229

\bibitem[{{Turnpenney} {et~al.}(2017){Turnpenney}, {Nichols}, {Wynn}, \&
  {Casewell}}]{turnpenney2017}
{Turnpenney}, S., {Nichols}, J.~D., {Wynn}, G.~A., \& {Casewell}, S.~L. 2017,
  \mnras, 470, 4274

\bibitem[{{Ulmschneider}(2003)}]{ulmschneider2003}
{Ulmschneider}, P. 2003, in Lecture Notes in Physics, Berlin Springer Verlag,
  Vol. 619, Lectures on Solar Physics, ed. H.~M. {Antia}, A.~{Bhatnagar}, \&
  P.~{Ulmschneider}, 232

\bibitem[{{Umana} {et~al.}(1998){Umana}, {Trigilio}, \& {Catalano}}]{umana1998}
{Umana}, G., {Trigilio}, C., \& {Catalano}, S. 1998, \aap, 329, 1010

\bibitem[{{Valenti} {et~al.}(1995){Valenti}, {Marcy}, \& {Basri}}]{valenti1995}
{Valenti}, J.~A., {Marcy}, G.~W., \& {Basri}, G. 1995, \apj, 439, 939

\bibitem[{{Vernazza} {et~al.}(1981){Vernazza}, {Avrett}, \&
  {Loeser}}]{vernazza1981}
{Vernazza}, J.~E., {Avrett}, E.~H., \& {Loeser}, R. 1981, \apjs, 45, 635

\bibitem[{{Vidotto} {et~al.}(2013){Vidotto}, {Jardine}, {Morin}, {Donati},
  {Lang}, \& {Russell}}]{vidotto2013}
{Vidotto}, A.~A., {Jardine}, M., {Morin}, J., {et~al.} 2013, \aap, 557, A67

\bibitem[{{Vrba} {et~al.}(2004){Vrba}, {Henden}, {Luginbuhl}, {Guetter},
  {Munn}, {Canzian}, {Burgasser}, {Kirkpatrick}, {Fan}, {Geballe},
  {Golimowski}, {Knapp}, {Leggett}, {Schneider}, \& {Brinkmann}}]{vrba2004}
{Vrba}, F.~J., {Henden}, A.~A., {Luginbuhl}, C.~B., {et~al.} 2004, \aj, 127,
  2948

\bibitem[{{Weinberger} {et~al.}(2016){Weinberger}, {Boss}, {Keiser},
  {Anglada-Escud{\'e}}, {Thompson}, \& {Burley}}]{weinberger2016}
{Weinberger}, A.~J., {Boss}, A.~P., {Keiser}, S.~A., {et~al.} 2016, \aj, 152,
  24

\bibitem[{{White} \& {Franciosini}(1995)}]{whiteFranciosini1995}
{White}, S.~M., \& {Franciosini}, E. 1995, \apj, 444, 342

\bibitem[{{White} {et~al.}(1989){White}, {Kundu}, \& {Jackson}}]{white1989}
{White}, S.~M., {Kundu}, M.~R., \& {Jackson}, P.~D. 1989, \aap, 225, 112

\bibitem[{{Williams} \& {Berger}(2015)}]{williams2015}
{Williams}, P.~K.~G., \& {Berger}, E. 2015, \apj, 808, 189

\bibitem[{{Williams} {et~al.}(2013){Williams}, {Berger}, \&
  {Zauderer}}]{williams2013}
{Williams}, P.~K.~G., {Berger}, E., \& {Zauderer}, B.~A. 2013, \apjl, 767, L30

\bibitem[{{Williams} {et~al.}(2015){Williams}, {Casewell}, {Stark},
  {Littlefair}, {Helling}, \& {Berger}}]{williams2015b}
{Williams}, P.~K.~G., {Casewell}, S.~L., {Stark}, C.~R., {et~al.} 2015, \apj,
  815, 64

\bibitem[{{Williams} {et~al.}(2014){Williams}, {Cook}, \&
  {Berger}}]{williams2014}
{Williams}, P.~K.~G., {Cook}, B.~A., \& {Berger}, E. 2014, \apj, 785, 9

\bibitem[{{Williams} {et~al.}(2017){Williams}, {Gizis}, \&
  {Berger}}]{williams2017}
{Williams}, P.~K.~G., {Gizis}, J.~E., \& {Berger}, E. 2017, \apj, 834, 117

\bibitem[{{Wilson} {et~al.}(2014){Wilson}, {Rajan}, \& {Patience}}]{wilson2014}
{Wilson}, P.~A., {Rajan}, A., \& {Patience}, J. 2014, \aap, 566, A111

\bibitem[{{Wright} \& {Drake}(2016)}]{wrightDrake2016}
{Wright}, N.~J., \& {Drake}, J.~J. 2016, \nat, 535, 526

\bibitem[{{Yadav} {et~al.}(2015){Yadav}, {Christensen}, {Morin}, {Gastine},
  {Reiners}, {Poppenhaeger}, \& {Wolk}}]{yadav2015}
{Yadav}, R.~K., {Christensen}, U.~R., {Morin}, J., {et~al.} 2015, \apjl, 813,
  L31

\bibitem[{{Yadav} {et~al.}(2016){Yadav}, {Christensen}, {Wolk}, \&
  {Poppenhaeger}}]{yadav2016}
{Yadav}, R.~K., {Christensen}, U.~R., {Wolk}, S.~J., \& {Poppenhaeger}, K.
  2016, \apjl, 833, L28

\bibitem[{{Zarka}(1998)}]{zarka1998}
{Zarka}, P. 1998, \jgr, 103, 20159

\end{thebibliography}

\end{document}